\documentclass[11pt,english]{article}

\renewcommand{\Im}{{\rm Im \, }}

\renewcommand{\vec}[1]{\mathbf{#1}}
\newcommand{\CO}{{\cal O}}
\newcommand{\CD}{{\cal D}}
\newcommand{\CL}{{\cal L}}

\newcommand{\CI}{{\cal I}}

\newcommand{\so}{{\mathfrak{so}}}
\newcommand{\SO}{{\text{SO}}}
\newcommand{\CJ}{{\cal J}}

\makeatletter
\newcommand*{\rom}[1]{\expandafter\@slowromancap\romannumeral #1@}
\makeatother

\usepackage[latin9]{inputenc}
\usepackage{geometry}
\usepackage{verbatim}
\usepackage{prettyref}
\usepackage[numbers,sort&compress]{natbib}
\usepackage{amsmath}
\usepackage{amssymb}
\usepackage{lmodern}
\usepackage{cases}
\usepackage{ytableau}
\usepackage{graphicx}
\usepackage{caption}
\usepackage{subcaption}
\usepackage{tikz}

 \usepackage{slashed}
\usepackage{esvect}
\usepackage{dsfont}

\usepackage{color}
\definecolor{darkgreen}{rgb}{0,0.5,0}
\definecolor{darkblue}{rgb}{0,0,0.6}
\definecolor{purple}{rgb}{0.4,.2,0.7}
\usepackage[colorlinks=true,citecolor=darkgreen,linkcolor=purple,urlcolor=purple]{hyperref}

\numberwithin{equation}{section}
\numberwithin{figure}{section}
\numberwithin{table}{section}

\def\CH{{\cal H}}

\def\CD{{\cal D}}
\def\tr{\,{\rm tr}\,}

\DeclareMathOperator{\Tr}{Tr}

\DeclareFontShape{OT1}{cmr}{mx}{n}{<->cmr10}{}

\begin{document}

\fontseries{mx}\selectfont

\begin{center}
\LARGE \bf AdS one-loop partition functions from bulk and edge characters
\end{center}

\vskip1cm

\begin{center}
Zimo Sun
\vskip5mm
{
\it{\footnotesize  Department of Physics, Columbia University}\\
}
\end{center}

\vskip2cm

\vskip0mm

\begin{abstract}
We show that the one-loop partition function of any higher spin field in $(d+1)$-dimensional Anti-de Sitter spacetime can be expressed as an integral transform of an $\text{SO}(2,d)$ bulk character and an $\text{SO}(2,d-2)$ edge character. We apply this character integral formula to various higher-spin Vasiliev gravities and find  miraculous (almost) cancellations between bulk and edge  characters that lead to  agreement with the predictions of HS/CFT holography. We also discuss about the relation between the  character integral representation and
Rindler-AdS thermal partition function. %subject to corrections of edge modes living on horizon captured by the edge character.
\end{abstract}

\newpage

\tableofcontents

\section{Introduction}
The conjectured dualities \cite{Klebanov:2002ja, Sezgin:2003pt} between free/critical CFTs in $\text{U}(N)$ (or $\text{O}(N)$) fundamental representation and Vasiliev higher spin gravities in AdS have stimulated a lot of nontrivial tests (for a recent review of the higher spin/CFT duality see \cite{Giombi:2016ejx}). These tests can be roughly divided into two classes: match bulk tree level three-point functions \cite{Giombi:2012ms, Giombi:2009wh} and match bulk one-loop free energy \cite{Giombi:2013fka, Giombi:2014iua, Giombi:2016pvg, Gunaydin:2016amv, Brust:2016zns, Bae:2016rgm, Skvortsov:2017ldz, Basile:2018zoy, Basile:2018acb}. 

The first  test on one-loop free energy,  for (non)minimal type-A higher spin theory in $\text{AdS}_4$, was carried out by Giombi and Klebanov in \cite{Giombi:2013fka} where the authors used the zeta function regularization for (i) computing the  free energy of a single field and (ii)  summing over free energies of all field content. Then the same method was applied to higher dimensional AdS \cite{Giombi:2014iua}, type-B higher spin gravity \cite{Giombi:2016pvg} and partially massless higher spin fields \cite{Brust:2016zns}. However, one has to work dimension by dimension using this method, in particular for even dimensional AdS, due to the technical difficulty in summing over field content. This difficulty was bypassed in \cite{Basile:2018zoy, Basile:2018acb} where the authors, inspired by earlier works \cite{Bae:2016rgm, Skvortsov:2017ldz}, expressed the higher spin spectral zeta functions, cf. (\ref{spectralzeta}), as an integral transformation of the corresponding characters of the AdS isometry group and apply it to a large class of higher spin theories.

On the other hand, a character integral representation of one-loop free energies on de Sitter spacetime is found in \cite{Anninos:2020hfj} recently by using the heat kernel regularization. For example, for a massive vector field in $\text{dS}_{d+1}$ of mass $m=\sqrt{\frac{(d-2)^2}{4}+\nu^2}$, the one-loop free energy can be expressed as (suppressing the regularization)
\begin{align}\label{example1}
\log Z=\log Z_{\text{bulk}}-\log Z_{\text{edge}}=\int_0^\infty \, dt \, \frac{1}{2 \, u}\frac{1+e^{-u}}{1-e^{-u}}\left(\chi^{\text{dS}}_{\text{bulk}}(u)-\chi^{\text{dS}}_{\text{edge}}(u)\right)
\end{align}
where 
\begin{align}
\chi^{\text{dS}}_{\text{bulk}}(u)=d \frac{e^{-(\frac{d}{2}+i \nu)u}+e^{-(\frac{d}{2}-i \nu)u}}{(1-e^{-u})^d}, \,\,\,\,\,\chi^{\text{dS}}_{\text{edge}}(u)=\frac{e^{-(\frac{d-2}{2}+i \nu)u}+e^{-(\frac{d-2}{2}-i \nu)u}}{(1-e^{-u})^{d-2}}
\end{align}
The bulk character $\chi^{\text{dS}}_{\text{bulk}}(u)$ is the Harish-Chandra character of the spin-$1$, $\Delta\!=\!\frac{d}{2}\!+\!i\nu$ representation of $\text{SO}(1, d\!+\!1)$ and the edge character $\chi^{\text{dS}}_{\text{edge}}(u)$ is the Harish-Chandra character of the spin-$0$, $\Delta\!=\!\frac{d\!-\!2}{2}\!+\!i\nu$ representation of $\text{SO}(1, d\!-\!1)$.
Apart from the standard meaning as the one-loop path integral on $S^{d+1}$, which is the Euclidean Wick rotation of $\text{dS}_{d+1}$,  it is also pointed out in the same paper that the character integral representation (\ref{example1}) can alternatively be interpreted as the bulk thermal free energy in the static patch of $\text{dS}_{d+1}$ subject to possible corrections from edge modes living on the cosmological horizon.  The appearance of edge modes here should not be surprising. In fact, it was  already noticed in \cite{Donnelly:2014fua,Donnelly:2015hxa,Blommaert:2018rsf} that edge modes can be used to explain the long-standing discrepancy between the Euclidean path integral and bulk canonical definitions of the entanglement entropy for Maxwell field in Rindler space \cite{Kabat:1995eq} . 

 In this paper, we extend this dS character integral representation to AdS spacetime. Essentially this extension amounts to replacing the bulk $\text{SO}(1, d\!+\!1)$  characters by the corresponding $\text{SO}(2, d)$ characters and the replacing the edge $\text{SO}(1, d\!-\!1)$  characters by the corresponding $\text{SO}(2, d-2)$ characters. For example, for a massive vector field in $\text{AdS}_{d+1}$, the unregularized free energy (the properly regularized version will be derived in section \ref{Contour}) can be expressed as
\begin{align}\label{evenZZ}
\log Z=\log Z_{\text{bulk}}-\log Z_{\text{edge}}=\int_0^\infty \, du \, \frac{1}{2t}\frac{1+e^{-u}}{1-e^{-u}}\left(\chi^{\text{AdS}}_{\text{bulk}}(u)-\chi^{\text{AdS}}_{\text{edge}}(u)\right)
\end{align}
when $d$ is odd and 
\begin{align}\label{oddZZ}
\log Z=\log Z_{\text{bulk}}-\log Z_{\text{edge}}=\log R\, \oint_{C} \frac{du}{2\pi i}\,\left( \frac{1}{2u}\frac{1+e^{-u}}{1-e^{-u}}\left(\chi^{\text{AdS}}_{\text{bulk}}(u)-\chi^{\text{AdS}}_{\text{edge}}(u)\right)\right)
\end{align}
when $d$ is even, where $R$ is an IR cutoff for the infinite AdS volume and $C$ is a small counterclockwise circle around $u=0$.  In this case, the AdS bulk character, corresponding to the spin-$1$, $\Delta=\frac{d}{2}+\nu$ representation of $\text{SO}(2,d)$ is 
\begin{align}
\chi^{\text{AdS}}_{\text{bulk}}(u)=d \frac{e^{-(\frac{d}{2}+\nu)u}}{(1-e^{-u})^d}
\end{align}
and the AdS edge character, corresponding to spin-$0$, $\Delta=\frac{d-2}{2}+\nu$ representation of $\text{SO}(2,d-2)$ is 
\begin{align}
\chi^{\text{AdS}}_{\text{edge}}(u)=\frac{e^{-(\frac{d-2}{2}+\nu)u}}{(1-e^{-u})^{d-2}}
\end{align}
Compared to the character integral representation found in \cite{Basile:2018zoy, Basile:2018acb}, eq. (\ref{evenZZ}) or (\ref{oddZZ}) seems quite different and considerably simpler because there is no angular integrals associated to the $\text{SO}(d)$ Cartan subgroup.  However, in all cases we have checked, explicit evaluation of the angular integrals in \cite{Basile:2018zoy, Basile:2018acb} reproduces (the UV-finite part of) the character integral representations derived in this paper. In particular, when applied to Vasiliev higher spin theories, the two methods completely agree, though in our setup the (almost) vanishing of total one-loop free energy is a result of the  miraculous (almost) cancellations between the total bulk characters and the total edge characters.  For nonmiminal type-A theory, the cancellation between the total bulk character and the total edge character is {\it exact}. For example, in $\text{AdS}_4$, the total bulk character is 
\small
\begin{align}
\chi^{\text{AdS}_4}_{\text{bulk}}(u)=\frac{e^{-u}}{(1-e^{-u})^3}+\sum_{s \ge 1}\frac{(2s+1)e^{-(1+s)u}-(2s-1)e^{-(s+2)u}}{(1-e^{-u})^3}=\left[\frac{e^{-\frac{1}{2}u}+e^{-\frac{3}{2}u}}{(1-e^{-u})^2}\right]^2
\end{align}
\normalsize
and the total edge character is 
\small
\begin{align}
\chi^{\text{AdS}_4}_{\text{edge}}(u)=\sum_{s \ge 1}\frac{s(s+1)(2s+1)}{6}\frac{e^{-s\, u}}{1-e^{-u}}-\sum_{s \ge 2}\frac{s(s-1)(2s-1)}{6}\frac{e^{-(s+1)\, u}}{1-e^{-u}}=\left[\frac{e^{-\frac{1}{2}u}+e^{-\frac{3}{2}u}}{(1-e^{-u})^2}\right]^2
\end{align}
\normalsize
For minimal type-A theory, the  cancellation between bulk and edge characters yields an $\SO(1, d)$ Harish-Chandra character corresponding  a conformally couple  scalar, cf. (\ref{Amin}) 
\begin{align}
\log Z^{\text{AdS}_{d+1}}_{\text{A}^{\text{min}}}=\int^\infty_0 \frac{du}{2u}\frac{1+e^{-u}}{1-e^{-u}}\frac{e^{-\left(\frac{d-1}{2}+\frac{1}{2}\right)u}+e^{-\left(\frac{d-1}{2}-\frac{1}{2}\right)u}}{(1-e^{-u})^{d-1}}
\end{align}
The dS character integral representation   of sphere partition functions obtained in \cite{Anninos:2020hfj} directly
identifies $Z^{\text{AdS}_{d+1}}_{\text{A}^{\text{min}}}$ as the one-loop partition function of a conformally coupled scalar on $S^d$, without having to evaluate the integral explicitly. Thus, by combing the AdS and dS character integral representations, the boundary theory interpretation of $Z^{\text{AdS}_{d+1}}_{\text{A}^{\text{min}}}$ becomes manifest.
Similarly, for nonminimal type-B theory, the cancellation between the bulk and edge characters manifestly exhibits a double-trace deformation structure, cf. (\ref{genB})
\begin{align}
\log Z^{\text{AdS}_{d+1}}_{\text{B}}=(-)^{\frac{d-1}{2}}\int^\infty_0\frac{du}{2u}\frac{1+e^{-u}}{1-e^{-u}}\frac{e^{-\frac{d-1}{2}u}-e^{-\frac{d+1}{2}u}}{(1-e^{-u})^{d}}
\end{align}
Apart from these technical advantages, our character integral representation admits a thermal interpretation (at least for even dimensional AdS). More explicitly, we argue that the bulk part of the one-loop path integral on $\text{EAdS}_{d+1}$ can be interpreted as the quasi-canonical partition function in the Rindler patch  of Lorentzian AdS \cite{Parikh:2012kg} (consult the appendix \ref{coodAdS} for details about the Rindler patch which is also called Rindler-AdS) and the edge part is associated to edge modes localized on the horizon of the Rindler patch.

The organization of the paper is as follows. In section \ref{review}, we review the one-loop partition function and heat kernels in $\text{AdS}_{d+1}$. 
In section \ref{WU} and \ref{HSchar}, we show a very simple but not rigorous derivation of the unregularized character integral representation in the case of even dimensional AdS. In section \ref{Contour}, we give a rigorous version of the regularized character integral formula that works for both even and odd dimensional AdS by adding a UV-regulator and specifying the appropriate integral contour.  An explicit evaluation of the regularized character integral is done in section \ref{rigovalue}. In section \ref{Double}, we derive the effect of (higher spin) double trace deformation on one-loop free energy. In section \ref{Vasiliev}, we review various Flato-Fronsdal theorems which fix the free spectrum of higher spin theory group theoretically and apply the character integral formula to (non)minimal type-A/B Vasiliev theories. In section \ref{th}, we comment on the thermal interpretation of the character integral representation. Finally, appendices contain various technical results, some of which are of  independent interest. For example, in appendix \ref{Planch}, we give a physical derivation of the $\SO(1, d+1)$ Plancherel measure supported on the scalar principal series. In appendix \ref{comchar}, we compute the $\SO(2,1)$ Harish-Chandra character corresponding to unitary highest-weight representations and in appendix \ref{GG}, we explain some of the physics it encodes, including the quasinormal mode spectrum and normal mode density of Rindler-AdS.

\section{One-loop partition functions and heat kernels on AdS}\label{review}
The one-loop partition function $Z$ of a (real bosonic) quantum field theory is given by a functional determinant
\begin{align}
\log Z=-\frac{1}{2}\log \det(\CD)
\end{align}
where $\CD$ is the ``Laplacian'' in the quadratic Lagrangian $\CL_2=\frac{1}{2}\phi\CD\phi$. (There might be some nontrivial factors that are not captured by the functional determinant  due to subtleties like zero modes   when the quantum field theory is defined on a compact manifold. We ignore these subtleties in this general discussion as they will not appear in this paper). When the theory has a gauge symmetry, we should also subtract the functional determinant of the corresponding ghost field. In general, the functional determinant is UV-divergent and  needs to be regularized. The regularization scheme for $\log Z$ we'll use in this paper is 
\begin{align}\label{KtoZ}
\log Z=\frac{1}{2}\int_0^\infty\, \frac{dt}{t} \,e^{-\frac{\epsilon^2}{4t}} \, K_\CD(t)
\end{align}
where $K_\CD(t)\equiv \Tr e^{-t \CD}$ is the heat kernel of $\CD$. In terms of the spectrum of $\CD$, the heat kernel is formally \footnote{If the spectrum of $\CD$ is continuous, the sum over $n$ gets replaced by an integral and the degeneracy $d_n$ gets replaced by the density of eigenmodes.} defined as 
\begin{align}
K_\CD(t)=\sum_{n\ge 0} d_n e^{-t\lambda_n}
\end{align}
where $\lambda_n$ is the eigenvalue of $\CD$ of degeneracy $d_n$. Performing a Mellin transformation for the  heat kernel $K_{\CD}(t)$ yields the spectral zeta function 
\begin{align}\label{spectralzeta}
\zeta_D(z)\equiv \sum_{n\ge 0}\frac{d_n}{\lambda_n^z}=\frac{1}{\Gamma(z)}\int_0^\infty \,\frac{dt}{t}\, t^z K_{\CD}(t)
\end{align}
which is also a common tool to regularize partition function.

Given a free field in $\text{AdS}_{d+1}$ carrying a generic  unitary irreducible representation (UIR) $[\Delta, \vec s]$ of $\text{SO}(2, d)$ with $\Delta=\frac{d}{2}+\nu$, the corresponding heat kernel $K_{\vec s, \nu}(t)$ is constructed explicitly in \cite{Gopakumar:2011qs} by a group theoretical method. Alternatively, we can infer heat kernel from the associated spectral zeta function \cite{ Basile:2018zoy} by an inverse Mellin transformation: 
\begin{align}\label{genZ}
\log Z_{\vec s, \nu}=\frac{\text{Vol}(\text{AdS}_{d+1})}{\text{Vol}(S^d)}\frac{D_{\vec s}^{d}}{2^{d-1}\Gamma(\frac{d+1}{2})^2}\int_0^\infty\, \frac{dt}{2t}\, e^{-\frac{\epsilon^2}{4t}}\, \int_0^\infty\, d\lambda\,  \mu^{(d)}_{\vec s}(\lambda) e^{-t(\lambda^2+\nu^2)}
\end{align}
Explanations of  the various notations appearing in eq. (\ref{genZ}) are given as follows:
\begin{itemize}
\item $\text{Vol}(S^d)$: the volume of a $d$-dimensional sphere. 
\begin{align}
\text{Vol}(S^d)=\frac{2\pi^{\frac{d+1}{2}}}{\Gamma(\frac{d+1}{2})}
\end{align}
\item $\text{Vol}(\text{AdS}_{d+1})$: the regularized volume of a $(d+1)$-dimensional Euclidean AdS \cite{Diaz:2007an, Casini:2011kv, Casini:2010kt, Giombi:2013yva}. For example, in \cite{Diaz:2007an} the authors used unit ball realization of Euclidean AdS and computed the volume by dimensional regularization. 
\begin{align}
\text{Vol}(\text{AdS}_{d+1})=\begin{cases} \frac{2(-\pi )^{\frac{d}{2}}}{\Gamma(\frac{d}{2}+1)}\log R,\,\,\,\, &d \,\, \text{even}
\\ \pi^{\frac{d}{2}}\Gamma(-\frac{d}{2}),\,\,\,\, &d\,\, \text{odd} \end{cases}
\end{align}
\item $D_{\vec s}^{d}$: the dimension of  $\so(d)$ representation of highest weight vector $\vec s=(s_1, s_2, \cdots s_{[\frac{d}{2}]})$ (our convention for highest weight vectors is $s_1\ge s_2\ge \cdots\ge s_{[\frac{d}{2}]-1}\ge |s_{[\frac{d}{2}]}|$). For example, when $\vec s=(s, 0, \cdots, 0)$, it gives the dimension of spin-$s$ representation
\begin{align}
D^{d}_s=\frac{(d+2 s-2) \Gamma (d+s-2)}{\Gamma (d-1) \Gamma (s+1)}
\end{align}
and when $\vec s=(n, s, 0, \cdots, 0)$, it gives
\begin{align}\label{Dns}
D^{d}_{n, s}=\frac{(d+2 n-2) (d+2 s-4) (n-s+1) (d+n-4)! (d+s-5)! (d+n+s-3)}{(d-4)! (d-2)! (n+1)! s!}
\end{align}
When we interpret $D^{d}_{n, s}$ as the dimension of an $\so(d)$ representation, $n$ cannot be smaller than $s$. On the other hand, $D^{d}_{n, s}$ as a function of $n, s$ defined by (\ref{Dns}), admits a meromorphic continuation to $\mathbb{C}^2$. Using the second point of view, we can find some interesting and useful relations like $D^{d}_{n-1, s}=-D^{d}_{s-1, n}$ which holds for all $n, s\in \mathbb{C}$.

\item $\mu^{(d)}_{\vec s}(\lambda)$: the spin-$\vec s$ spectral density up to normalization \cite{Camporesi:1994ga} (but we still call it the ``spectral density'' for simplicity)
\begin{align}\label{Plancherel measure}
\mu^{(d)}_{\vec s}(\lambda)=\begin{cases}\prod_{j=1}^r(\lambda^2+\ell_j^2), \,\,\,\, & d=2r\\
\prod_{j=1}^r(\lambda^2+\ell_j^2)\lambda \tanh^\epsilon(\pi\lambda), \,\,\,\, & d=2r+1
\end{cases}
\end{align}
where $\ell_j\equiv s_j+\frac{d}{2}-j$, and $\epsilon=\pm 1$ for bosonic/fermionic fields. The $\so(d)$ spin label $\vec s$ will be dropped when we deal with a scalar field. We focus on bosonic fields in the main text of this paper and an example about Dirac spinors can be found in the appendix \ref{spinor}. The spectral density $\mu^{(d)}_{\vec s}(\lambda)$ with a Wick rotation  $\lambda\to i\lambda$, is also the Plancherel measure on the principal series of $\text{SO}(1, d+1)$ \cite{Dobrev:1977qv}. The simplest way to derive it \cite{Gopakumar:2011qs}  is based on an analytical continuation of the $\text{SO}(d+2)$ Plancherel measure, which is proportional to the dimension of $\text{SO}(d+2)$ representation (we  also provide a physical derivation/interpretation of the spectral density in the appendix \ref{Planch}). As a result of this analytical continuation, the polynomial part $P_{\vec s}(\lambda)\equiv\lambda\prod_{j=1}^r(\lambda^2+\ell_j^2)$ of the spectral density also appears in the following formula that relates $D_{(s_0, \vec s)}^{d+2}$ and $D_{\vec s}^{d}$ \cite{Basile:2018zoy}
\begin{align}\label{d2tod}
D^{d+2}_{(s_0, \vec s)}=\frac{2 D^{d}_{\vec s}}{d!}\left(s_0+\frac{d}{2}\right)\prod_{j=1}^r\left[\left(s_0+\frac{d}{2}\right)^2-\ell_j^2\right]
\end{align}
with $s_0$ replaced by $-\frac{d}{2}+i\lambda$.
This equation is extremely useful when we compare the character integral representations for AdS and dS in appendix \ref{dSAdS}.
\end{itemize}

The various $\Gamma$-functions in  (\ref{genZ}) can be greatly simplified for integer dimension $d$ and we're left with
\begin{align}\label{logZv2}
\log Z_{\vec s, \nu}=\frac{D_{\vec s}^{d}}{d!}\int_0^\infty\, \frac{dt}{t}\, e^{-\frac{\epsilon^2}{4t}}\, \int_0^\infty \, d\lambda\,  \mu^{(d)}_{\vec s}(\lambda)\, e^{-t(\lambda^2+\nu^2)}\times \begin{cases} \frac{(-)^{\frac{d}{2}}\log R}{\pi}, \,\,\,\, &d \,\, \text{even}\\ \frac{(-)^{\frac{d+1}{2}}}{2}, \,\,\,\, & d \,\, \text{odd}\end{cases}
\end{align}
Starting from this equation, we'll show that all partition functions $\log Z_{\vec s, \nu}$ can be expressed as a integral transformations of $\text{SO}(2,d)$ characters up to edge mode corrections.

\section{Warm-up example: scalar fields in even dimensional AdS}\label{WU}
\subsection{Scalar fields in $\text{AdS}_2$}\label{warmup}
As a warm-up, let's consider a scalar field $\varphi$ of mass $m^2=\nu^2-\frac{1}{4}$ in $\text{AdS}_2$ which corresponds to the scaling dimension $\Delta_+=\frac{1}{2}+\nu$  representation of the isometry group  $\text{SO}(2, 1)$. Applying (\ref{logZv2}) to this field yields the partition function
\begin{align}
\log Z_\nu=-\frac{1}{4}\int_0^\infty\frac{dt}{t}e^{-\frac{\epsilon^2}{4t}}\int_{-\infty}^\infty d\lambda\, \lambda \tanh(\pi\lambda)e^{-t (\lambda^2+\nu^2)}
\end{align}
where we've extended the integration domain of $\lambda$ to the whole real line.  To perform the integral over $\lambda$, we use the Hubbard-Stratonovich trick:
\begin{align}\label{LogZ}
\log Z_\nu&=-\frac{1}{4}\int_{-\infty}^\infty du\int_0^\infty\frac{dt}{2\,\sqrt{\pi}\, t^{3/2}}e^{-\frac{\epsilon^2+u^2}{4t}-t\nu^2}\int_{-\infty}^\infty d\lambda\, \lambda \tanh(\pi\lambda)e^{i\lambda u}\nonumber\\
&=-\frac{1}{2}\int_{0}^\infty \frac{du}{\sqrt{\epsilon^2+u^2}}e^{-\nu\sqrt{\epsilon^2+u^2}}\, W(u)
\end{align}
where 
\begin{align}\label{W1}
W(u)\equiv\int_{-\infty}^\infty d\lambda\, \lambda \tanh(\pi\lambda)e^{i\lambda u}
\end{align}
The naive Fourier transformation (\ref{W1}) is ill-defined. However for our practical purpose i.e. to get the character integral formula as soon as possible, we pretend that it's well-defined and  the contour can be closed at infinity. A more rigorous treatment is postponed until section \ref{Contour} and \ref{rigovalue} where we give a fully regularized character integral and  justify our naive result obtained here is indeed reasonable and sufficient for most applications, in particular the total partition function of Vasiliev theories.  From now on, keeping the above comments in mind, we're free to close the $\lambda$-contour in $W(u)$ in either upper or lower half-plane depending on the sign of $u$ and we also write the partition function $\log Z_\nu$ in the unregularized form by putting $\epsilon=0$ formally. When $u>0$, we close the contour in the upper half-plane picking up simple poles at $\lambda=i(n+\frac{1}{2}), n\in\mathbb {N}$ and when $u<0$, we close the contour in the lower half-plane picking up simple poles at $\lambda=-i(n+\frac{1}{2}), n\in\mathbb {N}$.  By summing over residues in both cases, we find 
\begin{align}
W(u)=-\frac{1+e^{-u}}{1-e^{-u}}\frac{e^{-\frac{1}{2}u}}{1-e^{-u}}
\end{align}
Thus the unregularized partition function is given by
\begin{align}\label{Zn2}
\log Z_\nu&=\int_0^\infty \frac{du}{2u}\,\frac{1+e^{-u}}{1-e^{-u}}\,\chi_{\Delta_+}^{\text{AdS}_2}(u), \,\,\,\, \chi_{\Delta_+}^{\text{AdS}_2}(u)=\frac{e^{-\Delta_+ u}}{1-e^{-u}}
\end{align}
where $ \chi_{\Delta_+}^{\text{AdS}_2}(u)$ is the character of scaling dimension $\Delta_+$ representation of $\text{SO(2,1)}$. Before moving to the higher dimensional examples, let's notice that when we evaluate the $t$-integral in eq. (\ref{LogZ}), we implicitly assume  $\nu>0$, so $\Delta_+=\frac{1}{2}+\nu$ corresponds to  the ``standard quantization'' in bulk. On the other hand, we can safely send $\nu$ to $-\nu$  in eq. (\ref{Zn2}) by analytic continuation, as long as $0<\nu<\frac{1}{2}$. Altogether we  conclude that, with the standard boundary condition, the  partition function of $\varphi$ is 
\begin{align}
\log Z_\nu&=\int_0^\infty \frac{du}{2u}\,\frac{1+e^{-u}}{1-e^{-u}}\,\chi_{\Delta_+}^{\text{AdS}_2}(u)\end{align}
and with the alternate boundary condition, the partition function is 
\begin{align}
\log Z_{-\nu}&=\int_0^\infty \frac{du}{2u}\,\frac{1+e^{-u}}{1-e^{-u}}\,\chi_{\Delta_-}^{\text{AdS}_2}(u)\end{align}
where
$\Delta_-=\frac{1}{2}-\nu$.

\subsection{Scalar fields in  $\text{AdS}_{2r+2}$}
The  computation above in $\text{AdS}_2$ can be  generalized straightforwardly to scalar field in  any $\text{AdS}_{d+1}$ with $d=2r+1$ odd. Assuming that the scalar field has scaling dimension $\Delta_+=\frac{d}{2}+\nu$, its partition function is given by 
\begin{align}\label{SiAdS}
\log Z_\nu=\frac{(-)^{r+1}}{2\, d!}\int_0^\infty\frac{dt}{t}e^{-\frac{\epsilon^2}{4t}}\int_0^\infty d\lambda\, \mu^{(d)}(\lambda)\,e^{-t(\lambda^2+\nu^2)}
\end{align}
where the scalar spectral density is 
\begin{align}\label{scalarspecden}
\mu^{(d)}(\lambda)=\prod_{j=0}^{r-1}\left[\lambda^2+\left(j+\frac{1}{2}\right)^2\right]\lambda\tanh(\pi\lambda)
\end{align}
Following the same steps as in section \ref{warmup}, we obtain 
\begin{align}\label{Zd}
\log Z_\nu=\frac{(-)^{r+1}}{\, d!} \int_{0}^\infty\frac{du}{2u}\, W^{(d)}(u)\, e^{-\nu\, u}
\end{align}
where $W^{(d)}(u)\equiv \int_{-\infty}^\infty d\lambda\, \mu^{(d)}(\lambda)e^{i\lambda u}$.
The ``Fourier transform''  $W^{(d)}(u)$ can be evaluated according to the comments below eq. (\ref{W1})
\begin{align}\label{Wd}
W^{(d)}(u)=(-)^{r+1}d!\frac{e^{-\frac{d}{2}u}}{(1-e^{-u})^d}\frac{1+e^{-u}}{1-e^{-u}}
\end{align}
Plugging (\ref{Wd}) into (\ref{Zd}) yields a character integral expression for the unregularized $\log Z_\nu$
\begin{align}\label{scalar}
\log Z_\nu=\int_0^\infty\frac{du}{2u}\,\frac{1+e^{-u}}{1-e^{-u}}\,\chi^{\text{AdS}_{d+1}}_{\Delta_+}(u), \,\,\,\,\chi^{\text{AdS}_{d+1}}_{\Delta_+}(u)=\frac{e^{-\Delta_+ u}}{(1-e^{-u})^d}
\end{align}
where $ \chi_{\Delta_+}^{\text{AdS}_{d+1}}(u)$ is the character of scaling dimension $\Delta_+$ representation of $\text{SO(2,d)}$. Before turning to the higher spin case, let's comment on the relation between the character integral and the original heat kernel integral. The heat kernel in AdS is defined through the spectral density $\mu(\lambda)$ whose explicit construction is given in \cite{Camporesi:1994ga}. Briefly speaking, the authors of \cite{Camporesi:1994ga} found a complete set of $\delta$-function normalizable eigenfunctions of the Laplacian operator $-\nabla^2$ in $\text{EAdS}_{d+1}$:
\begin{align}
-\nabla^2 h^{(\lambda \sigma)}(x)&=\left(\frac{d^2}{4}+\lambda^2\right) h^{(\lambda \sigma)}(x)\label{hlm1}\\
\langle h^{(\lambda \sigma)}, h^{(\lambda \sigma')}\rangle&\equiv \int d^{d+1}x\sqrt{g}\, h^{(\lambda \sigma) *}(x) h^{(\lambda \sigma')}(x)=\delta_{\sigma \sigma'}\delta(\lambda-\lambda')\label{hlm2}
\end{align}
where $\sigma$ is a discrete label for distinguishing eigenfunctions of the same $\lambda$. (Note that the inner product for $ h^{(\lambda \sigma)}$ involves an integration over the whole EAdS rather than a spatial slice as in the standard Klein-Gordon inner product). The spectral density is defined via these eigenfunctions: $\mu(\lambda)\propto \sum_\sigma h^{(\lambda \sigma) *}h^{(\lambda \sigma')}(0)$. Therefore the original heat kernel method involves an integral over the whole continuous spectrum labeled by $(\lambda, \sigma)$. On the other hand, the character can be expanded into a discrete sum
\begin{align}
\chi^{\text{AdS}_{d+1}}_{\Delta_+}(u)=\sum_{n\ge 0}\binom{d+n-1}{d-1}e^{-(\Delta_++n)u}
\end{align}
This expansion encodes a whole tower of solutions to the equation of motion $(-\nabla^2+\Delta_+(\Delta_+-d))\phi=0$ in global AdS that furnish a representation of $\text{SO}(2, d)$. More explicitly, $\phi_0=\frac{e^{-i\Delta_+ t}}{(1+r^2)^{\Delta_+/2}}$ is the primary mode, i.e. ground state, in the global coordinate: $ds^2=-(1+r^2)dt^2+\frac{dr^2}{1+r^2}+r^2 d\Omega^2$.  It solves the equation of motion, falls like $r^{-\Delta_+}$ at the boundary but its Wick rotation under $t\to -i\tau$ is not normalizable in the sense of (\ref{hlm2}). By acting the conformal algebra $\so(2,d)$ on $\phi_0$ repeatedly, we get a collection of modes that also solve the equation of motion and have the same boundary condition. At each frequency $\omega_n=\Delta_++n$, the degeneracy of these modes are exactly $\binom{d+n-1}{d-1}$. Therefore while switching from the heat kernel integral to the character integral, we effectively turn a {\it continuous} spectrum into a {\it discrete} spectrum  and curiously both of them encode the information of partition function. This observation is the main point of \cite{Keeler:2014hba}. Actually the character integral representation we found is equivalent to the  ``zero mode method'' used in that paper. For example, using the unregularized expression (\ref{scalar}), we  obtain formally
\begin{align}
\log Z_\nu=\sum_{n\ge 0} D^{d+2}_n \int_{0}^\infty \frac{du}{2u} e^{-(\Delta_++n)u}=-\frac{1}{2}\sum_{n\ge 0} D^{d+2}_n \, \log (\Delta_++n)
\end{align}
which recovers the result in \cite{Keeler:2014hba} up to some holomorphic function denoted by $\text{Pol}(\Delta_+)$ there.  $\text{Pol}(\Delta_+)$ is a polynomial in $\Delta_+$ and depends on the UV-cutoff. In section \ref{rigovalue}, we'll show that it can also be recovered if we use the fully regularized character integral.

\section{Higher spin fields in $\text{AdS}_{2r+2}$}\label{HSchar}
In this section, we turn to the character integral representation of higher spin fields in $\text{AdS}_{d+1}$ with $d=2r+1$\footnote{We will focus on the $r\ge 1$ case because there is a discrete spectrum for each higher spin STT Laplacian in $\text{AdS}_2$ which corresponds to the discrete series of $\SO(2,1)$ and doesn't have any analogue in higher dimensions \cite{Camporesi:1994ga, Banerjee:2010qc}.}. Unlike scalar fields, a spin-$s$ field $\varphi_{\mu_1\cdots\mu_s}$ in AdS can carry either massive or massless irreducible representation \cite{Basile:2016aen} depending on the scaling dimension. When $\Delta=\Delta_{s, t}\equiv d\!+\!t\!-\!1$ with $t\in\{0, 1, \cdots, s-1\}$, $\varphi_{\mu_1\cdots\mu_s}$ is called a partially massless (PM) field of depth $t$ and it has a gauge symmetry $\delta\varphi_{\mu_1\cdots\mu_s}=\nabla_{(\mu_{t+1}\cdots \mu_s}\xi_{\mu_1\cdots\mu_t)}+\cdots$ \cite{Brust:2016zns}. In this case, we should include the contribution of the ghost field, which has  spin-$t$ and scaling dimension $\Delta_{t, s}=d+s-1$, in the one-loop partition function. When $\Delta=\frac{d}{2}+\nu$ is not in the discrete set $\{\Delta_{s,t}\}$, the field $\varphi_{\mu_1\cdots\mu_s}$ falls into the massive representations and doesn't have gauge symmetry. Due to the emergence of gauge symmetry, the characters corresponding to massive and massless representations take very different forms
\begin{align}
&\text{massive}\,\,\,\, [\Delta, s]: \chi^{\text{AdS}_{d+1}}_{[\Delta, s]}(u)=D_s^{d}\frac{e^{-\Delta u}}{(1-e^{-u})^d}\nonumber\\
&\text{massless}\,\,\,\, [\Delta_{s, t}, s]: \chi^{\text{AdS}_{d+1}}_{[\Delta_{s, t}, s]}(u)=\frac{D_s^{d} e^{-(d+t-1)u}-D_t^{d}e^{-(d+s-1)u}}{(1-e^{-u})^d}
\end{align}

Let's start from computing the partition function of a spin-$s$ field in the massive representations
\begin{align}
\log Z_{s, \nu}=\frac{(-)^{\frac{d+1}{2}}}{2}\frac{D_s^{d}}{ d!}\int_0^\infty \frac{dt}{t} \,e^{-\frac{\epsilon^2}{4t}}\int^\infty_0 d\lambda\, \mu^{(d)}_{s}(\lambda)\,e^{-t(\lambda^2+\nu^2)}
\end{align}
where the spin-$s$ spectrum density is
\begin{align}
\mu^{(d)}_s(\lambda)=\prod_{j=0}^{r-2}\left[\lambda^2+\left(j+\frac{1}{2}\right)^2\right]\left[\lambda^2+\left(s+r-\frac{1}{2}\right)^2\right]\lambda\tanh(\pi\lambda)
\end{align}
Following the same steps as in the scalar case, we obtain 
\begin{align}\label{red}
\log Z_{s, \nu}=\frac{(-)^{\frac{d+1}{2}}}{ d!}D^{d}_s\int^\infty_{0}\frac{du}{2u}\,W_s^{(d)}(u)\,e^{-\nu \,u}
\end{align}
where $W^{(d)}_s(u)\equiv \int^\infty_{-\infty} d\lambda \, \mu_s^{(d)}(\lambda)\, e^{i\lambda u}$ is an even function in $u$ and when $u>0$, it is
\begin{align}\label{Wdd}
W_s^{(d)}(u)=(-)^{\frac{d-1}{2}}(d-2)!\frac{e^{-(\frac{d}{2}-1)u}}{(1-e^{-u})^d}\frac{1+e^{-u}}{1-e^{-u}}\left[s(s+d-2)(1-e^{-u})^2-d(d-1)e^{-u}\right]
\end{align}
Plugging  (\ref{Wdd}) into (\ref{red})  yields a new character integral 
\begin{align}\label{edgefirst}
\log Z_{s, \nu}=\int^\infty_0 \frac{du}{2u}\frac{1+e^{-u}}{1-e^{-u}}\left(\chi^{\text{AdS}_{d+1}}_{[\Delta, s]}(u)-D^{d+2}_{s-1}\frac{e^{-(\frac{d-2}{2}+\nu)u}}{(1-e^{-u})^{d-2}}\right)
\end{align}
The second term in the bracket  corresponds to subtracting the partition function of $D^{d+2}_{s-1}$ scalars on $\text{EAdS}_{d-1}$ with scaling dimension $\frac{d-2}{2}+\nu$ since it involves an $\text{SO}(2, d-2)$ character of scalar representation.  As in the de Sitter case \cite{Anninos:2020hfj}, we tend to identify these scalar degrees of freedom as edge modes living on the horizon of Rindler-AdS \cite{Parikh:2012kg} which is a Lorentzian Wick rotation of EAdS and has a $\text{EAdS}_{d-1}$ shaped horizon. More discussions about Rindler-AdS to will be left to section \ref{th} and appendix \ref{coodAdS}.  We can also write $\log Z_{s, \nu}$ in terms of an $\text{SO}(2,d)$ character and an $\text{SO}(2,d+2)$ character
 \begin{align}\label{use}
 \log Z_{s, \nu}=\int^\infty_0 \frac{du}{2u}\frac{1+e^{-u}}{1-e^{-u}}\left(\chi^{\text{AdS}_{d+1}}_{[\Delta, s]}(u)-\left(e^{\frac{u}{2}}-e^{-\frac{u}{2}}\right)^4\chi^{\text{AdS}_{d+3}}_{[\Delta+1, s-1]}(u)\right)
 \end{align}
This form of character integral representation doesn't have the physically meaningful edge character structure but it turns out to be much more convenient than (\ref{edgefirst}) computationally  when we sum over all field content in Vasiliev higher spin gravities.

Finally,  let's move to our main interest: (partially) massless fields. Due to gauge symmetry, the  partition function of a PM field with spin-$s$ and depth $t\in\{0, 1,\cdots, s-1\}$ is given by
\begin{align}\label{PM}
\log Z_{s, \nu_t}=\frac{(-)^{\frac{d+1}{2}}}{d!}\int^\infty_0\frac{du}{u}\,\left(D^{d}_s e^{-(\frac{d}{2}+t-1) u}\,W_s^{(d)}(u)\!-\!D^{d}_t e^{-(\frac{d}{2}+s-1) u}\,W_t^{(d)}(u)\right)
\end{align}
where the first term corresponds to the spin-$s$ gauge field and the second term arises from the spin-$t$ ghost field. By using the explicit expression of $W^{(d)}_s$ derived in eq. (\ref{Wdd}), we can rewrite (\ref{PM}) in the same form as (\ref{use})
 \begin{align}\label{pow}
 \log Z_{s, \nu_t}=\int^\infty_0 \frac{du}{2u}\frac{1+e^{-u}}{1-e^{-u}}\left(\chi^{\text{AdS}_{d+1}}_{[d+t-1, s]}(u)-\left(e^{\frac{u}{2}}-e^{-\frac{u}{2}}\right)^4\chi^{\text{AdS}_{d+3}}_{[d+t, s-1]}(u)\right)
 \end{align}
 where the characters of PM fields are 
 \begin{align}
 \chi^{\text{AdS}_{d+1}}_{[d+t-1, s]}(u)&=\frac{D^{d}_se^{-(d+t-1)u}-D^{d}_t e^{-(d+s-1)u}}{(1-e^{-u})^d}\\
  \chi^{\text{AdS}_{d+3}}_{[d+t, s-1]}(u)&=\frac{D^{d+2}_{s-1}e^{-(d+t)u}-D^{d+2}_{t-1}e^{-(d+s)u}}{(1-e^{-u})^{d+2}}
 \end{align}
 Note $[d+t, s-1]$ is a massless representation of $\text{SO}(2, d+2)$ with spin-$(s\!-\!1)$ and depth-$(t\!-\!1)$.

\section{Regularization, contour prescription and odd dimensional AdS}\label{Contour}
In the previous sections, we've derived a {\it formal}  character integral formula for the {\it unregularized} one-loop partition functions of both scalar fields and higher spin fields in even dimensional AdS. However, to make sense of the character integral mathematically and apply it to actual computation of renormalized partition functions, we have to use a well-defined and efficient regularization scheme. In this section, we'll sort out this issue. Surprisingly the resolution turns out to have a very important byproduct: a character integral representation that works in odd dimensional AdS.

\subsection{Regularization and contour prescription}
We use a real scalar field to illustrate the regularization scheme. But it will be clear in the end that the same regularization also works for higher spin fields.
\subsubsection{$d=2r+1$}\label{con1}
It's mentioned in section \ref{warmup} that $W^{(d)}(u)$,  Fourier transformation of the scalar spectral density $\mu^{(d)}(\lambda)$, is not well-defined. As a manifestation of this point, $W^{(d)}(u)$ is singular at $u=0$. This singularity may lead to two inequivalent definitions of inverse Fourier transformation of $W^{(d)}(u)$ since the contour can either go above or  below $u=0$: 
\begin{align}\label{tildemu}
\tilde\mu^{(\pm)}_d(\lambda)\equiv \int_{\mathbb{R}\pm i\delta}\frac{du}{2\pi}\, W^{(d)}(u)\,e^{-i\lambda u}=(-)^{r+1}d!\int_{\mathbb{R}\pm i\delta}\frac{du}{2\pi}\frac{1+e^{-u}}{1-e^{-u}}\frac{e^{-(\frac{d}{2}+i \lambda)u}}{(1-e^{-u})^d}
\end{align}
where $\delta$ is a small positive number. (At this stage, the size of $\delta$ is not important as long as it's smaller than $2\pi$. We'll later impose a more stringent constraint on it). It's clear that the two definitions of inverse Fourier transformation differ by the residue of the integrand at $u=0$. In terms of the notation $H_{d, \nu}(u)\equiv \frac{1+e^{-u}}{1-e^{-u}}\frac{e^{-(\frac{d}{2}+\nu) u}}{(1-e^{-u})^d}$ introduced in the appendix \ref{residues}, the function $\tilde\mu^{(\pm)}_d(\lambda)$ can also expressed as 
\begin{align}
\tilde\mu^{(\pm)}_d(\lambda)=(-)^{r+1}d!\int_{\mathbb{R}\pm i\delta}\frac{du}{2\pi}\, H_{d, i\lambda}(u)
\end{align}
We use the function $H_{d, i\lambda}(u)$ here because its residue at $u=0$ is given by eq. (\ref{resH}) and the residues at other poles $u=2\pi i n, n\in \mathbb{Z}$ can be easily inferred by using its quasi-periodicity $H_{d, i\lambda}(u+2\pi in)=(-e^{2\pi\lambda})^n H_{d, i\lambda}(u)$. With the information of residues known, we close the contour at infinity and get
\begin{align}\label{mupm}
\tilde\mu^{(\pm)}_d(\lambda)&=\prod_{j=0}^{r-1}\left(\lambda^2+\left(j+\frac{1}{2}\right)^2\right)(\lambda\tanh(\pi\lambda)\pm\lambda)
\end{align}
Therefore in order to recover the spectral density $\mu^{(d)}$, cf. (\ref{scalarspecden}), the contour prescription should be the average of $\int_{\mathbb{R}\pm i\delta}$:
\begin{align}\label{conto}
\mu^{(d)}(\lambda)=\frac{1}{2}(\tilde\mu^{(+)}_d+\tilde\mu_d^{(-)})=\frac{1}{2}\left(\int_{\mathbb{R}+i\delta}+\int_{\mathbb{R}-i\delta}\right)\frac{du}{2\pi}\, W^{(d)}(u)\,e^{-i\lambda u}
\end{align}
Plugging this equation into the scalar partition function (\ref{SiAdS}) , we obtain
\begin{align}\label{avgcon}
\log Z_\nu&=\frac{(-)^{r+1}}{4\, d!}\left(\int_{\mathbb{R}+i\delta}+\int_{\mathbb{R}-i\delta}\right)\frac{du}{4\pi}\, W^{(d)}(u)\int^\infty_0\frac{dt}{t}\,e^{-\frac{\epsilon^2}{4t}-t\nu^2}\int_{-\infty}^\infty d\lambda\, e^{-t\lambda^2-i\lambda u}\nonumber\\
&=\frac{1}{4}\left(\int_{\mathbb{R}+i\delta}+\int_{\mathbb{R}-i\delta}\right)\frac{du}{2\sqrt{u^2+\epsilon^2}}\frac{1+e^{-u}}{1-e^{-u}}\frac{e^{-\frac{d}{2}u-\nu\sqrt{\epsilon^2+u^2}}}{(1-e^{-u})^d}
\end{align}
where $\delta$ has to be smaller than $\epsilon$, otherwise the contours would cross the branch cut of $\sqrt{u^2+\epsilon^2}$, which has two disconnected pieces with one piece going upwards from $i\epsilon$ to $i\infty$ and the other going downwards from $-i\epsilon$ to $-i\infty$.  Due to the manifest $u\to -u$ symmetry of the integrand for odd $d$, the two contours $\mathbb{R}\pm i\delta$ are actually equivalent and it suffices to use one of them
\begin{align}\label{regint}
\log Z_\nu=\frac{1}{2}\int_{\mathbb{R}+i\delta}\frac{du}{2\sqrt{u^2+\epsilon^2}}\frac{1+e^{-u}}{1-e^{-u}}\frac{e^{-\frac{d}{2}u-\nu\sqrt{\epsilon^2+u^2}}}{(1-e^{-u})^d}
\end{align}
The computation above also works for higher spin fields. It suffices to show that the higher spin generalization of (\ref{conto}) holds. Notice that  $W^{(d)}_s(u)$ corresponding to a spin-$s$ field is related to the scalar version $W^{(d)}(u)$  by
\begin{align}
W^{(d)}_s(u)=W^{(d)}(u)+s(s+d-2)W^{(d-2)}(u)
\end{align}
Applying the integral (\ref{conto}) to this relation indeed yields
\begin{align}\label{WdSF}
\frac{1}{2}\left(\int_{\mathbb{R}+i\delta}+\int_{\mathbb{R}-i\delta}\right)\frac{du}{2\pi}\, W^{(d)}_s(u)\,e^{-i\lambda u}=\mu^{(d)}_s
\end{align}
Therefore the regularized character integral for a spin-$s$ field in massive representations is 
\begin{align}\label{fullreg}
\log Z_{s,\nu}=\int_{\mathbb{R}+i\delta}\frac{du}{4\sqrt{u^2+\epsilon^2}}\frac{1+e^{-u}}{1-e^{-u}}\left(\frac{D^{d}_s \,e^{-\frac{d}{2}u-\nu\sqrt{\epsilon^2+u^2}}}{(1-e^{-u})^d}-\frac{D^{d+2}_{s-1} \,e^{-\frac{d-2}{2}u-\nu\sqrt{\epsilon^2+u^2}}}{(1-e^{-u})^{d-2}}\right)
\end{align}
Starting from the eq. (\ref{fullreg}), we'll derive a complete expression for the regularized partition function $Z_{s,\nu}$ in section \ref{rigovalue}.

\subsubsection{$d=2r$}\label{cont2}
\begin{figure}
\centering
\begin{subfigure}{.4\textwidth}
  \centering
  \includegraphics[width=1.0\linewidth]{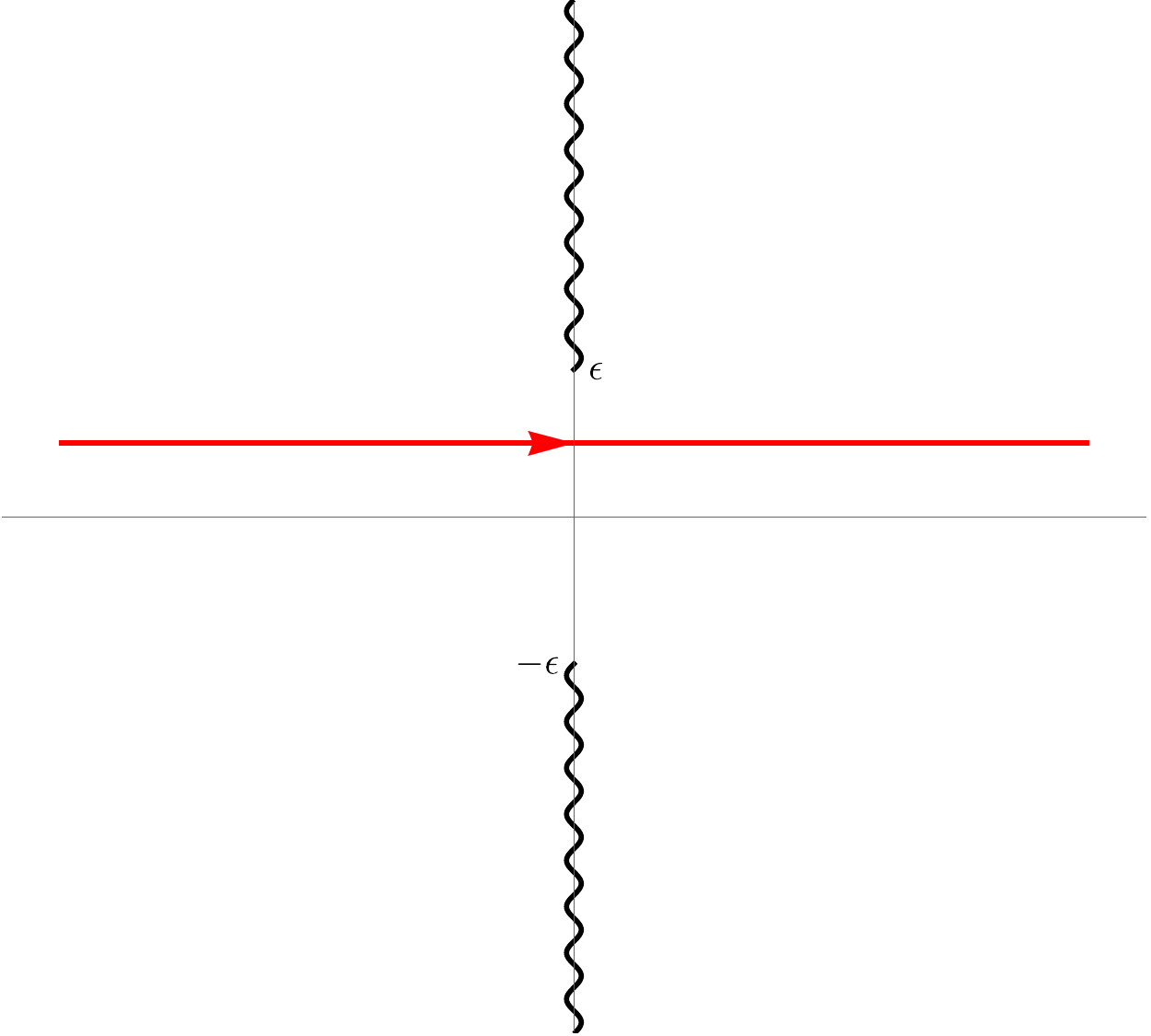}
  \caption{odd $d$}
  \label{fig:sub1}
\end{subfigure}\,\,\,\,\,
\begin{subfigure}{.4\textwidth}
  \centering
  \includegraphics[width=1.0\linewidth]{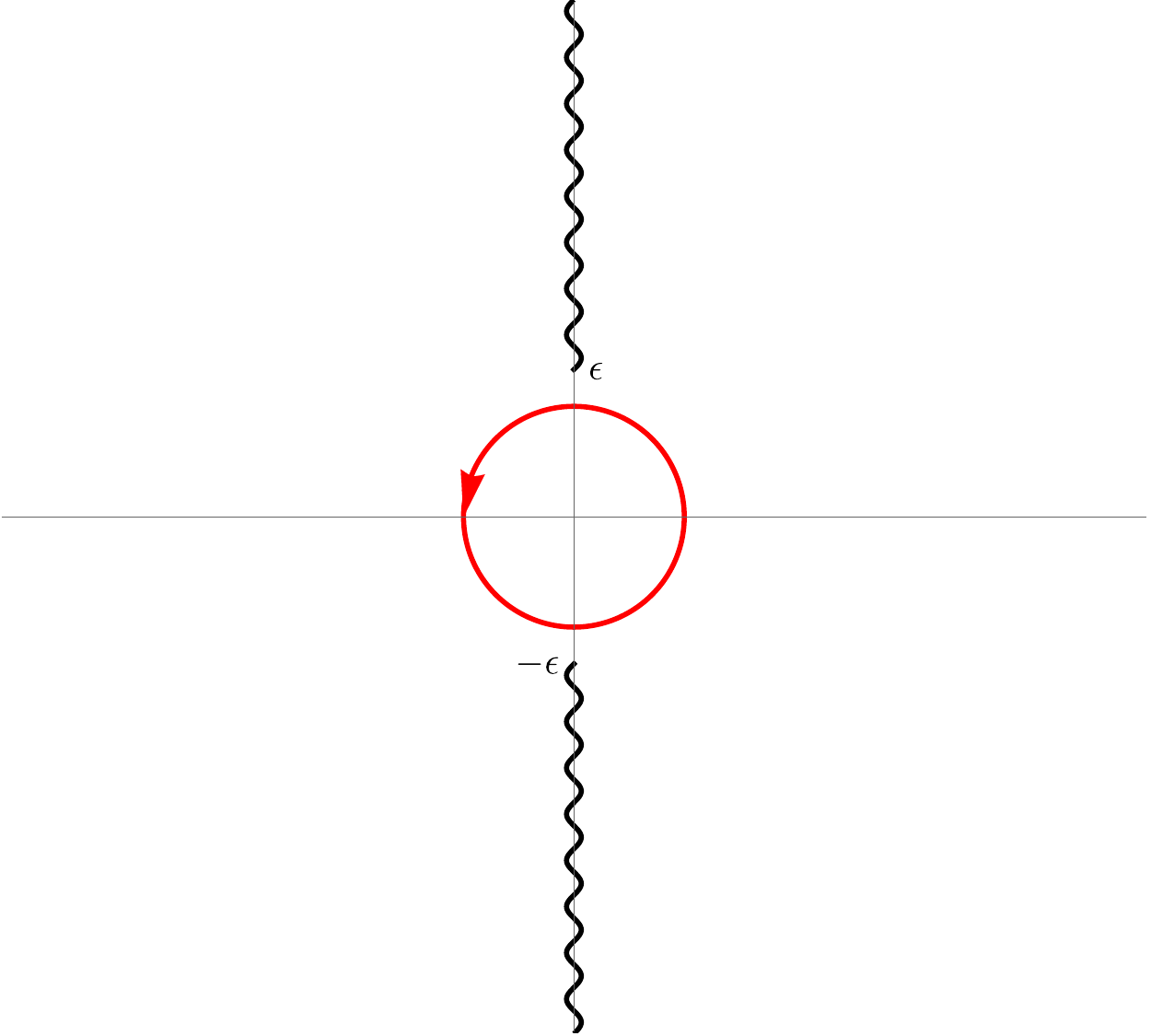}
  \caption{even $d$}
  \label{fig:sub2}
\end{subfigure}
\caption{$u$-contours for even dimensional AdS (left) and odd dimensional AdS (right). The black wavy lines represent branch cuts. }
\label{fig:test}
\end{figure}

The odd $d$ case above tells us the correct strategy to get a regularized character integral. First, we pick a  function $W^{(d)}(u)\propto \frac{1+e^{-u}}{1-e^{-u}}\frac{e^{-\frac{d}{2}u}}{(1-e^{-u})^d}$ and compute its inverse Fourier transformations defined by  two different contour choices. Then one proper linear combination of these choices can give the correct spectral density. Plug this integral expression of spectral density into the original partition function (\ref{logZv2}) and we finally obtain the regularized character integral.  Now let's apply this strategy to the $d=2r$ case where we choose $W^{(d)}(u)$ to be 
\begin{align}
W^{(d)}(u)=(-)^{r+1} d!\frac{1+e^{-u}}{1-e^{-u}}\frac{e^{-\frac{d}{2}u}}{(1-e^{-u})^d}
\end{align}
The two possible inverse Fourier transformations defined as in eq. (\ref{tildemu}) are 
\begin{align}
\tilde\mu^{(\pm)}_d(\lambda)&=i \prod_{j=0}^{r-1}(\lambda^2+j^2)(\coth(\pi\lambda)\pm1)
\end{align}
Therefore the spectral density $\mu^{(d)}(\lambda)=\prod_{j=0}^{r-1}(\lambda^2+j^2)$ in $\text{AdS}_{2r+1}$ is given by
\begin{align}\label{conte}
\mu^{(d)}(\lambda)=\frac{i}{2}(\tilde\mu_d^{(-)}-\tilde\mu_d^{(+)})=\frac{i}{2}\left(\int_{\mathbb{R}-i\delta}-\int_{\mathbb{R}+i\delta}\right)\frac{du}{2\pi}\, W^{(d)}(u)\,e^{-i\lambda u}
\end{align}
Substituting this equation into (\ref{logZv2}) and performing the $t, \lambda$ integrals, we end up with
\begin{align}\label{simpZ}
\frac{\log Z_\nu}{\log R}=\frac{i(-)^r}{2\pi d!}\left(\int_{\mathbb{R}-i\delta}-\int_{\mathbb{R}+i\delta}\right)\frac{du}{2\sqrt{u^2+\epsilon^2}}W^{(d)}(u) e^{-\nu\sqrt{u^2+\epsilon^2}}
\end{align}
Since there is no poles in the strip bounded by $\mathbb{R}\pm i\delta$, we can further deform the contour to be a small circle $C_0$ around $u=0$. Then the integral is equivalent to evaluating the residue at $u=0$:
\begin{align}\label{sdsd}
\frac{\log Z_\nu}{\log R}&=\text{Res}_{u\to 0}\left(\frac{e^{-\nu\sqrt{u^2+\epsilon^2}}}{2\sqrt{u^2+\epsilon^2}}\frac{1+e^{-u}}{1-e^{-u}}\frac{e^{-\frac{d}{2}u}}{(1-e^{-u})^d}\right)
\end{align}
Similarly for higher spin fields, by using appropriate square root regularization, we can also get
\begin{align}
\frac{\log Z_{s,\nu}}{\log R}=\text{Res}_{u\to 0}\left[\frac{e^{-\nu\sqrt{\epsilon^2+u^2}}}{2\sqrt{u^2+\epsilon^2}}\frac{1+e^{-u}}{1-e^{-u}}\left(\frac{D^{d}_s \,e^{-\frac{d}{2}u}}{(1-e^{-u})^d}-\frac{D^{d+2}_{s-1}\, e^{-\frac{d-2}{2}u}}{(1-e^{-u})^{d-2}}\right)\right]
\end{align}
\\
\noindent{}\textbf{Conclusion}
\\
Altogether, we can conclude that the regularized one-loop partition function (with the UV regularization introduced by $e^{-\frac{\epsilon^2}{4t}}$ in the original definition (\ref{genZ})) of a field in the $[\frac{d}{2}+\nu, s]$ massive representation of $\text{SO}(2, d)$ is 
\begin{align}\label{REG1}
\log Z_{s, \nu}=\frac{A_d}{2}\int_{C_d} du\frac{e^{-\nu\sqrt{\epsilon^2+u^2}}}{\sqrt{u^2+\epsilon^2}}\frac{1+e^{-u}}{1-e^{-u}}\left(\frac{D^{d}_s \, e^{-\frac{d}{2}u}}{(1-e^{-u})^d}-\frac{D^{d+2}_{s-1} \, e^{-\frac{d-2}{2}u}}{(1-e^{-u})^{d-2}}\right)
\end{align}
where the overall constant and contour choice,  fig. (\ref{fig:test}), are
\begin{align}
A_d=\begin{cases} \frac{\log R}{2\pi i} \,\,\,\, &d\,\, \text{even}\\ \frac{1}{2} \,\,\,\, &d\,\, \text{odd}\end{cases}\,\,\,\,\,\,\,\, C_d=\begin{cases} \text{a counterclockwise circle around}\,\, 0 \,\,\,\, &d\,\, \text{even}\\ \mathbb{R}+i\delta \,\,\,\, &d\,\, \text{odd}\end{cases}
\end{align}
For a field in massless representations, we need to include the corresponding ghost contribution
\small
\begin{align}\label{REG2}
\log Z_{s, \nu_t}&=\frac{A_d}{2}\int_{C_d}\frac{du}{\sqrt{u^2+\epsilon^2}}\frac{1+e^{-u}}{1-e^{-u}}\frac{e^{-\frac{d}{2}u}}{(1-e^{-u})^d}\left(D_s^{d}\,e^{-\nu_t\sqrt{u^2+\epsilon^2}}-D_t^{d}\,e^{-\nu_s\sqrt{u^2+\epsilon^2}}\right)\nonumber\\
&-\frac{A_d}{2}\int_{C_d}\frac{du}{\sqrt{u^2+\epsilon^2}}\frac{1+e^{-u}}{1-e^{-u}}\frac{e^{-\frac{d-2}{2}u}}{(1-e^{-u})^{d-2}}\left(D_{s-1}^{d+2}\,e^{-\nu_t\sqrt{u^2+\epsilon^2}}-D_{t-1}^{d+2}\,e^{-\nu_s\sqrt{u^2+\epsilon^2}}\right)
\end{align}
\normalsize
where $\nu_t=\frac{d}{2}+t-1$ and $\nu_s=\frac{d}{2}+s-1$. 

\section{Evaluation of the regularized character integrals}\label{rigovalue}
In this section, we give an efficient and general recipe to compute the regularized character integral formula following the appendix C of  \cite{Anninos:2020hfj}. For the simplicity of notation, we'll  use scalar fields as an illustration of this recipe. But our reasoning and final result can be easily generalized to fields of arbitrary spin. In addition, the result can be used to  justify that the unregularized character integral is sufficient for the application to Vasiliev gravities. First, let's briefly review the standard heat kernel method in computing one-loop partition functions
\begin{align}
\log Z_\nu(\epsilon)=\int_0^\infty \frac{dt}{2t}e^{-\frac{\epsilon^2}{4t}}K_\nu(t), \,\,\,\, \, K_\nu(t)=\frac{\text{Vol}(\text{AdS}_{d+1})}{2^d\pi^{\frac{d+1}{2}}\Gamma(\frac{d+1}{2})}\int_0^\infty d\lambda\, \mu^{(d)}(\lambda)\,e^{-t(\lambda^2+\nu^2)}
\end{align}
where I've plugged in the explicit form of  $\text{Vol}(S^d)$ compared to eq. (\ref{genZ}). By using the small $t$ expansion of heat kernel, say $K_\nu(t)=\sum_{k=0}^{d+1} \alpha_k\,t ^{-\frac{d+1-k}{2}}+\CO(t)$, we can separate it into UV and IR parts without ambiguity
\begin{align}
K^{\text{uv}}_\nu(t)\equiv \sum_{k=0}^{d+1}\alpha_k\,t^{-\frac{d+1-k}{2}}, \,\,\,\,\, K^{\text{ir}}_\nu(t)=K_\nu(t)-K^{\text{uv}}_\nu(t)
\end{align}
The UV-part of the heat kernel expansion in AdS is fairly simple. When $d$ is even, $K_\nu(t)$ can be evaluated exactly because it's a Gaussian integral in $\lambda$. When $d$ is odd, using $\tanh{\pi\lambda}=1-\frac{2}{1+e^{2\pi\lambda}}$ we can split the heat kernel into two parts. The first part is a simple Gaussian integral as in the even $d$ case. The second part is of  $\CO(t^0)$  and hence we can put $t=0$ which yields an exactly solvable integral. In addition, by direct computation, one can show that $\alpha_k$ is nonvanishing only for even $k$. For example, in $d=3$, we obtain the nonzero heat kernel coefficients: $\alpha_0=\frac{1}{12}, \alpha_2=\frac{1-4\nu^2}{48}$ and $\alpha_4=-\frac{17}{5760}-\frac{\nu^2}{48}+\frac{\nu^4}{24}$. 

Introducing an infinitesimal IR cutoff $\kappa\to 0$, we can further separate $\log Z_{\nu}(\epsilon)$ into two parts
\begin{align}
\log Z^{\text{uv}}_\nu(\epsilon)=\int_0^\infty \frac{dt}{2t}K^{\text{uv}}_\nu(t)e^{-\frac{\epsilon^2}{4t}-\kappa^2 t}, \,\,\,\,\,\log Z^{\text{ir}}_\nu=\int_0^\infty \frac{dt}{2t}K^{\text{ir}}_\nu(t)e^{-\kappa^2 t}
\end{align}
where the UV regulator has been dropped in the IR integral because it's by construction UV finite. The IR regulator $e^{-\kappa^2 t} $ is inserted in the UV integral because the integrand has a $\frac{1}{t}$ term when $\alpha_{d+1}\not= 0$, i.e. when $d$ is odd. In the end, the $\log\kappa$ terms in $\log Z^{\text{uv}}_\nu$ and $\log Z^{\text{ir}}_\nu$ will cancel out and we're left with
\small
\begin{align}\label{regHK}
\log Z_\nu(\epsilon)=\frac{1}{2}\sum_{k=0}^d\alpha_k\,\Gamma\left(\frac{d+1-k}{2}\right)\left(\frac{2}{\epsilon}\right)^{d+1-k}+\alpha_{d+1}\log\left(\frac{2}{e^\gamma\epsilon}\right)+\frac{1}{2}\zeta_\nu'(0)
\end{align}
\normalsize
where $\zeta_\nu(z)\equiv \frac{1}{\Gamma(z)}\int_0^\infty \frac{dt}{t}t^z K_\nu(t)$ is the spectral zeta function and $\alpha_{d+1}=\zeta_\nu(0)$. Next, we'll apply this UV-IR separation idea to the evaluation of partition function in the regularized character integral formalism.

\subsection{Even dimensional $\text{AdS}_{2r+2}$}
In section \ref{con1}, we've found that the square-root regularized partition function in $\text{AdS}_{2r+2}$ is given by
\begin{align}
\log Z_\nu(\epsilon)=\frac{1}{2}\int_{\mathbb{R}+i\delta}\frac{du}{2\sqrt{u^2+\epsilon^2}}\frac{1+e^{-u}}{1-e^{-u}}e^{-\frac{d}{2}u-\nu\sqrt{u^2+\epsilon^2}}
\end{align}
where $d=2r+1$. Putting $\epsilon=0$ we recover the formal UV-divergent character formula:
\begin{align}
\log Z_{\nu}(\epsilon=0)=\int_{0}^{\infty}\frac{du}{2u}H_{d, \nu}(u), \,\,\,\, H_{d, \nu}(u)=\frac{1+e^{-u}}{1-e^{-u}}\frac{e^{-(\frac{d}{2}+\nu)u}}{(1-e^{-u})^d}
\end{align}
The unregularized integrand $\frac{1}{2u}H_{d, \nu}(u)$ admits a Laurent expansion around $u=0$ with coefficients  being polynomials in $\nu$:
\begin{align}
\frac{1}{2u} H_{d, \nu}(u)=\frac{1}{u}\sum_{k\ge 0} b_k(\nu) u^{-(d+1-k)}, \,\,\,\, b_k(\nu)=\sum_{\ell=0}^{k}b_{k\ell}\nu^\ell
\end{align}
where $b_{k\ell}$ vanishes when $k+\ell$ is odd due to the symmetry $H_{d, -\nu}(-u)=H_{d, \nu}(u)$ for odd $d$. The terms corresponding to  $0\le k\le d+1$ in $H_{d, \nu}(u)$ are UV divergent in $\log Z_{\nu}(\epsilon=0)$. Therefore we  separate them from the remaining UV-finite terms
\begin{align} \label{expF}
H_{d, \nu}^{\text{uv}}(u)\equiv 2\sum_{k\ge d+1} b_k(\nu) \, u^{-(d+1-k)}, \,\,\,\,\,H_{d, \nu}^{\text{ir}}(u)\equiv H_{d, \nu}(u)-H_{d, \nu}^{\text{uv}}(u)
\end{align}
Similarly using an infinitesimal IR regulator $\kappa\to 0^+$, we obtain a UV-IR separation for the regularized partition function $\log Z_\nu(\epsilon)=\log Z^{\text{uv}}_\nu(\epsilon)+\log Z_\nu^{\text{ir}}$:
\begin{align}
\log Z^{\text{uv}}_\nu(\epsilon)=\frac{1}{4}\int_{\mathbb{R}+i\delta}\frac{du}{r u} H_{d, r\nu}^{\text{uv}}(u)e^{-\kappa |u|}, \,\,\,\, \log Z_\nu^{\text{ir}}=\int_{0}^\infty\frac{du}{2u}H_{d, \nu}^{\text{ir}}(u)e^{-\kappa u}
\end{align}
where $r=\sqrt{u^2+\epsilon^2}/u$. In the IR part of the partition function, we've deformed the contour and safely put $\epsilon=0$.
The $\log \kappa$ terms will drop out at the end when summing up $\log Z^{\text{uv}}_\nu$ and $\log Z^{\text{ir}}_\nu$.

\vspace{10pt}

\noindent{}\textbf{Evaluation of UV part}

\begin{figure}
\centering
\begin{subfigure}{.4\textwidth}
  \centering
  \includegraphics[width=1.0\linewidth]{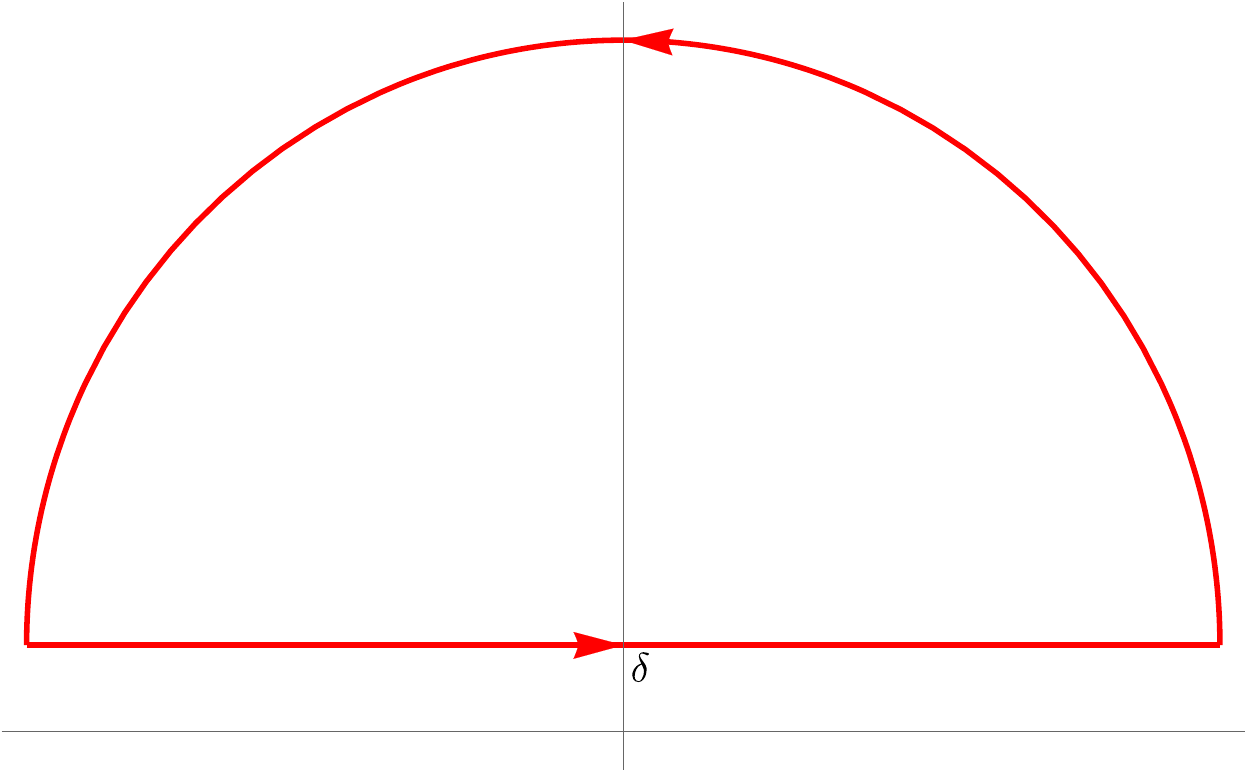}
  \caption{odd $\ell$}
  \label{fig:subb1}
\end{subfigure}\,\,\,\,\,
\begin{subfigure}{.4\textwidth}
  \centering
  \includegraphics[width=1.0\linewidth]{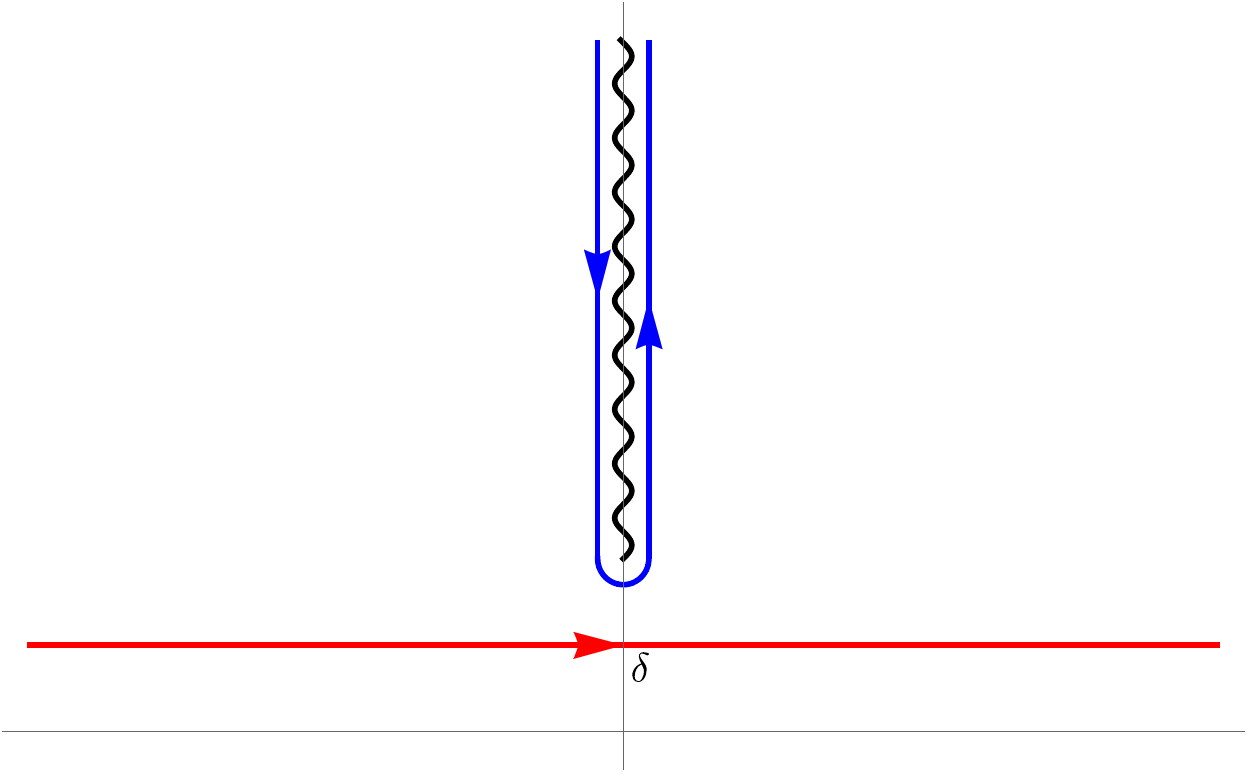}
  \caption{even $\ell$}
  \label{fig:subb2}
\end{subfigure}
\caption{Contour for the integral $\CI_\epsilon(k,\ell)$. When $\ell$ is odd,  the contour is closed at infinity. When $\ell$ is even, the red contour in deformed to the blue one running along the branch cut of $\sqrt{u^2+1}$.}
\label{fig:Iint}
\end{figure}

\noindent{}Using the expansion  (\ref{expF}), the UV part of $\log Z_\nu(\epsilon)$ can be written as 
\begin{align}\label{ZUV}
\log Z^{\text{uv}}_\nu(\epsilon)=\frac{1}{2}\sum_{k=0}^d\sum_{\ell=0}^k b_{k\ell}\,\nu^{\ell}\int_{\mathbb{R}+i\delta}\frac{du}{ r u}\,r^\ell u^{-(d+1-k)}+\frac{1}{2}\sum_{\ell=0}^{d+1} b_{d+1, \ell}\,\nu^\ell\int_{\mathbb{R}+i\delta}\frac{du}{ r u}\,r^\ell e^{-\kappa |u|}
\end{align}
where  the IR regulator $\kappa$  is not necessary for $0\le k\le d$. Then it suffices to evaluate the following two $u$-integrals
\begin{align}
&\CI_\epsilon(k, \ell)\equiv \int_{\mathbb{R}+i\delta}\frac{du}{ 2r u}\,r^\ell u^{-(d+1-k)}=\epsilon^{-(d+1-k)}\frac{1}{2}\int_{\mathbb{R}+i\delta}du\, \frac{(1+u^2)^{\frac{\ell-1}{2}}}{u^{d+\ell+1-k}}\nonumber\\
& \CJ_\epsilon(\ell)\equiv \int_{\mathbb{R}+i\delta}\frac{du}{ 2 r u}\,r^\ell e^{-\kappa |u|}=\frac{1}{2}\int_{\mathbb{R}+i\delta}du\,\frac{(1+u^2)^{\frac{\ell-1}{2}}}{u^{\ell}}e^{-\epsilon\kappa |u|}
\end{align} 
When $\ell$ is odd in $\CI_\epsilon(k, \ell)$, we can close the contour in the upper half plane (fig. \ref{fig:subb1}) and the integral vanishes because $(1+u^2)^{\frac{\ell-1}{2}}$ has no pole in this case. Thus $\CI_\epsilon(k, \ell)$ is nonvanishing only when $\ell$ is even. On the other hand, the coefficient of $\CI_\epsilon(k, \ell)$, i.e. $b_{k\ell}$, vanishes for $k+\ell$ odd. Therefore, only terms with even $k, \ell$  survive in  $\log Z^{\text{uv}}_\nu(\epsilon)$. When $\ell$ is even in $\CI_\epsilon(k, \ell)$,  the previous ``no poles'' argument doesn't hold any more due to the presence of a branch cut from $i$ to $i\infty$ in the upper half plane. However, we can deform the contour to integrate along the branch cut (fig. \ref{fig:subb2}) which yields
\small
\begin{align}\label{CI}
\CI_\epsilon(k, \ell)=\frac{1}{2} (-)^{\frac{d+1-k}{2}} B\left(\frac{d+1-k}{2}, \frac{1+\ell}{2}\right)
\end{align} 
\normalsize
where the $B$-function is $B(a,b)=\frac{\Gamma(a)\Gamma(b)}{\Gamma(a+b)}$.

For the integral $\CJ_\epsilon(\ell)$, we split it into an IR-divergent part and an IR-finite part. In the IR-finite part, we can drop the IR regulator $e^{-\kappa u}$.
\begin{align}
\CJ_\epsilon(\ell)&=\frac{1}{2}\int_{\mathbb{R}+i\delta}\frac{du}{\sqrt{1+u^2}}\left((1+u^{-2})^{\frac{\ell}{2}}-1\right)+\frac{1}{2}\int_{\mathbb{R}+i\delta}\frac{du}{\sqrt{1+u^2}}e^{-\epsilon\kappa |u|}\nonumber\\
&=\frac{1}{2}\int_{\mathbb{R}+i\delta}\frac{du}{\sqrt{1+u^2}}\left((1+u^{-2})^{\frac{\ell}{2}}-1\right)+\int_{0}^\infty \frac{du}{\sqrt{1+u^2}}e^{-\epsilon\kappa u}
\end{align} 
Since the coefficient of $\CJ_\epsilon(\ell)$ is $b_{d+1, \ell}$, it suffices to consider even $\ell$ which implies that $(1+u^{-2})^{\frac{\ell}{2}}$ can be expanded into a polynomial of $u^{-2}$ and for each term in the polynomial we can use the same contour trick as in the $\CI_\epsilon$ case to evaluate the integral. The IR-divergent part can be analytically evaluated in mathematica and it has a very simple small $\epsilon$ behavior. Altogether, the final result for $\CJ_\epsilon(\ell)$ is
\begin{align}\label{CJ}
\CJ_\epsilon(\ell)&=\frac{1}{2}\sum_{n=1}^{\ell/2}(-)^n\binom{\frac{\ell}{2}}{n}B\left(n,\frac{1}{2}\right)+\log\frac{2}{e^\gamma \epsilon\kappa}\nonumber\\
&=-\left(H_\ell-\frac{1}{2}H_{\ell/2}+\log\frac{e^\gamma \epsilon\kappa}{2}\right)
\end{align}
 where $H_\ell$ is the harmonic number of order $\ell$. Plugging (\ref{CI}) and (\ref{CJ}) into (\ref{ZUV}) yields the regularized UV-part of the partition function
 \small
\begin{align}\label{uv}
\log Z^{\text{uv}}_\nu(\epsilon)&=\frac{1}{2}\sum_{k=0}^{r}(-)^{r+1-k}\epsilon^{-(d+1-2k)}\sum_{\ell=0}^{2k}b_{2k, \ell}B\left(\frac{d+1}{2}-k, \frac{\ell+1}{2}\right)\nu^\ell\nonumber\\
&-\sum_{\ell=0}^{d+1}b_{d+1,\ell}\left(H_\ell-\frac{1}{2}H_{\ell/2}\right)\nu^\ell-b_{d+1}(\nu)\log\frac{e^\gamma \epsilon\kappa}{2}
\end{align}
\normalsize
For example, for $d=3$, we get
\small
\begin{align}\label{regzuv}
\log Z^{\text{uv}}_\nu(\epsilon)=\frac{2}{3\,\epsilon^4}+\frac{1-4\nu^2}{24\,\epsilon^2}+\left(\frac{\nu^2}{48}-\frac{\nu^4}{18}\right)+\left(\frac{\nu ^4}{24}+\frac{\nu ^2}{48}-\frac{17}{5760}\right)\log\frac{2}{e^\gamma \epsilon\kappa}
\end{align}
\normalsize
which reproduces the $\text{Pol}(\Delta)$ part of $\log Z_\nu$ in \cite{Keeler:2014hba} \footnote{There is an overall $\frac{1}{2}$ factor difference because  \cite{Keeler:2014hba} computes the partition function of a {\it complex} scalar rather than a {\it real} scalar. } without using the heat kernel coefficients $\alpha_k$. In fact, by comparing (\ref{regzuv}) and (\ref{regHK}), we can express the nonzero heat kernel coefficients $\alpha_k$ in terms of $b_{kl}$
\small
\begin{align}
\alpha_{2k}=(-4)^{k-\frac{d+1}{2}}\sum_{\ell=0}^{2k}\frac{\Gamma(\frac{\ell+1}{2})}{\Gamma(\frac{d+\ell}{2}+1-k)}b_{2k, \ell}\, \nu^{\ell}
\end{align}
\normalsize
\\
\noindent{}\textbf{Evaluation of IR part}
\\
\noindent{}The IR part can be evaluated through certain zeta function method as in \cite{Anninos:2020hfj}
\begin{align}\label{ir}
\log Z^{\text{ir}}_\nu=\frac{1}{2}\bar\zeta'_\nu(0)+b_{d+1}(\nu)\log(\kappa), \,\,\,\,\, \bar\zeta_\nu(z)\equiv \frac{1}{\Gamma(z)}\int_0^\infty \frac{du}{u}\, u^z H_{d, \nu}(u)
\end{align}
Notice that the ``character zeta function'' $\bar\zeta_\nu(z)$ is originally defined by the  integral above for $z$ sufficiently large and  then  analytically continued to small $z$. $b_{d+1}(\nu)$ is related to the character zeta function as $b_{d+1}(\nu)=\frac{1}{2}\bar\zeta_\nu(0)$. Combing the UV part (\ref{uv}) and IR part (\ref{ir}) leads to
 \small
\begin{align}\label{uvir}
\log Z_\nu(\epsilon)&=\frac{1}{2}\sum_{k=0}^{r}(-)^{r+1-k}\epsilon^{-(d+1-2k)}\sum_{\ell=0}^{2k}b_{2k, \ell}B\left(\frac{d+1}{2}-k, \frac{\ell+1}{2}\right)\nu^\ell\nonumber\\
&-\sum_{\ell=0}^{d+1}b_{d+1,\ell}\left(H_\ell-\frac{1}{2}H_{\ell/2}\right)\nu^\ell+b_{d+1}(\nu)\log\left(\frac{2}{e^\gamma \epsilon}\right)+\frac{1}{2}\bar\zeta'_\nu(0)
\end{align}
\normalsize
Compared to the standard heat kernel results, we find the discrepancy between the {\it character zeta function} $\bar\zeta_\nu(z)$ and the {\it spectral zeta function}  $\zeta_\nu(z)=\frac{1}{\Gamma(z)}\int_0^\infty \frac{dt}{t}t^z K_\nu(t)$:
\begin{align}
\frac{1}{2}\zeta_\nu'(0)=\frac{1}{2}\bar\zeta_\nu'(0)-\sum_{\ell=0}^{d+1}b_{d+1, \ell}\left(H_\ell-\frac{1}{2}H_{\ell/2}\right)\nu^\ell
\end{align}
This difference is the so-called multiplicative anomaly, arising as a UV correction to the formal factorization $\log\det(A B)=\log \det A+\log\det B$, as reviewed in \cite{Dowker:2014tea}. Multiplicative anomaly is computed specifically for  fields in AdS in \cite{Basile:2018zoy}, where it's called ``secondary contribution''. Though the information about multiplicative anomaly is lost , the formal factorization makes the evaluation of $\bar\zeta_\nu(z)$ much simpler than $\zeta_\nu(z)$. For example, we can expand $H_{d,\nu}(u)$ with respect to $e^{-u}$ as $H_{d,\nu}(u)=\sum_{n\ge 0}D^{d+2}_{n} e^{-(\Delta+n)u}$ and for each fixed $n$, the $u$-integral yields $(n+\Delta)^{-z}$. Then we can immediately express $\bar\zeta_\nu(z)$ as a finite sum of Hurwitz zeta functions
\begin{align}
\bar\zeta_\nu(z)=\sum_{n\ge 0}\frac{D^{d+2}_{n}}{(n+\Delta)^z}=D^{d+2}_{\delta-\Delta}\,\zeta(z,\Delta)
\end{align}
where $\delta$ is an operator defined as $\delta^n \zeta(z, \Delta)\equiv \zeta(z-n, \Delta)$. When $d=3$, we have $D^{d+2}_{\delta-\Delta}=\frac{\delta^3}{3}-\nu  \delta^2+\frac{12 \nu ^2-1}{12} \delta+\frac{\nu -4 \nu ^3}{12}$ and thus 
\begin{align}
\bar\zeta_\nu(z)=\frac{1}{3}\zeta(z-3,\Delta)-\nu \zeta(z-2,\Delta)+\left(\nu^2-\frac{1}{12}\right) \zeta(z-1,\Delta)+\frac{\nu -4 \nu ^3}{12}\zeta(z,\Delta)
\end{align}
Altogether, the full one-loop partition function of a real scalar field with scaling dimension $\Delta=\frac{3}{2}+\nu$ in $\text{AdS}_4$ is 
\begin{align}
\log Z_\nu(\epsilon)&=\frac{2}{3\,\epsilon^4}+\frac{1-4\nu^2}{24\,\epsilon^2}+\left(\frac{\nu^2}{48}-\frac{\nu^4}{18}\right)+\left(\frac{\nu ^4}{24}+\frac{\nu ^2}{48}-\frac{17}{5760}\right)\log\frac{2}{e^\gamma \epsilon}\nonumber\\
&+\frac{1}{6}\zeta'(-3,\Delta)-\frac{\nu}{2} \zeta'(-2,\Delta)+\frac{1}{2}\left(\nu^2-\frac{1}{12}\right) \zeta'(-1,\Delta)+\frac{\nu -4 \nu ^3}{24}\zeta'(0,\Delta)
\end{align}
consistent with  \cite{Keeler:2014hba}.
%As a quick check, putting $z=0$ and simplifying the zeta functions, we actually recover the previously claimed relation $\frac{1}{2}\bar\zeta_\nu(0)=b_4(\nu)=\frac{\nu ^4}{24}-\frac{\nu ^2}{48}-\frac{17}{5760}$.

\subsection{Odd dimensional $\text{AdS}_{2r+1}$}
The UV-regularized  partition function in $\text{AdS}_{2r+1}$ can be written in terms of the following character integral 
\begin{align}
\frac{\log Z_\nu(\epsilon)}{\log R}=\frac{1}{2\pi i}\int_{C_0}\frac{du}{2\sqrt{u^2+\epsilon^2}}\frac{1+e^{-u}}{1-e^{-u}}e^{-\frac{d}{2}u-\nu\sqrt{u^2+\epsilon^2}}
\end{align}
where $d=2r$. As in the even dimensional AdS case, we can separate $H_{d,\nu}(u)$  into a UV-part and an IR-part. But the resulting IR partition function $\log Z^{\text{ir}}_\nu=\text{Res}_{u\to 0}\frac{1}{2u}H_{d,\nu}^{\text{ir}}(u)$ vanishes because $\frac{1}{2u}H_{d,\nu}^{\text{ir}}(u)$ doesn't have a pole at $u=0$ by our prescription for the UV-IR separation. Therefore, it suffices to compute the UV-part of the partition function
\begin{align}
\frac{\log Z_\nu(\epsilon)}{\log R}&=\frac{\log Z^{\text{uv}}_\nu(\epsilon)}{\log R}=\sum_{k=0}^{d+1}\sum_{\ell=0}^k \, b_{k\ell}\nu^{\ell}\frac{1}{2\pi i}\int_{C_0}\frac{du}{u}\frac{r^{\ell-1}}{u^{d+1-k}}\nonumber\\
&=\sum_{k=0}^{d+1}\sum_{\ell=0}^k \, b_{k\ell}\,\nu^{\ell}\,\epsilon^{-(d+1-k)}\text{Res}_{u\to 0}\frac{(1+u^2)^{\frac{\ell-1}{2}}}{u^{d+\ell+1-k}}
\end{align}
where $b_{k\ell}$ is nonvanishing only when $k+\ell$ is even. By evaluating the residue explicitly, we find
\begin{align}\label{odds}
\frac{\log Z_\nu(\epsilon)}{\log R}&=\sum_{k=0}^{\frac{d}{2}}\epsilon^{-(d+1-2k)}\sum_{\ell=0}^{2k} b_{2k, \ell}\binom{\frac{\ell-1}{2}}{\frac{d-\ell+2k}{2}}\nu^\ell+b_{d+1}(\nu)\nonumber\\
&=\sum_{k=0}^{\frac{d}{2}}\epsilon^{-(d+1-2k)}\sum_{\ell=0}^{2k} b_{2k, \ell}\binom{\frac{\ell-1}{2}}{\frac{d-\ell+2k}{2}}\nu^\ell+\text{Res}_{u\to 0}\left(\frac{1}{2u}H_{d, \nu}(u)\right)
\end{align}
where we rewrite $b_{d+1}(\nu)$ as a residue. As expected, there is no $\log\epsilon$ divergence or multiplicative anomaly and hence we can define a renormalized partition function  by subtracting all the divergent terms unambiguously 
\begin{align}\label{sddd}
\frac{\log Z^{\text{ren}}_\nu}{\log R}=\text{Res}_{u\to 0}\left(\frac{1}{2u}H_{d, \nu}(u)\right)
\end{align}
where the residue is evaluated in appendix \ref{residues}:
\begin{align}
\frac{\log Z^{\text{ren}}_\nu}{\log R}=-\frac{1}{(2r)!}\sum_{n=1}^{r}\frac{a_n(r)}{2n+1}\nu^{2n+1}, \,\,\,\,\, \prod_{j=0}^{r-1}(x-j^2)=\sum_{n=1}^r\, a_n(r) \,x^n
\end{align}
 For example, according to this expression, the renormalized scalar partition function in $\text{AdS}_3$ is $-\frac{\nu^3}{6}$, consistent with \cite{Giombi:2008vd}.

\subsection{Summary}
The computations in this section provide a well-defined and efficient rule to obtain a regularized
partition function using only the unregularized character integral formula derived in section \ref{WU} and section \ref{HSchar}. Here we summarize this rule for both even and odd dimensional AdS.

\vspace{10pt}
\noindent{\textbf{Odd dimensional AdS}: $d=2r$}
\\
Given the total character $\chi_{\text{tot}}(u)=\chi_{b}(u)-\chi_e(u)$, where $\chi_{b}(u), \chi_e(u)$ are bulk character and edge character respectively, the renormalized partition function (with all negative powers of $\epsilon$ dropped) is given by
\begin{align}\label{evendZ}
\log Z^{\text{ren}}=\text{Res}_{u\to 0}\left(\frac{1}{2u}\frac{1+e^{-u}}{1-e^{-u}}\chi_{\text{tot}}(u)\right)\,\log R
\end{align}
For example, a massive spin-$s$ field with $\Delta=2+\nu$ on $\text{AdS}_5$ has bulk character $\chi_b(u)=D^{4}_s\frac{e^{-(2+\nu)u}}{(1-e^{-u})^4}$ and edge character $\chi_e(u)=D^{6}_{s-1}\frac{e^{-(1+\nu)u}}{(1-e^{-u})^2}$. Therefore according to (\ref{evendZ})
\begin{align}
\log Z^{\text{ren}}_{s,\nu}=\frac{(s+1)^2\nu ^3 }{360} \left(5 (s+1)^2-3 \nu ^2\right)\log R
\end{align}
This result agrees with the computation of \cite{Giombi:2014iua} based on direct spectral zeta function regularization. Another interesting example is  linearized gravity in $\text{AdS}_3$. In this case, the bulk character is $\chi_b(u)=2\frac{e^{-2u}-e^{-3 u}}{(1-e^{-u})^2}$, where the overall factor 2 is  spin degeneracy for any field of nonzero spin, and the edge character is $\chi_e(u)=4\,e^{-u}-e^{-2 u}$. Therefore, according to eq. (\ref{evendZ}) the renormalized partition function of 3D gravity is $\frac{13}{3}\log R$, consistent with \cite{Giombi:2008vd}.

\vspace{10pt}

\noindent{\textbf{Even dimensional AdS}: $d=2r+1$}
\\ 
\noindent{}For odd $d$, we need to consider massive and massless representations separately because in massless representations both gauge field and ghost field can contribute to the total multiplicative anomaly. Given a massive representation $\left[\frac{d}{2}+\nu, s\right]$, the unregularized partition function is
\begin{align}
\log Z_{s, \nu}=\int_{0}^\infty \frac{du}{2u} \, \left(f^b_{s,\nu}(u)-f^e_{s,\nu}(u)\right)\end{align}
where 
\begin{align}
f^b_{s,\nu}(u)=D^{d}_s\frac{1+e^{-u}}{1-e^{-u}}\frac{e^{-(\frac{d}{2}+\nu) u}}{(1-e^{-u})^d}, \,\,\,\,\, f^e_{s,\nu}(u)=D^{d+2}_{s-1}\frac{1+e^{-u}}{1-e^{-u}}\frac{e^{-(\frac{d-2}{2}+\nu) u}}{(1-e^{-u})^{d-2}}
\end{align}
$f^b_{s,\nu}(u)$ induces the {\it bulk} character zeta function $\bar\zeta^b_{s, \nu}(z)\equiv \frac{1}{\Gamma(z)}\int^{\infty}_{0}\frac{du}{u}u^z f^b_{s,\nu}(u)$ and $f^e_{s,\nu}(u)$ induces the {\it edge} character zeta function $\bar\zeta^e_{s, \nu}(z)\equiv \frac{1}{\Gamma(z)}\int^{\infty}_{0}\frac{du}{u}u^z f^e_{s,\nu}(u)$. The renormalized partition function (with all negative powers of $\epsilon$ dropped) is 
\small
\begin{align}\label{oddmassive}
\log Z^{\text{ren}}_{s, \nu}&=\frac{1}{2}\left(\bar\zeta^b_{s,\nu}(0)-\bar\zeta^e_{s,\nu}(0)\right)\log(2e^{-\gamma}/\epsilon)+\frac{1}{2}\left(\bar\zeta^{b'}_{s, \nu}(0)-\bar\zeta^{e'}_{s, \nu}(0)\right)-\sum_{\ell=0}^{d+1} b_{\ell}(s)\left(H_\ell-\frac{1}{2}H_{\ell/2}\right)\nu^\ell
\end{align}
\normalsize
where $\sum_{\ell=0}^{d+1}b_\ell(s)\nu^\ell=\frac{1}{2}(\bar\zeta^b_{s,\nu}(0)-\bar\zeta^e_{s,\nu}(0))$. Using the expansion $f^b_{s,\nu}(u)=\sum_{n\ge 0}P^b_{s,\nu}(n)e^{-(\Delta+n)t}$, where $\Delta=\frac{d}{2}+\nu$, the bulk character zeta function $\bar\zeta^b_{s, \nu}(z)$ can be written as a finite sum of Hurwitz zeta functions 
\begin{align}
\bar\zeta^b_{s, \nu}(z)=P^b_{s,\nu}(\delta-\Delta)\zeta(z, \Delta)
\end{align}
and similarly for $\bar\zeta^e_{s, \nu}(z)$. In $\text{AdS}_4$, for example, the bulk and edge character zeta functions are 
\begin{align}
&\bar\zeta^b_{s, \nu}(z)=(2s+1)\left(\frac{1}{3}\zeta(z-3,\Delta)-\nu\, \zeta(z-2,\Delta)+(\nu^2-\frac{1}{12})\zeta(z-1,\Delta)+\frac{\nu(1-4\nu^2)}{12}\zeta(z,\Delta)\right)\nonumber\\
&\bar\zeta^e_{s, \nu}(z)=\frac{1}{3}s(s+1)(2s+1)\left(\zeta(z-1,\Delta-1)-\nu \, \zeta(z,\Delta-1)\right)
\end{align}

For a massless representation of spin-$s$ and depth-$t$, the unregularized partition consists of four parts
\begin{align}
\log Z_{s, \nu_t}=\int_{0}^\infty \frac{du}{2u} \, \left(f^b_{s,\nu_t}(u)-f^e_{s,\nu_t}(u)-f^b_{t,\nu_s}(u)+f^e_{s,\nu_t}(u)\right)
\end{align}
The renormalized partition function can be expressed as
\small
\begin{align}\label{oddmassless}
\log Z^{\text{ren}}_{s, \nu_t}&=\frac{1}{2}\left(\bar\zeta^b_{s,\nu_t}(0)-\bar\zeta^e_{s,\nu_t}(0)-\bar\zeta^b_{t,\nu_s}(0)+\bar\zeta^e_{t,\nu_s}(0)\right)\log(2e^{-\gamma}/\epsilon)+\frac{1}{2}\left(\bar\zeta^{b'}_{s,\nu_t}(0)-\bar\zeta^{e'}_{s,\nu_t}(0)-\bar\zeta^{b'}_{t,\nu_s}(0)+\bar\zeta^{e'}_{t,\nu_s}(0)\right)\nonumber\\
&-\left(\sum_{\ell=0}^{d+1} b_{\ell}(s)\left(H_\ell-\frac{1}{2}H_{\ell/2}\right)\nu_t^\ell-\sum_{\ell=0}^{d+1} b_{\ell}(t)\left(H_\ell-\frac{1}{2}H_{\ell/2}\right)\nu_s^\ell\right)
\end{align}
\normalsize
For example, the renormalized partition function of linearized gravity in $\text{AdS}_4$ is 
\begin{align}\label{gravity}
\log Z^{\text{ren}}_{\text{gravity}}=\frac{47 \log \, A}{6}+\frac{1}{3} \zeta '(-3)+\frac{1957}{288}-\frac{\log 2}{2}-5 \log (2 \pi )-\frac{571}{90}\log(2e^{-\gamma}/\epsilon)
\end{align}
where $A$ is the Glaisher-Kinkelin constant. (\ref{gravity}) reproduces the $s=2$ result in \cite{Giombi:2013fka}.

\textbf{Comment}: Apart from the multiplicative anomaly, the remaining part of the partition function for both massive and massless  representations is completely captured by $\frac{1}{\Gamma(z)}\int^{\infty}_{0}\frac{du}{u^{1-z}} \frac{1+e^{-u}}{1-e^{-u}}\chi_{\text{tot}}(u)$, where $\chi_{\text{tot}}(u)$ consists of bulk and edge characters. Thus, by inserting a UV regulator  $u^z$ in the unregularized partition function, we recover the correct $\log$-divergence and finite part up to  multiplicative anomaly, which was proved to vanish when summing up the whole spectrum of Vasiliev theories \cite{Basile:2018zoy}. (We'll also show in section \ref{Vasiliev} the vanishing of multiplicative anomaly for type-A Vasiliev gravity using the character integral formalism).
Keeping this comment in mind, we are free to use the unregularized character integral formula to compute the full partition function of Vasiliev theories in section \ref{Vasiliev}.

\section{Double trace deformation}\label{Double}
A large $N$ $\text{CFT}_{d}$ where a primary operator $\CO$ has scaling dimension $\Delta_\CO$ can flow to another $\text{CFT}_{d}$ where $\CO$ has the shadow scaling dimension $\bar\Delta_\CO\equiv d-\Delta_\CO$ by turning on a double trace deformation $\CO^2$ in the Lagrangian \cite{Witten:2001ua, Gubser:2002vv}. On the AdS side, this RG flow is equivalent to  switching the boundary conditions when we quantize the dual bulk field. Due to  AdS/CFT duality, the effect of this RG flow on the partition function of the large $N$ $\text{CFT}_{d}$ living on $S^d$ can be computed from both boundary and bulk sides   \cite{Gubser:2002vv, Gubser:2002zh,Hartman:2006dy,Diaz:2007an,Klebanov:2011gs, Giombi:2013yva}.  In particular, in \cite{Giombi:2013yva} the authors thoroughly computed the effect of any higher spin currents. Let $\CO_{\mu_1\cdots\mu_s}$ be a spin-$s$ current and they found the change of free energy induced by $\CO_{\mu_1\cdots\mu_s}\CO^{\mu_1\cdots\mu_s}$ has $log$-divergence when $\Delta_{\CO}=d+s-2$, i.e. $\CO$ is a conserved current, and the change is of order 1 when $\Delta_{\CO}$ takes other values. In this section, we'll reproduce the main results of \cite{Giombi:2013yva} on bulk side by using character integral formula (\ref{fullreg}). As we've just mentioned that the double trace deformation induces the dual boundary condition,  it suffices to compute $ \log Z_{s, \nu}-\log Z_{s, -\nu}$.  We'll focus on the odd $d$ case (the even $d$ case can be analyzed similarly) and see that it's extremely convenient to use the character integral representation to do this computation because flipping the boundary condition is equivalent to switching to the character of the dual representation, i.e. $\chi^{\text{AdS}_{d+1}}_{[\Delta, s]}\to\chi^{\text{AdS}_{d+1}}_{[\bar\Delta, s]}$.  

We'll start from  considering  a scalar field with complex scaling dimension $\Delta=\frac{d}{2}+i\nu$ and then Wick rotate $\Delta$ to a real number. Before performing any actual computation, we want to mention the following observation which is based on the explicit evaluation in last section,  that the UV-divergent part including multiplicative anomaly of $\log Z_{\nu}$ is an even function in $\nu$ when $d$ is odd (see equation (\ref{regzuv}) as an explicit example). This observation implies that $\log Z_{i\nu}-\log Z_{-i\nu}$  is UV finite. In addition, in the difference
\begin{align}
\log Z_{i\nu}(\epsilon)-\log Z_{-i\nu}(\epsilon)=\frac{1}{2}\int_{\mathbb{R}+i\delta}\frac{du}{2 \sqrt{u^2+\epsilon^2}}\,\frac{1+e^{-u}}{1-e^{-u}}\frac{e^{-\frac{d}{2}u}(e^{-i \nu \sqrt{u^2+\epsilon^2}}-e^{+i \nu \sqrt{u^2+\epsilon^2}})}{(1-e^{-u})^d}
\end{align}
the integrand is indeed a single-valued function because when $u$ goes around one of the branch points of $\sqrt{u^2+\epsilon^2}$, the combination $\frac{e^{-i \nu \sqrt{u^2+\epsilon^2}}-e^{+i \nu \sqrt{u^2+\epsilon^2}}}{\sqrt{u^2+\epsilon^2}}$ keeps invariant though $\sqrt{u^2+\epsilon^2}$ itself picks an extra minus sign. Then we are free to shift the $u$-contour upwards such that $\delta>\epsilon$. With this new contour and using the UV-finiteness of $\log Z_{i\nu}-\log Z_{-i\nu}$, we can safely put $\epsilon\to 0$ which  amounts to sending $\sqrt{u^2+\epsilon^2}$ to $u$:
\begin{align}\label{deltaZnu1}
\log Z_{i\nu}-\log Z_{-i\nu}=\frac{1}{2}\int_{\mathbb{R}+i\delta}\frac{du}{2\,u}\,\frac{1+e^{-u}}{1-e^{-u}}\frac{e^{-\frac{d}{2}u}(e^{-i \nu u}-e^{+i \nu u})}{(1-e^{-u})^d}
\end{align}
To proceed further, we introduce a new integral that can eliminate $u$ in the denominator of (\ref{deltaZnu1}) and then switch the order of integrals
\begin{align}
\log Z_{i\nu}-\log Z_{-i\nu}=-\frac{i}{4}\int_{-\nu}^\nu d\lambda\,\int_{\mathbb{R}+i\delta} du\,\frac{1+e^{-u}}{1-e^{-u}}\frac{e^{-\frac{d}{2}u-i\lambda u}}{(1-e^{-u})^d}
\end{align}
 After these manipulations, the $u$-integral is essentially the definition of $\tilde\mu^{(+)}_d(\lambda)$, cf. (\ref{tildemu})
\begin{align}
\log Z_{i\nu}-\log Z_{-i\nu}=\frac{i\pi(-)^\frac{d-1}{2}}{2 \, d!}\int_{-\nu}^\nu d\lambda\,\tilde\mu^{(+)}_d(\lambda)=\frac{i (-)^r}{ d!}\int_{0}^\nu d\lambda\,\mu^{(d)}(\lambda)
\end{align}
where we've used that the even part of $\tilde\mu^{(+)}_d(\lambda)$ is the spectral density $\mu^{(d)}(\lambda)$. It's straightforward to generalize this method to higher spin fields by including the edge-mode  contribution and using eq. (\ref{WdSF})
\begin{align}\label{smu}
\log Z_{s, i\nu}-\log Z_{s, -i\nu} = \frac{i\pi (-)^r}{ d!}D^{d}_s\int_{0}^\nu d\lambda\, \mu^{(d)}_{s}(\lambda)
\end{align}  
Surprisingly, the change of free energy triggered by the higher spin double trace deformation at large $N$ is completely encoded in the higher spin spectral density. Given the eq. (\ref{smu}), we make a Wick rotation $\nu\to i\nu$ to obtain result for real scaling dimension $\Delta=\frac{d}{2}+\nu$
\begin{align}\label{DTD}
\log Z_{s, \nu}-\log Z_{s, -\nu}&=\frac{\pi (-)^r}{ d!}D^d_s\int_{0}^\nu d\lambda\, \mu_s^{(d)}(i \lambda)
\end{align} 
For example, at $d=3$ we get
\begin{align}\label{DTD1}
\log Z_{s, \nu}-\log Z_{s, -\nu}=\frac{(2s+1)\pi}{6}\int^{\Delta}_{\frac{3}{2}}
dx(x-\frac{3}{2})(x+s-1)(x-s-2)\cot(\pi x)
\end{align}
where we've changed variable $\lambda=x-\frac{3}{2}$.  This equation agrees with the result in \cite{Giombi:2013yva}. When $\nu$ reaches some half integer number, say $\nu_{s-1}=\frac{d}{2}+s-2$, which corresponds to a double trace deformation triggered by a spin-$s$ conserved current in free $U(N)$ vector model, the integral (\ref{DTD}) is divergent no matter what contour we use because the singularity $\nu=\nu_{s-1}$ is at the end point of the integration contour. To extract the leading divergence, we need the singular behavior of $\mu_s^{(d)}(i\lambda)$ near $\lambda=\nu_{s-1}$
\begin{align}
\frac{\pi (-)^r}{ d!}D^{d}_{s} \,\mu^{(d)}_s(i\lambda)&=-\frac{D^{d+2}_{s-1, s-1}}{2(\lambda-\nu_{s-1})}+\CO((\lambda-\nu_{s-1})^0)
\end{align}
which is a consequence of eq. (\ref{Dns}) and $D^{d+2}_{p-1, s}=-D^{d+2}_{s-1, p}$. Notice that $D^{d+2}_{s-1, s-1}\equiv n^{\text{KT}}_{s-1}$ is  the number of spin-$(s\!-\!1)$ Killing tensors on $S^{d+1}$ and also the number of spin-$(s-1)$ conformal Killing tensors on $S^d$ \cite{Giombi:2013yva}. Therefore if we truncate the integral (\ref{DTD}) at $\nu=\nu_s-\epsilon$, the change of $\log Z$ induced by a spin-$s$ conserved current has a $\log$-divergence part $\frac{1}{2}n^{\text{KT}}_{s-1}\log(\epsilon)$. 

%This suggests that, for $\Delta=d+s-2$
%\begin{align}
%\log Z_{s, \nu_{s-1}}-\log Z_{s, -\nu_{s-1}}=\frac{1}{2}n^{\text{KT}}_{s-1}\log N+\CO(N^0)
%\end{align}
%in the large $N$ limit. Thus the conclusion is that the change of free energy caused by $\CO_{\mu_1\cdots\mu_s}\CO^{\mu_1\cdots\mu_s}$ is of order $\log(N)$ when $\CO_{\mu_1\cdots\mu_s}$ is a conserved current and of order 1 otherwise.

\section{Application to Vasiliev theories}\label{Vasiliev}
With the character integral method developed in the previous sections, we're finally able to compute partition function of Vasiliev theories in all even dimensional AdS (The odd dimensional AdS case can be analyzed similarly and is indeed much simpler). We'll use (non)minimal type-$\text{A}_{\ell}$ theory and type-$\text{B}$ Vasiliev theory, which are reviewed below, to illustrate  the application of the character integral method. Before that we want to stress again, due to the comment at the end of the section \ref{rigovalue}, the unregularized version of character integral formula is sufficient.

\subsection{A brief review of Vasiliev theories and  Flato-Fronsdal theorems}
The simplest and best understood higher spin theory is the nonminimal type-A Vasiliev theory in $\text{AdS}_{d+1}$, which contains a $\Delta=d-2$ real scalar and a tower of massless higher spin gauge fields.  This theory is believed to be dual to a free $U(N)$ vector model on boundary described by Lagrangian $\CL=\frac{1}{2}\phi^*_i\,\square\, \phi_i, 1\le i\le N$.  The $U(N)$ fundamental field $\phi_i$ is in the scalar singleton representation $\left[\frac{d-2}{2}, 0\right]\equiv \text{Rac}$  of $\text{SO}(2, d)$. One direct result of the duality is a one-to-one correspondence between  the field content in bulk and the single-trace operators in $U(N)$ vector model. In representation theory, this is confirmed by Flato-Fronsdal theorem \cite{Flato:1978qz}: 
\begin{align}\label{sp1}
\text{Rac}\otimes\text{Rac}=\bigoplus_{s=0}^\infty [d+s-2, s]
\end{align}
which can be proved by using the following identity of characters 
\begin{align}\label{FF1}
\left(\chi^{\so(2,d)}_{\text{Rac}}(u)\right)^2=\sum_{s=0}^\infty \chi^{\text{AdS}_{d+1}}_{[d+s-2, s]}(u)\equiv \chi^{\text{AdS}_{d+1}}_{\text{A}}, \,\,\,\,\,\, \chi^{\so(2,d)}_{\text{Rac}}=\frac{e^u-e^{-u}}{(e^{\frac{u}{2}}-e^{-\frac{u}{2}})^d}
\end{align}
There are a lot of variants of the original type-A Vasiliev theory. For example, if we relax the requirement of unitarity, we can take the boundary CFT to be $\CL=\frac{1}{2}\phi^*_i\,\square^\ell\, \phi_i$, where each $\phi_i$ is in the representation $[\frac{d-2\ell}{2}, 0]\equiv \text{Rac}_\ell$  of $\text{SO}(2, d)$. For more details about the $\square^\ell$ theory, we refer readers to \cite{Brust:2016zns, Bekaert:2013zya}.  The bulk dual of this nonunitary CFT is called the type-$\text{A}_\ell$ higher spin gravity with  field content  given by a generalized Flato-Fronsdal theorem 
\cite{Bekaert:2013zya, Basile:2014wua, Grigoriev:2014kpa}
\begin{align}\label{sp2}
\text{Rac}_\ell\otimes\text{Rac}_\ell=\bigoplus_{p=1,3,\cdots}^{2\ell-1}\bigoplus_{s=0}^{\infty} [d+s-p-1, s]
\end{align}
where $[d+s-p-1, s]$ corresponds to a PM field of spin-$s$ and depth-$(s-p)$ for $p\le s$. At the level of characters, this tensor product decomposition is equivalent to 
\begin{align}\label{FF2}
\left(\chi^{\so(2,d)}_{\text{Rac}_\ell}(u)\right)^2=\sum_{p=1,3,\cdots}^{2\ell-1}\sum_{s=0}^\infty \chi^{\text{AdS}_{d+1}}_{[d+s-2, s]}(u)\equiv \chi^{\text{AdS}_{d+1}}_{\text{A}_\ell}, \,\,\,\,\,\, \chi^{\so(2,d)}_{\text{Rac}_\ell}=\frac{e^{\ell u}-e^{-\ell u}}{(e^{\frac{u}{2}}-e^{-\frac{u}{2}})^d}
\end{align}
In  type-$\text{A}_\ell$ theory,  we can further replace the complex scalars by real scalars that are in the fundamental representation of $O(N)$. The resulting AdS dual is called the minimal type-$\text{A}_\ell$ theory and its field content can be extracted from the symmetrized tensor product of two $\text{Rac}_\ell$:
\begin{align}\label{sp3}
\text{Rac}_\ell\odot\text{Rac}_\ell=\bigoplus_{p=1,3,\cdots}^{2\ell-1}\bigoplus_{s=0,2,\cdots}^{\infty} [d+s-p-1, s]
\end{align}
where only fields even spin exist. Summing over the characters for representations appearing in the tensor product decomposition (\ref{sp3}) leads to 
\begin{align}\label{FF3}
\chi^{\text{AdS}_{d+1}}_{\text{A}^{\text{min}}_\ell}\equiv \sum_{p=1,3,\cdots}^{2\ell-1}\sum_{s=0,2,\cdots}^\infty \chi^{\text{AdS}_{d+1}}_{[d+s-2, s]}(u)=\frac{1}{2}\left(\chi^{\so(2,d)}_{\text{Rac}_\ell}(u)\right)^2+\frac{1}{2}\chi^{\so(2,d)}_{\text{Rac}_\ell}(2u)
\end{align}
Another important variant of the original nonminimal type-A theory is the so-called type-B theory. It is the AdS-dual of free $U(N)$ Dirac fermions restricted to $U(N)$ singlet sector. Each Dirac fermion carries the spinor singleton representation $[\frac{d-1}{2}, \mathbf{\frac{1}{2}}]\equiv \text{Di}$ of $\text{SO}(2,d)$, where $\mathbf{\frac{1}{2}}$ denotes the spin-$\frac{1}{2}$ representation of $\so(d)$. The bulk field content is given by $\text{Di}\otimes \text{Di}$, which takes the following for odd $d$ \cite{Flato:1978qz}
\begin{align}\label{sp4}
\text{Di}\otimes \text{Di}=[d-1,0]\bigoplus\bigoplus_{m=0}^{\frac{d-3}{2}}\bigoplus_{s=1}^\infty [d-2+s, (s, 1^m)]
\end{align}
where $(s,1^m)$ is a shorthand notation for an $\so(d)$ Young diagram with $s$ boxes in the first row and one box in the following $m$ rows. When $d=3$, all $(s,1^m)$ are reduced to a spin-$s$ representation of $\so(3)$. Thus in $\text{AdS}_4$, the spectra of the type-$\text{A}$ and type-$\text{B}$ theory are the same except that the $m^2=-2$ scalar is quantized with $\Delta_-=1$ in the former and $\Delta_+=2$ in the latter. However, for higher $d$, the spectrum type-$\text{B}$ theory is much more complicated due to the presence of fields of mixed symmetry. Let's call the collection of fields of ``spin'' $(s, 1^m)$ the $m$-sector. The $m=0$ sector is almost the same as spectrum of type-A theory except the scaling dimension of the scalar. For the $m\ge 1$ sectors, fields with $s\ge 2$ are massless gauge fields with the corresponding ghost fields in the representation $[d\!-\!1\!+\!s, (s\!-\!1,1^m)]$ of $\text{SO}(2, d)$ while the $s=1$ fields are massive and totally antisymmetric.

\subsection{Type-A higher spin gravity}
\noindent{}\textbf{Nonminimal theory}:
Field content of the nonmonimal type-A higher spin gravity is given by eq. (\ref{sp1}). Using (\ref{scalar}) for the $\Delta=d-2$ scalar and  (\ref{pow}) for the massless gauge fields, we get
\begin{align}\label{loZA}
\log Z^{\text{AdS}_{d+1}}_{\text{A}}=\int^\infty_0 \frac{du}{2u}\frac{1+e^{-u}}{1-e^{-u}}\left(\sum_{s\ge 0}\chi^{\text{AdS}_{d+1}}_{[d+s-2, s]}(u)-\left(e^{\frac{u}{2}}-e^{-\frac{u}{2}}\right)^4\sum_{s\ge 0}\chi^{\text{AdS}_{d+3}}_{[d+2+s-2, s]}(u)\right)
\end{align}
where $s$ is shifted by 1 in the edge character to match with the bulk part. Due to the Flato-Fronsdal theorem (\ref{FF1}), both sums in  (\ref{loZA}) give the square of a Rac-character. Plugging in the explicit form of these Rac-characters, it's clear that the bulk and edge contributions exactly cancel out and hence the total one-loop free energy of type-A theory vanishes:
\begin{align}
\log Z^{\text{AdS}_{d+1}}_{\text{A}}=\int^\infty_0 \frac{du}{2u}\frac{1+e^{-u}}{1-e^{-u}}\left[\left(\chi^{\so(2,d)}_{\text{Rac}}(u)\right)^2-\left(e^{\frac{u}{2}}-e^{-\frac{u}{2}}\right)^4\left(\chi^{\so(2,d+2)}_{\text{Rac}}(u)\right)^2\right]=0
\end{align}
\\
Before moving to the minimal case, we want to check that the total multiplicative anomaly indeed vanishes. To do this, we should use the fully regularized character integral formula, with which the exact cancellation between bulk and edge contributions  doesn't hold any more at the integrand level. Instead, we get 
\begin{align}\label{regsum}
\log Z^{\text{AdS}_{d+1}}_{\text{A}}=4^{-d}\int_{\mathbb{R}+i\delta} \frac{du}{ r u}\frac{1+e^{-u}}{1-e^{-u}}\frac{(1+\cosh(ru))(\cosh(ru)-\cosh(u))}{(\sinh\frac{u}{2}\sinh\frac{ru}{2})^d}
\end{align}
where $r=\sqrt{u^2+\epsilon^2}/u$. The multiplicative anomaly, if exists, should appear as the coefficient of $\epsilon^0$ in the small $\epsilon$ expansion of (\ref{regsum}), which can be realized by a change of variable $u\to \epsilon\, u$ and expanding the integrand around small $\epsilon$:
\small
\begin{align}\label{regsum1}
\log Z^{\text{AdS}_{d+1}}_{\text{A}}=4^{-d}\int_{\mathbb{R}+i\delta} \frac{du}{\sqrt{u^2+1}}\frac{1+e^{-\epsilon u}}{1-e^{-\epsilon u}}\frac{(1+\cosh(\epsilon\sqrt{1+u^2}))(\cosh(\epsilon\sqrt{1+u^2})-\cosh(\epsilon u))}{(\sinh\frac{\epsilon u}{2}\sinh\frac{\epsilon\sqrt{1+u^2}}{2})^d}
\end{align}
\normalsize
Notice that the integrand of (\ref{regsum1}) is an odd function of $\epsilon$ and hence cannot have any $\epsilon^0$ term in small $\epsilon$ expansion. This observation leads to the vanishing of the total multiplicative anomaly in nonminimal type-A theory. 
\vspace{10pt}

\noindent{}\textbf{Minimal theory}: Since minimal type-A theory contains only  fields of even spins, its total partition function can be written as
\begin{align}\label{minZ}
\log Z^{\text{AdS}_{d+1}}_{\text{A}^{\text{min}}}=\int^\infty_0 \frac{du}{2u}\frac{1+e^{-u}}{1-e^{-u}}\left(\sum_{s \,\,\,\text{even}}\chi^{\text{AdS}_{d+1}}_{[d+s-2, s]}(u)-\left(e^{\frac{u}{2}}-e^{-\frac{u}{2}}\right)^4\sum_{s\,\,\, \text{odd}}\chi^{\text{AdS}_{d+3}}_{[d+2+s-2, s]}(u)\right)
\end{align}
where the spin in edge characters is shifted by 1. The sum over all bulk characters lead to $\chi^{\text{AdS}_{d+1}}_{\text{A}^{\text{min}}}$ 
\begin{align}\label{minchar}
\chi^{\text{AdS}_{d+1}}_{\text{A}^{\text{min}}}(u)=\frac{1}{2}\chi^{\text{AdS}_{d+1}}_{\text{A}}(u)+\frac{1}{2}\chi^{\so(2,d)}_{\text{Rac}}(2u)
\end{align}
where we've used eq. (\ref{FF1}) and (\ref{FF3}).
The sum of edge characters, since only odd spin fields are involved, yields the difference between the nonminimal character and minimal character in $\text{AdS}_{d+3}$:
\begin{align}\label{oddchar}
\chi^{\text{AdS}_{d+3}}_{\text{A}}(u)-\chi^{\text{AdS}_{d+3}}_{\text{A}^{\text{min}}}(u)=\frac{1}{2}\chi^{\text{AdS}_{d+3}}_{\text{A}}(u)-\frac{1}{2}\chi^{\so(2,d+2)}_{\text{Rac}}(2u)
\end{align}
Plugging eq. (\ref{minchar}) and (\ref{oddchar}) into (\ref{minZ}), the type-A characters cancel out as in the nonminimal theory and thus  the remaining term is  
\begin{align}\label{Amin}
\log Z^{\text{AdS}_{d+1}}_{\text{A}^{\text{min}}}&=\frac{1}{2}\int^\infty_0 \frac{du}{2u}\frac{1+e^{-u}}{1-e^{-u}}\left(\chi^{\so(2,d)}_{\text{Rac}}(2u)+\left(e^{\frac{u}{2}}-e^{-\frac{u}{2}}\right)^4\chi^{\so(2,d+2)}_{\text{Rac}}(2u)\right)
\nonumber\\
&=\int^\infty_0 \frac{du}{2u}\frac{1+e^{-u}}{1-e^{-u}}\frac{e^{-\left(\frac{d-1}{2}+\frac{1}{2}\right)u}+e^{-\left(\frac{d-1}{2}-\frac{1}{2}\right)u}}{(1-e^{-u})^{d-1}}
\end{align}
where in the second line $u$ has been rescaled $u\to \frac{u}{2}$. Notice that  $\frac{e^{-\frac{d}{2} u}+e^{-(\frac{d}{2}-1)u}}{(1-e^{-u})^{d-1}}$ is the Harish-Chandra character of the $\Delta=\frac{d-2}{2}$ representation of $\text{SO}(1, d)$. Then according to the character integral representation of the sphere partition functions found in \cite{Anninos:2020hfj},   the partition function of minimal type-A theory on $\text{AdS}_{d+1}$ is the same as the partition function of a conformally coupled  scalar on $S^d$. This result  agrees with \cite{Giombi:2013fka}, where the appearance of this scalar partition function  is interpreted as an $N\to N-1$ shift in the identification of Newton's constant $G_N\sim\frac{1}{N}$. Again, starting from the square root regularized character and following the same argument as in the nonminimal case, one can also show the vanishing of total multiplicative anomaly for minimal type-A theory. Let's also mention that when $d=2r$ is even, the analogue of (\ref{Amin}) implies the coefficient of $\log R$  of minimal type-A theory in $\text{AdS}_{2r+1}$ matches the Weyl anomaly of a conformally couple scalar on the boundary, which is a $d$-dimensional sphere of radius $R$. 

\subsection{Type-$\text{A}_\ell$ higher spin gravities}
\noindent{}\textbf{Nonminimal theory}:
Given the spectrum of nonminimal type-$\text{A}_\ell$ theory (\ref{sp2}), the total partition function can be written as 
\small
\begin{align}
\log Z^{\text{AdS}_{d+1}}_{\text{A}_\ell}=\int^\infty_0 \frac{du}{2u}\frac{1+e^{-u}}{1-e^{-u}}\left(\sum_{p=1,3}^{2\ell-1}\sum_{s\ge 0}\chi^{\text{AdS}_{d+1}}_{[d+s-p-1, s]}(u)-\left(e^{\frac{u}{2}}-e^{-\frac{u}{2}}\right)^4\sum_{p=1,3}^{2\ell-1}\sum_{s\ge 0}\chi^{\text{AdS}_{d+3}}_{[d+2+s-p-1, s]}(u)\right)
\end{align}
\normalsize
where the spin label $s$ is shifted by 1 in the sum of edge characters. Using the generalized Flato-Fronsdal theorem (\ref{FF2}), it's straightforward to show that the bulk and edge contributions exactly cancel out
\begin{align}
\log Z^{\text{AdS}_{d+1}}_{\text{A}_\ell}=\int^\infty_0 \frac{du}{2u}\frac{1+e^{-u}}{1-e^{-u}}\left[\left(\chi^{\so(2,d)}_{\text{Rac}_\ell}(u)\right)^2-\left(e^{\frac{u}{2}}-e^{-\frac{u}{2}}\right)^4\left(\chi^{\so(2,d+2)}_{\text{Rac}_\ell}(u)\right)^2\right]=0
\end{align}
Therefore the total free energy of nonminimal type-$\text{A}_\ell$ also vanishes.

\vspace{10pt}

\noindent{}\textbf{Minimal theory}: Following the same steps as in the minimal type-A case, we can directly write down the result for minimal type-$\text{A}_\ell$ theory
\begin{align}\label{minAl}
\log Z^{\text{AdS}_{d+1}}_{\text{A}_\ell^{\text{min}}}&=\frac{1}{2}\int^\infty_0 \frac{du}{2u}\frac{1+e^{-u}}{1-e^{-u}}\left(\chi^{\so(2,d)}_{\text{Rac}_\ell}(2u)+\left(e^{\frac{u}{2}}-e^{-\frac{u}{2}}\right)^4\chi^{\so(2,d+2)}_{\text{Rac}_\ell}(2u)\right)
\nonumber\\
&=\int^\infty_0 \frac{du}{2u}\frac{1+e^{-u}}{1-e^{-u}}\frac{e^{-\frac{d}{2}u}\,(e^{\ell u}-e^{-\ell u})}{(1-e^{-u})^{d}}
\end{align}
where $u$ is rescaled in the second line. Naively speaking, eq. (\ref{minAl}) doesn't look like any character integral. But using the expansion $1-e^{-2\ell u}=(1-e^{-u})\sum_{n=0}^{2\ell-1}e^{-n u}$,  we can transform it into a finite sum of character integrals in the dS sense
\begin{align}\label{Zmin3}
\log Z^{\text{AdS}_{d+1}}_{\text{A}_\ell^{\text{min}}}=\sum_{n=1}^\ell\int^\infty_0 \frac{du}{2u}\frac{1+e^{-u}}{1-e^{-u}}\frac{e^{-(\frac{d-1}{2}+(n-\frac{1}{2}))u}+e^{-(\frac{d-1}{2}-(n-\frac{1}{2}))u}}{(1-e^{-u})^{d-1}}
\end{align}
where $\frac{e^{-(\frac{d-1}{2}+(n-\frac{1}{2}))u}+e^{-(\frac{d-1}{2}-(n-\frac{1}{2}))u}}{(1-e^{-u})^{d-1}}$ is the Harish-Chandra character of the $\Delta=\frac{d-1}{2}+n-\frac{1}{2}$ representation of $\SO(1, d)$ and the corresponding character integral represents the one-loop partition function of a scalar field of mass $m_n^2=(\frac{d}{2}-n)(\frac{d-2}{2}+n)$ on $S^d$ \cite{Anninos:2020hfj}.  Defining a collection of scalar Laplacians $\{-\nabla^2+m^2_n\}_{1\le n\le \ell}$, the partition function $\log Z^{\text{AdS}_{d+1}}_{\text{A}_\ell^{\text{min}}}$ can be rewritten as
\begin{align}
\log Z^{\text{AdS}_{d+1}}_{\text{A}_\ell^{\text{min}}}&=\sum_{n=1}^\ell\log \det(-\nabla^2+m^2_n)^{-\frac{1}{2}}=\log \det\left(\square^\ell_{S^d}\right)^{-\frac{1}{2}}, \,\,\,\,\, \square^\ell_{S^d}\equiv \prod_{n=1}^\ell(-\nabla^2+m^2_n)
\end{align}
where we are allowed to put the Laplacian into a product form because  there is no multiplicative anomaly on an odd dimensional manifold. Notice that $\square^\ell_{S^d}$, the Weyl-covariant generalization of $\square^\ell$, is a GJMS operator on $S^d$ \cite{Juhl:2009, Juhl:2011ua, Fefferman:2012, Beccaria:2015vaa}  and  when $\ell=1$ it is reduced to the conformal Laplacian on $S^d$. Therefore, the one-loop partition function of minimal type-$\text{A}_\ell$ theory on $\text{AdS}_{d+1}$ is the same as the one-loop partition function of the $\square^\ell$-theory on $S^d$. This is again consistent with the $N\to N-1$ interpretation.

\subsection{Type-B higher spin gravities}
\noindent{}$\mathbf{\textbf{AdS}_4}$:
Let's start considering the type-B theory in $\text{AdS}_4$. Its has the same spectrum as  type-A theory except the boundary condition imposed on the scalar is flipped. Using $\log Z^{\text{AdS}_4}_{\text{A}}=0$, we are left with 
\begin{align}\label{B1}
\log Z^{\text{AdS}_4}_{\text{B}}&=\int^\infty_0\frac{du}{2u}\frac{1+e^{-u}}{1-e^{-u}}\left(\chi_{\Delta=2}^{\text{AdS}_4}(u)-\chi_{\Delta=1}^{\text{AdS}_4}(u)\right)=-\int^\infty_0\frac{du}{2u}\frac{1+e^{-u}}{1-e^{-u}}\frac{e^{-u}}{(1-e^{-u})^2}
\end{align}
which represents the change of partition function induced by a double-trace deformation. 
To evaluate this integral, we can either regularize it by inserting $u^z$ and express it in terms of Hurwitz zeta function or directly use eq. (\ref{DTD1})
\begin{align}
\log Z^{\text{AdS}_4}_{\text{B}}=\frac{\pi}{6}\int_{\frac{3}{2}}^2(x-1)(x-\frac{3}{2})(x-2)\cot(\pi x)=\frac{\zeta(3)}{8\pi^2}
\end{align}

\vspace{10pt}

\noindent{}$\mathbf{\textbf{AdS}_6}$ \textbf{and higher}:
The spectrum of type-B theory in $\text{AdS}_6$ can be divided into the $m=0$ sector and the $m=1$ sector.
\begin{align}\label{s6}
\underbrace{\left([4, 0]\bigoplus\bigoplus_{s\ge 1}[3+s, (s,0)]\right)}_{m=0\,\, \text{sector}}\bigoplus\underbrace{\left(\bigoplus_{s\ge 1}[3+s, (s,1)]\right)}_{m=1\,\, \text{sector}}
\end{align}
In the $m=0$ sector we can turn to  the result of type-A theory because the only difference is  scaling dimension of the scalar field. Using $\log Z^{\text{AdS}_6}_{\text{A}}=0$, we obtain the following result without any extra effort
\begin{align}\label{P1}
\log Z^{\text{AdS}_6}_{m=0}=\int^\infty_0\frac{du}{2u}\frac{1+e^{-u}}{1-e^{-u}}\frac{e^{-4 u}-e^{-3 u}}{(1-e^{-u})^5}=-\int^\infty_0\frac{du}{2u}\frac{1+e^{-u}}{1-e^{-u}}\frac{e^{-3 u}}{(1-e^{-u})^4}
\end{align}
Unlike in $\text{AdS}_4$, the $m=0$ partition function $\log Z^{\text{AdS}_6}_{m=0}$ itself doesn't have the double trace-deformation interpretation because $\Delta=3$ and $\Delta=4$ are not conjugate scaling dimensions. We'll see that the double-trace deformation pattern can be restored with the $m=1$ sector taken into account. The $m=1$ sector is  more  involving since it consists of fields with mixed symmetry. Following the same steps as in section \ref{HSchar}, we derive the character integral formula for fields in massive representation $\left[\frac{5}{2}+\nu, (s,1)\right]$
\small
\begin{align}\label{m=1}
\log Z_{(s,1), \nu}=\int_0^\infty\frac{du}{2u}\frac{1+e^{-u}}{1-e^{-u}}\left[D^{5}_{s,1} \frac{e^{-(\frac{5}{2}+\nu)u}}{(1-e^{-u})^5}-D^{7}_{s-1}\left(\frac{3\,e^{-(\frac{3}{2}+\nu)u}}{(1-e^{-u})^3}-\frac{e^{-(\frac{1}{2}+\nu)u}}{1-e^{-u}}\right)\right]
\end{align}
\normalsize
Unlike the spin-$s$ case, the edge part of $\log Z_{(s,1), \nu}$ should be interpreted as the 1-loop path integral of $D^{7}_{s-1}$ massive spin-1 fields of scaling dimension $\frac{d}{2}+\nu$ living on $\text{EAdS}_4$, which is the horizon of the Rindler patch of $\text{AdS}_6$. Though this new observation \footnote{More generally, we find that for a massive field with hook-like spin $(s, 1^m)$ and scaling dimension $\Delta=\frac{d}{2}+\nu$ in $\text{AdS}_{d+1}$, the edge part of the 1-loop partition function corresponds to $D^{d+2}_{s-1}$ new massive fields with totally antisymmetric  spin $(1^m)$ and scaling dimension $\Delta=\frac{d-2}{2}+\nu$ living on  $\text{EAdS}_{d-1}$.} of edge modes is intriguing and may help to sharpen the understanding about edge modes,  we'll not try to provide a precise interpretation  for it.
Summing over all the fields in the $m=1$ sector, including the ghosts associated with the $s\ge 2$ ones, we end up with a very simple expression 
\begin{align}\label{P2}
\log Z^{\text{AdS}_6}_{m=1}=\int^\infty_0\frac{du}{2u}\frac{1+e^{-u}}{1-e^{-u}}\frac{e^{-2 u}+e^{-3 u}}{(1-e^{-u})^4}
\end{align}
Combing (\ref{P1}) and (\ref{P2}) leads to the total partition function  of type-B theory in $\text{AdS}_6$
\begin{align}\label{FB5}
\log Z^{\text{AdS}_6}_{\text{B}}=\log Z^{\text{AdS}_6}_{m=0}+\log Z^{\text{AdS}_6}_{m=1}=\int^\infty_0\frac{du}{2u}\frac{1+e^{-u}}{1-e^{-u}}\frac{e^{-2 u}-e^{-3 u}}{(1-e^{-u})^5}
\end{align}
which apparently has the interpretation of double-trace deformation of a conformally coupled scalar field on $S^5$.
In higher dimensions, the partition function of $m=0$ sector is still trivial. For $m\ge 1$, the partition function restricted to the $m$ sector is given by
\begin{align}\label{Pm}
\log Z^{\text{AdS}_{d+1}}_{m}=(-)^{m-1}\int^\infty_0\frac{du}{2u}\frac{1+e^{-u}}{1-e^{-u}}\frac{e^{-(d-1-m)u}+e^{-(d-2-m)u}}{(1-e^{-u})^{d-1}}
\end{align}
which we've  checked up to $\text{AdS}_{16}$ by mathematica. Summing over all $\log Z^{\text{AdS}_{d+1}}_{m}$, we recover the structure of double-trace deformation of a conformally coupled scalar on $S^d$ up to a sign \cite{Giombi:2016pvg}
\begin{align}\label{genB}
\log Z^{\text{AdS}_{d+1}}_{\text{B}}=\sum_{m=0}^{\frac{d-3}{2}}\log Z^{\text{AdS}_{d+1}}_{m}=(-)^{\frac{d-1}{2}}\int^\infty_0\frac{du}{2u}\frac{1+e^{-u}}{1-e^{-u}}\frac{e^{-\frac{d-1}{2}u}-e^{-\frac{d+1}{2}u}}{(1-e^{-u})^{d}}
\end{align}
More explicitly, when $\frac{d-1}{2}$ is odd, eq. (\ref{genB}) means that we turn on a double-trace deformation $\CO^2_{\Delta_-}, \Delta_-=\frac{d-1}{2}$ which induces an RG flow from the original UV fixed point to a new IR fixed point and when $\frac{d-1}{2}$ is even, it means that we turn on a double-trace deformation $\CO^2_{\Delta_+}, \Delta_+=\frac{d+1}{2}$ which triggers an RG flow from the original IR fixed point to a new UV fixed point. Using eq. (\ref{DTD}) for $\nu=\frac{1}{2}$ and $s=0$,  $\log Z^{\text{AdS}_{d+1}}_{\text{B}}$ is alternatively expressed as
\begin{align}\label{ZB2}
\log Z^{\text{AdS}_{d+1}}_{\text{B}}=\frac{\pi}{d!}\int_0^{\frac{1}{2}}\,d\lambda\,\prod_{j=0}^{\frac{d-3}{2}}\left[\left(j+\frac{1}{2}\right)^2-\lambda^2\right]\,\lambda\tanh(\pi\lambda)
\end{align}
For some lower dimensions, say $d=3, 5, 7$, eq. (\ref{ZB2}) yields
\small
\begin{align}
\log Z^{\text{AdS}_{4}}_{\text{B}}=\frac{\zeta(3)}{8\pi^2}, \,\,\,\,\,\log Z^{\text{AdS}_{6}}_{\text{B}}=\frac{\zeta (3)}{96 \pi ^2}+\frac{\zeta (5)}{32 \pi ^4} ,\,\,\,\,\,\log Z^{\text{AdS}_{8}}_{\text{B}}=\frac{\zeta (3)}{720 \pi ^2}+\frac{\zeta (5)}{192 \pi ^4}+\frac{\zeta (7)}{128 \pi ^6}
\end{align}
\normalsize
consistent with \cite{Giombi:2014iua}.
For completeness, let's also give the explicit result for any even dimensional AdS
\begin{align}
\log Z^{\text{AdS}_{2r+2}}_{\text{B}}=\sum_{n=1}^r \frac{(-)^{n+r}}{2(2\pi)^{2n}}\frac{(2n)!}{(2k)!}a_n(r)\zeta(2n+1)
\end{align}
where $\{a_n(r)\}$ are defined as $\prod_{j=0}^{r-1}(x-j^2)=\sum_{n=1}^r\, a_n(r) \,x^n$. This result contracts with the proposed boundary duality which predicts vanishing one-loop free energy and meanwhile it is too complicated to be accommodated by a shift of $N$.

\section{Comments on thermal interpretations}\label{th}
In \cite{Anninos:2020hfj}, it's argued by using the character integral representations like eq. (\ref{example1}) that the one-loop partition function $Z^{(1)}_{\text{PI}}$ of a field $\varphi$ on $S^{d+1}$ is related to the bulk quasi-canonical partition function of $\varphi$ in the static patch of $\text{dS}_{d+1}$, subject to possible edge corrections localized on the dS cosmological horizon. In this section, we will first briefly review this argument and then explore the generalization to the path integral on $\text{EAdS}_{d+1}$.

\subsection{Review the thermal picture in dS}
For an inertial observer in $\text{dS}_{d+1}$, the perceived universe is the (southern) static patch (the de Sitter radius is taken to be 1)
\begin{align}
ds^2=-(1-r^2)dt^2+\frac{dr^2}{1-r^2}+r^2 d\Omega^2_{d-1}
\end{align}
which has a cosmological horizon at $r=1$ of temperature $T=\frac{1}{2\pi}$. Given certain field content in bulk, the field quanta are in thermal equilibrium with the horizon and we can compute the bulk quasi-canonical partition function as $\Tr_S e^{-2\pi H_S}$,  where $H_S$ is the Hamiltonian which generates times translation in the southern static patch and $\Tr_S$ denotes trace over the {\it southern}  multi-particle Hilbert space. At a very formal  level, $\Tr_S e^{-2\pi H_S}$ is supposed to be ``equal to'' \footnote{It's well known that when the spacetime manifold has a direct product structure, i.e. $\mathbb R\times M_{d}$, the thermal partition function $\Tr e^{-\beta H}$ is {\it equal to} the Euclidean path integral on $S^1_{\beta}\times M_d$. However, as de Sitter is not a direct product and  the flow generated by $H_S$ is degenerate at the horizon, edge corrections to the identification $\Tr_S e^{-2\pi H_S}= Z_{\text{PI}}$ are to be expected.} the Euclidean path integral on $S^{d+1}$, because operationally $\Tr_S e^{-2\pi H_S}$ means the path integral on a manifold obtained from the static patch by Wick rotation $t=-i\tau$ and identification $\tau\sim\tau+2\pi$, which is nothing but the unit sphere $S^{d+1}$. We can make sense of this formal argument by using the $\SO(1, d+1)$ Harish-Chandra characters at least at one-loop level. On the path integral side, the (unregularized) one-loop sphere partition function of a (bosonic) field $\varphi$ is given by \cite{Anninos:2020hfj}
\begin{align}
\log Z^{(1)}_{\text{PI}}=\int_0^\infty\, \frac{du}{2\,u} \frac{1+e^{-u}}{1-e^{-u}}\chi^{\text{dS}}_{\varphi}(u)+\text{edge corrections}
\end{align}
where $\chi^{\text{dS}}_{\varphi}(u)$ is the $\SO(1, d+1)$ Harish-Chandra character $\tr_G\,  e^{-i u H}$ corresponding to the UIR carried by $\varphi$. Here $\tr_G$ means tracing over the {\it global } single-particle Hilbert space of $\varphi$. In the thermal picture, we use the ideal gas approximation for thermal partition function at the one-loop level
\begin{align}\label{logTr}
\log \Tr_S e^{-2\pi H_S}=-\int_0^\infty\, d\omega \, \rho_S(\omega) \log \left( e^{\pi \omega}-e^{-\pi \omega}\right)
\end{align}
where $\rho_{S}(\omega)=\tr_S \, \delta(\omega-H_S)$ is the density of single-particle states of $\varphi$ in the southern Hilbert space. Define $\chi_S(u)\equiv\int_0^\infty\, d\omega \rho_S(\omega)(e^{i\omega u}+e^{-i\omega u})$ and then $\log \Tr_S e^{-2\pi H_S}$ can be expressed as 
\begin{align}
\log \Tr_S e^{-2\pi H_S}=\int_0^\infty\, \frac{du}{2\,u} \frac{1+e^{-u}}{1-e^{-u}}\,\chi_S(u)
\end{align}
The density $\rho_S(\omega)$ is badly divergent because $H_S$ has a continuous spectrum due to the infinite redshift near horizon. To make sense of $\rho_S(\omega)$, we need two steps:
\begin{itemize}
\item \textbf{Step 1}: Identify $\rho_S(\omega)$ with the {\it global} density of states  $\rho_G(\omega)=\tr_G \,\delta(\omega-H)$ for positive $\omega$. This identification holds because there exists a one-to-one map between southern and global single-particle states of the same $H=\omega>0$ induced by the Bogoliubov transformations \cite{Israel:1976ur}.
\item \textbf{Step 2}: $\rho_G(\omega)$ can be extracted from the Harish-Chandrea character $\chi^{\text{dS}}_{\varphi}(u)$ (with suitable UV-regularizations like Pauli-Villas, cf. appendix \ref{coarse},  or dimensional regularization which are suppressed here)
\begin{align}
\rho_G(\omega)=\int_{-\infty}^\infty \frac{du}{2\pi} \, \chi^{\text{dS}}_{\varphi}(u) e^{i u\omega}=\int_{0}^\infty \frac{du}{2\pi} \, \chi^{\text{dS}}_{\varphi}(u)
\left(e^{i \omega\, u}+e^{-i\omega\, u }\right)
\end{align}
where we've used $\chi^{\text{dS}}_{\varphi}(u)=\chi^{\text{dS}}_{\varphi}(-u)$.
\end{itemize}
Altogether, we have $\chi_S(u)=\chi^{\text{dS}}_{\varphi}(u)$ and  $\Tr_S e^{-2\pi H_S}$ is the same as the bulk part of $\log Z^{(1)}_{\text{PI}}$. This is the thermal interpretation of the one-loop sphere partition functions.

\subsection{Thermal picture in AdS}
To explore the thermal interpretation of the character integral representation of partition functions AdS, we first need to find coordinate systems of AdS that have a horizon structure. Such coordinates are summarized in the appendix \ref{coodAdS}.

\subsubsection{$\text{AdS}_2$}
\begin{figure}[h]
\centering
  \includegraphics[width=0.4\linewidth]{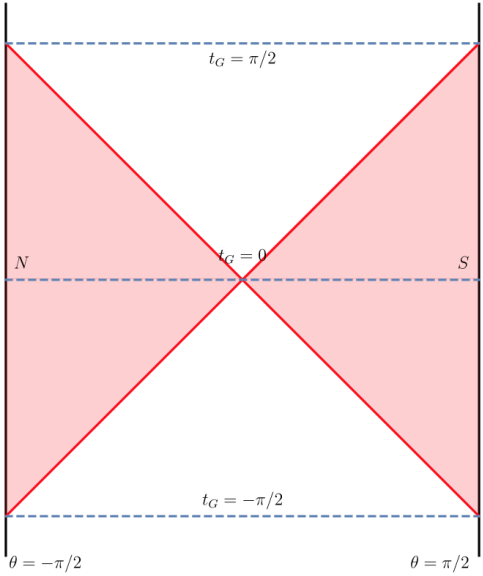}
  \caption{\small A portion of the periodic $\text{AdS}_2$ Penrose diagram near $t_G=0$. The two vertical lines $\theta=\pm\frac{\pi}{2}$ are the boundaries of $\text{AdS}_2$. The black hole solution in eq. (\ref{BH1}) corresponds to the red region denoted by ``S''. The region ``N'' is the image of ``S'' under the map $X^1\to -X^1$. The red lines represent the bifurcate Killing horizon.}
  \label{BH2}
\end{figure}

In the 2D Lorentzian AdS, there exist a black hole solution  \cite{Keeler:2014hba} with coordinates, cf. (\ref{BHcoord})
\begin{align}\label{BH1}
X^0=\rho, \,\,\,\,\, X^1=\sqrt{\rho^2-1}\cosh \, t_S,\,\,\,\,\, X^2=\sqrt{\rho^2-1}\sinh \, t_S\end{align}
and metric $ds^2=-(\rho^2-1)dt^2_S+\frac{d\rho^2}{\rho^2-1}$, which shows a point-like horizon at $\rho=1$ of temperature $T=\frac{1}{2\pi}$. Wick rotation $t_S\to -i\tau$ and identification $\tau\sim\tau+2\pi$ yield the 2D Euclidean AdS. 
Compared to the conformal global coordinate of $\text{AdS}_2$ 

\begin{align}
X^0=\frac{\cos t_G}{\cos\theta}, \,\,\,\,\, X^1=\tan\theta, \,\,\,\,\, X^2=\frac{\sin t_G}{\cos\theta}
\end{align}
the black hole solution (\ref{BH1}) covers the region: $\theta\in (0,\frac{\pi}{2})$ and $\sin\theta>|\sin t_G|$, cf. fig (\ref{BH2}). 

This scenario is very similar to its dS counter part and hence we're  allowed to use the dS argument to claim that the thermal partition function of a field $\varphi$ in the black hole patch of $\text{AdS}_2$ is given by
\begin{align}
\log \Tr_S e^{-2\pi L_{21}}=\int_0^\infty\, \frac{du}{2\,u} \frac{1+e^{-u}}{1-e^{-u}}\,\chi_S(u)
\end{align}
where the noncompact Lorentz generator $L_{21}\in\so(2,1)$ generates time translation  $t_S\to t_S+\text{const}$ and $\chi_S(u)$ is the ``character'' defined with respect to the single-particle Hilbert space in the black hole patch. Using the Bogoliubov transformations \cite{Israel:1976ur}, $\chi_S(u)$ can be replaced by the $\SO(2,1)$ Harish-Chandra character $\chi_\varphi^{\text{HC}}(u)=\tr_G \, e^{-i u L_{21}}$ which is traced over the {\it global} single-particle Hilbert space:
\begin{align}
\log \Tr_S e^{-2\pi L_{21}}=\int_0^\infty\, \frac{du}{2\,u} \frac{1+e^{-u}}{1-e^{-u}}\,\chi_\varphi^{\text{HC}}(u)
\end{align}
At this stage, we want to emphasize that the Harish-Chandra character $\chi_\varphi^{\text{HC}}(u)$, by definition, is completely different from the characters we've used in the previous sections, like $\chi^{\text{AdS}_2}_\Delta(u)=\frac{e^{-\Delta}}{1-e^{-u}}$. The latter are defined as $\tr_G\, e^{-u H}$ for {\it positive} $u$, where $H$ is the global Hamiltonian generating global time translation $t_G\to t_G+\text{const}$. Indeed, these characters are {\it not}  group characters. So it seems that we cannot naively identify the thermal partition function $\Tr_S e^{-2\pi L_{21}}$ as the one-loop path integral on Euclidean $\text{AdS}_2$. However, in the appendix \ref{comchar}, we explicitly compute the Harish-Chandra character $\chi_\varphi^{\text{HC}}(u)$ when $\varphi$ is a scalar field of scaling dimension $\Delta$ and we find perhaps surprisingly
\begin{align}
\boxed{\chi_\varphi^{\text{HC}}(u)=\chi^{\text{AdS}_2}_\Delta(|u|)}
\end{align}
which yields
\begin{align}
\log \Tr_S e^{-2\pi L_{21}}=\int_0^\infty\, \frac{du}{2\,u} \frac{1+e^{-u}}{1-e^{-u}}\frac{e^{-\Delta u}}{1-e^{-u}}
\end{align}
in agreement with the path integral result cf. (\ref{Zn2}). Therefore the one-loop path integral on $\text{EAdS}_2$ can be interpreted as the quasi-canonical partition function $\Tr_S e^{-2\pi L_{21}}$ in the black hole patch. To further understand why the Harish-Chandra character $\chi_\varphi^{\text{HC}}(u)$ appears in the quasi-canonical partition function $\Tr_s e^{-2\pi L_{21}}$, we explore the underlying physical meanings of $\chi_\varphi^{\text{HC}}(u)$ in appendix \ref{GG}. In section \ref{QNM}, we show that $\chi_\varphi^{\text{HC}}(u)$ encodes the quasinormal spectrum of $\varphi$ in the black hole patch of $\text{AdS}_2$ and in section \ref{coarse}, we extract a well-defined single-particle density of states in the black patch from $\chi_\varphi^{\text{HC}}(u)$ and show numerically that it can be realized as the continuous limit of the density of states in some simple model with a  finite dimensional Hilbert space.

\subsubsection{Higher dimensions}
In higher dimensional $\text{AdS}_{d+1}$, the universe perceived by an accelerating observer is called (southern) Rindler-AdS due to the presence of a Rindler horizon \cite{Parikh:2012kg}. The Rindler-AdS admits a $\text{dS}_{d+1}$ foliation, cf.  appendix \ref{coodAdS}:
\begin{align}
ds^2=d\eta^2+\sinh \eta^2 \left(-(1-r^2)dt_S^2+\frac{dr^2}{1-r^2}+r^2 d\Omega^2_{d-2}\right)
\end{align}
and hence has temperature $T=\frac{1}{2\pi}$. In the Rindler-AdS patch, we can use the dS type argument to show that the quasi-canonical partition function of $\varphi$ with spin-$s$ and scaling dimension $\Delta$ is
\begin{align}
\log \Tr_S e^{-2\pi L_{d+1,d}}=\int_0^\infty\, \frac{du}{2\,u} \frac{1+e^{-u}}{1-e^{-u}}\,\chi_\varphi^{\text{HC}}(u)
\end{align}
where the  noncompact Lorentz generator $L_{d+1,d}\in\so(2,d)$ generates time translation  $t_S\to t_S+\text{const}$ and $\chi_\varphi^{\text{HC}}(u)$ is the Harish-Chandra character $\tr_G\, e^{-i u L_{d+1,d}}$. In the appendix \ref{comchar}, we argue that $\chi_\varphi^{\text{HC}}(u)=\chi^{\text{AdS}_{d+1}}_\Delta(|u|)=\frac{e^{-\Delta|u|}}{(1-e^{-|u|})^d}$ when $\varphi$ is a scalar field and we believe $\chi_\varphi^{\text{HC}}(u)=\chi^{\text{AdS}_{d+1}}_{[\Delta,s]}(|u|)$ should still hold when $\varphi$ is a spin-$s$ field. Granting this relation, we are left with
\begin{align}\label{xxxx}
\boxed{\log \Tr_S e^{-2\pi L_{d+1,d}}=\int_0^\infty\, \frac{du}{2\,u} \frac{1+e^{-u}}{1-e^{-u}}\,\chi^{\text{AdS}_{d+1}}_{[\Delta,s]}(u)}
\end{align}
When $d$ is odd, (\ref{xxxx}) exhibits the the agreement between $ \Tr_S e^{-2\pi L_{d+1,d}}$ and the bulk part of the one-loop path integral on Euclidean AdS. However, when $d$ is even, (\ref{xxxx}) is different from the path integral result (\ref{evendZ}), including the volume dependence and the contour choice. We believe that the key of solving this difference is computing the properly IR regulated path integral on Euclidean AdS and understanding the mixing of UV and IR divergences. More explicitly, a functional determinant of an operator $\CD$ can be represented as an integral transformation of the corresponding (integrated) heat kernel $K_\CD(t)=\int_M\, d^D x\,\sqrt{g} \, K_\CD(t ;x, x)$, cf. (\ref{KtoZ}). If the base manifold $M$ is maximally symmetric and the operator $\CD$ also preserves the isometry group of $M$, then $K_{\CD}(t; x,x)$ is independent of $x$ and hence 
the integral $\int_M\, d^D x\,\sqrt{g} $  simply yields the volume of $M$. When $M$ is a compact manifold like sphere, the factorization $K_\CD(t)=\text{Vol}_M \, K_\CD(t ;x_0, x_0)$ is well-defined but when $M$ is a noncompact manifold like flat space and Euclidean AdS, the naive factorization suffers from an IR divergence $\text{Vol}_M\to\infty$. Such an IR divergence is not a big issue while computing the free energy of ideal gas in flat space if we only care about the leading large-volume behavior, i.e. the extensive part.   However, in the AdS case, as we want to extract the $R^0$ or $\log R$  piece \footnote{It's very likely to have a $R^0$ piece even when $d$ is even if we implement the IR regulator properly. Of course, the $R^0$ piece in this case is ambiguous because it is contaminated by the $\log R$ piece.}, it's apparently more appropriate to introduce a radial cutoff $R$ and impose certain boundary conditions on the cutoff surface. This procedure would spoil the $\SO(2,d)$ symmetry and discretize the spectrum of Laplacian operators. The symmetry breaking can lead to considerable technical difficulties in computing the heat kernel $K_\CD(t ;x, x)$.
Another approach to this difference is dimensional regularization which works for both the UV and IR divergences, along the line of 
\cite{Skvortsov:2017ldz, Diaz:2007an}. But the physical picture is not clear if we implement this formal regularization scheme. We will leave this to future work.

\section*{Acknowledgments} 
I'm grateful to Dionysios Anninos, Frederik Denef, Austin Joyce, Albert Law, Ruben Monten and Robert Penna for numerous stimulating
discussions, at different stages of this research project. I also thank 
Frederik Denef for reading the paper and providing precious comments. ZS was supported
in part by the U.S. Department of Energy grant de-sc0011941.
\appendix

\section{Partition function of  Dirac spinors}\label{spinor}
As an example of applying the character integral method to fermions, let's consider a complex Dirac spinor of scaling dimension $\Delta = \frac{d}{2}+ \nu$ in $\text{AdS}_{d+1}$ with $d=2r+1$. It carries a highest weight representation $\vec s=(\frac{1}{2}, \cdots, \frac{1}{2})\equiv\mathbf{\frac{1}{2}}$ of $\so(d)$ which has real dimension $2^{r+1}$. The one-loop partition function of this field is given by 
\begin{align}\label{general}
\log Z=\frac{(-2)^r}{d!}\int_0^\infty \frac{dt}{t} \,e^{-\frac{\epsilon^2}{4t}}\int^\infty_0 d\lambda\, \mu_{\mathbf{\frac{1}{2}}}(\lambda)\,e^{-t(\lambda^2+\nu^2)}
\end{align}
where the spinor spectral function $\mu_{\mathbf{\frac{1}{2}}}(\lambda)$ is 
\begin{align}
\mu_{\mathbf{\frac{1}{2}}}(\lambda)=\prod_{j=1}^{r}\left(\lambda^2+j^2\right)\frac{\lambda}{\tanh(\pi\lambda)}
\end{align}
After using the standard Hubbard-Stratonovich trick, the partition function is completely encoded in $W_{\mathbf{\frac{1}{2}}}(u)=\int_{-\infty}^\infty d\lambda \,\mu_{\mathbf{\frac{1}{2}}}(\lambda)\, e^{i\lambda u}$:
\begin{align}\label{B3}
\log Z=\frac{(-2)^r}{d!}\int_0^\infty\frac{du}{u}\,W_{\mathbf{\frac{1}{2}}}(u)\,e^{-\nu u}
\end{align}
To perform the $\lambda$-integral in $W_{\mathbf{\frac{1}{2}}}(u)$ , we can close the contour in the upper half plane and pick up the poles at $\lambda= i n, n\ge r+1$, where $\frac{\lambda}{\tanh(\pi\lambda)}$ has residue $\frac{i n}{\pi}$.
\begin{align}\label{B5}
W_{\mathbf{\frac{1}{2}}}(u)&=2(-)^{r+1}\sum_{n\ge r+1}n\prod_{j=1}^r(n^2-j^2) e^{-n u}=(-1)^{r+1} d!\frac{2 \, e^{-\frac{d+1}{2}u}}{(1-e^{-u})^{d+1}}
\end{align}
Plugging (\ref{B5}) into (\ref{B3}) yields the unregularized partition function
\begin{align}
\log Z=\int_0^\infty\frac{du}{u}\frac{-e^{-\frac{u}{2}}}{1-e^{-u}}\chi^{\text{AdS}_{d+1}}_{[\Delta, \mathbf{\frac{1}{2}}]}(u), \,\,\,\,\chi^{\text{AdS}_{d+1}}_{[\Delta, \mathbf{\frac{1}{2}}]}(u)=\frac{2^{r+1}e^{-\Delta u}}{(1-e^{-u})^{d}}
\end{align}
We can also easily write down the regularized version following the derivation in section \ref{Contour} 
\begin{align}
\log Z=-\frac{1}{2}\int_{\mathbb{R}+i\delta}\frac{du}{\sqrt{u^2+\epsilon^2}}\frac{e^{-\frac{u}{2}}}{1-e^{-u}}\frac{2^{r+1}e^{-\frac{d}{2}u-\nu\sqrt{u^2+\epsilon^2}}}{(1-e^{-u})^d}
\end{align}
Compared to the bosonic case,  the only difference is that the representation-independent factor $\frac{1+e^{-u}}{1-e^{-u}}$ gets replaced by $\frac{-2 e^{-\frac{u}{2}}}{1-e^{-u}}$. 

\section{Physical interpretation of spectral density/Plancherel measure}\label{Planch}
In an ordinary quantum mechanical system, given a Hamiltonian $H$, the associated density of state (DOS) is defined as $\rho(E)=\Tr \delta(H-E)$ where we trace over the whole Hilbert space. Using the well-known distributional identity $\frac{1}{x\pm i\epsilon}=\text{P}(\frac{1}{x})\mp i\pi \delta(x)$, the DOS can also be formally expressed 
\begin{align}\label{DOS0}
\rho(E)=\frac{1}{2\pi i}\left(R(E+i\epsilon)-R(E-i\epsilon)\right)
\end{align}
where $R(E)\equiv\Tr \frac{1}{H-E}$ is the so-called resolvent and the limit $\epsilon\to 0^+$ is understood. In this appendix, we will show that the (scalar) Plancherel measure given by eq. (\ref{Plancherel measure}) can be interpreted as a DOS in the sense of (\ref{DOS0}).

For a real scalar field in $\text{EAdS}_{d+1}$, we choose Hamiltonian $H$ to be the Laplace-Beltrami operator $-\nabla^2$ which has a continuous spectrum $E_\lambda\equiv \frac{d^2}{4}+\lambda^2$ for all $\lambda\in\mathbb{R}_{\ge 0}$ \cite{Camporesi:1994ga}. In this case, the operator  $\frac{1}{H-E}$ is nothing but a scalar Green function $G_{\Delta}(X, X')$ with mass  $m^2=-E \equiv \Delta(\Delta -d)$ \cite{Burgess:1984ti, Inami:1985wu,Burges:1985qq}:
\begin{align}\label{Green}
G_\Delta(X, X')=G_\Delta(P)=\frac{\Gamma(\Delta)(-2P)^{-\Delta}}{2\pi^{\frac{d}{2}}\Gamma(\Delta-\frac{d-2}{2})}F\left(\frac{\Delta}{2}, \frac{\Delta+1}{2}, \Delta-\frac{d-2}{2}, \frac{1}{P^2}\right), \,\,\,\,\, P= X\cdot X'
\end{align}
where $X, X'$ are points in the embedding space representation of $\text{EAdS}_{d+1}$. Plugging in $E=E_\lambda\pm i\epsilon$, the corresponding resolvent $R(E_\lambda\pm i\epsilon)$ is given by 
\begin{align}\label{Resol}
R(E_\lambda\pm i\epsilon)=\int_{\text{EAdS}_{d+1}} d^{d+1}X \, G_{\frac{d}{2}\mp i\lambda}(X, X)=\text{Vol}(\text{AdS}_{d+1})\, G_{\frac{d}{2}\mp i\lambda}(X, X)
\end{align}
Therefore, combining eq. (\ref{DOS0}) and (\ref{Resol}), we find that the DOS per volume of $-\nabla^2$ in $\text{EAdS}_{d+1}$ is simply
\begin{align}\label{DOS}
\frac{\rho_{d+1}(E_\lambda)}{\text{Vol}(\text{AdS}_{d+1})}=-\frac{1}{2\pi i}\lim_{P\to -1^-}\left(G_{\frac{d}{2}+i\lambda}(P)-G_{\frac{d}{2}-i\lambda}(P)\right)
\end{align}
where $P\to -1^-$ means that $P$ approaches $-1$ from the left. (Technically the direction of limit is important, and physically the direction is also fixed because $X\cdot X'\le -1$ for any two points on $\text{EAdS}_{d+1}$). Before showing the main result extracted from eq. (\ref{DOS}), let's digress a bit and discuss some properties of hypergeometric functions appearing in eq. (\ref{Green}). In general, the hypergeometric function $F(a, b, c; x)$ with $\text{Re}(c-(a+b))<0$ is singular around $x=1$ \cite{mathphys}
\begin{align}
\lim_{x\to 1-}\frac{F(a, b, c, x)}{(1-x)^{c-a-b}}=-\frac{\Gamma(c)\Gamma(a+b-c)}{\Gamma(a)\Gamma(b)}
\end{align}
which implies that  the two Green functions in (\ref{DOS})  have singularity as $P$ approaches $-1^-$. However, amazingly all the divergences cancel out in the end  when we take the difference (which of course is expected since the DOS per volume should be well-defined). We list 
some lower dimensional examples here
\begin{align}\label{oddex}
\frac{\rho_2(E_\lambda)}{\text{Vol}(\text{AdS}_{2})}&=\frac{1}{4\pi} \tanh(\pi \lambda), \,\,\,\,\, &d=1\nonumber\\
\frac{\rho_4(E_\lambda)}{\text{Vol}(\text{AdS}_{4})}&=\frac{1}{16\pi^2}\left(\lambda^2+\frac{1}{4}\right)\tanh(\pi \lambda), \,\,\,\,\, &d=3\nonumber\\
\frac{\rho_6(E_\lambda)}{\text{Vol}(\text{AdS}_{6})}&=\frac{1}{128\pi^3}\left(\lambda^2+\frac{1}{4}\right)\left(\lambda^2+\frac{9}{4}\right)\tanh(\pi \lambda), \,\,\,\,\, &d=5
\end{align}
and 
\begin{align}\label{evenex}
\frac{\rho_3(E_\lambda)}{\text{Vol}(\text{AdS}_{3})}&=\frac{1}{4\pi^2} \lambda, \,\,\,\,\, &d=2\nonumber\\
\frac{\rho_5(E_\lambda)}{\text{Vol}(\text{AdS}_{5})}&=\frac{1}{24\pi^3}\lambda(1+\lambda^2), \,\,\,\,\, &d=4\nonumber\\
\frac{\rho_7(E_\lambda)}{\text{Vol}(\text{AdS}_{7})}&=\frac{1}{240\pi^4}\lambda\left(\lambda^2+1\right)\left(\lambda^2+4\right), \,\,\,\,\, &d=6
\end{align}
Notice that what we've obtained here is the number of states per unit ``energy'' $E_\lambda$ rather than spectral density because the latter is the number of states per unit $\lambda$. However, they can be easily mapped to each other by a change of integral measure $d E_\lambda =2\lambda \,d\lambda$. This observation suggests us to define the spectral density as 
\begin{align}\label{Plan}
\tilde\rho_{d+1}(\lambda)\equiv \frac{ 2\lambda \, \rho_{d+1}(E_\lambda)}{\text{Vol}(\text{AdS}_{d+1})}=\frac{1}{2^{d-1}\Gamma(\frac{d+1}{2})^2 \text{Vol}(S^d)}\frac{|\Gamma(\frac{d}{2}+i\lambda)|^2}{|\Gamma(i\lambda)|}
\end{align} 
where the $\lambda$-dependent factor $\frac{|\Gamma(\frac{d}{2}+i\lambda)|^2}{|\Gamma(i\lambda)|}$ is exactly what we call $\mu^{(d)}(\lambda)$ in eq. (\ref{scalarspecden}). As a final consistency check, let's reconstruct the scalar heat kernel associated to $(-\nabla^2+\nu^2-\frac{d^2}{4})$ from its canonical definition, i.e. ``summing'' over all energy eigenfunctions 
\begin{align}
K_\nu(t)&\equiv \Tr e^{-(-\nabla^2+\nu^2-\frac{d^2}{4})}=\int_{\frac{d^2}{4}}^\infty \, dE_\lambda\, \rho_{d+1}\, (E_\lambda) e^{-(E_\lambda+\nu^2-\frac{d^2}{4})}\nonumber\\
&=\text{Vol}(\text{AdS}_{d+1}) \int_0^\infty d\lambda\, \tilde \rho_{d+1}(\lambda) e^{-t(\lambda^2+\nu^2)}\nonumber\\
&=\frac{\text{Vol}(\text{AdS}_{d+1})}{ \text{Vol}(S^d)}\frac{1}{2^{d-1}\Gamma(\frac{d+1}{2})^2}\int_0^\infty d\lambda\, \mu^{(d)}(\lambda) e^{-t(\lambda^2+\nu^2)}
\end{align}
Altogether, the computations in this appendix help us to identify the spectral density or the Plancherel measure of $\SO(1, d+1)$ which has a rigorous mathematical definition in the pure group theory setup  \cite{Dobrev:1977qv,Knapp}, as the density of states associated to the Hamiltonian $H=-\nabla^2$ in a unit volume of $\text{EAdS}_{d+1}$ up to some representation-independent normalization factors.

\section{Comparison with dS character integral}\label{dSAdS}
In sections \ref{WU} and \ref{HSchar}, we derived character integral formulae for one-loop partition functions of both scalars and spin-$s$ fields in even dimensional AdS. These formulae are very similar with their dS counterpart derived in \cite{Anninos:2020hfj} , where the unregularized one-loop partition function  of a {\it massive} spin-$s$ field with scaling dimension $\Delta=\frac{d}{2}+i\nu$ is given by (the following formula works for both even and odd $d$ in dS)
\begin{align}\label{dS}
\log Z_{\text{PI}}=\int^{\infty}_{0}\frac{du}{2u}(e^{-(\frac{d}{2}+i\nu) u}+e^{-(\frac{d}{2}-i\nu) u})\sum_{n\ge -1} D^{d+2}_{n, s} e^{-n u}
\end{align}
where extending the sum to $n=-1$ is a result of locality. Summing over $n$ yields a (bulk+edge) type contribution as in $W^{(d)}_s$:
\begin{align}
\log Z_{\text{PI}}=\int^{\infty}_{0}\frac{du}{2u}\frac{1+e^{-u}}{1-e^{-u}}\left(D_s^{d}\frac{e^{-(\frac{d}{2}+i\nu) u}+e^{-(\frac{d}{2}-i\nu)u}}{(1-e^{-u})^d}-D^{d+2}_{s-1}\frac{e^{-(\frac{d-2}{2}+i\nu) u}+e^{-(\frac{d-2}{2}-i\nu)u}}{(1-e^{-u})^{d-2}}\right)
\end{align}
 In this appendix, we'll show that the origin of such similarity between AdS and dS can be traced back to  the eq. (\ref{d2tod}).
 
On the AdS side, we know that the unregularized partition function of  a field carrying the {\it massive} representation $[\Delta=\frac{d}{2}+\nu, \vec s]$ of $\text{SO}(2, d)$ is given by (assuming $d=2r+1$)
\begin{align}
\log Z_{\vec s, \nu}&=\frac{(-)^{r+1}}{d!}\int^\infty_0\frac{du}{2u}\,D^{d}_{\vec s}\,W^{(d)}_{\vec s}(u)\, e^{-\nu\, u}
\end{align}
where  $W^{(d)}_{\vec s}(u)\equiv\int d\lambda\, \mu_{\vec s}^{(d)}(\lambda) e^{i\lambda u}$ can be written as a series by closing the contour at infinity:
\begin{align}
W^{(d)}_{\vec s}(u)=2(-)^{r+1}\sum_{n\ge 0}\left(n+\frac{1}{2}\right)\prod_{j=1}^r\left(\left(n+\frac{1}{2}\right)^2-\ell_j^2\right)\,e^{-(n+\frac{1}{2})u}
\end{align}
Using the eq. (\ref{d2tod}), we obtain  $\frac{(-)^{r+1}}{d!}D^{d}_{\vec s}W^{(d)}_{\vec s}(u)=\sum_{n\ge 0}D^{d+2}_{n, s} e^{-(n+\frac{1}{2})u}$ which yields 
\begin{align}\label{comp}
\log Z_{\vec s,\nu}&=\int^\infty_0\frac{du}{2u}\,e^{-\nu\, u}\sum_{n\ge 0} D^{d+2}_{n-r, \vec s} \,e^{-(n+\frac{d}{2})u}\nonumber\\
&=\int^\infty_0\frac{du}{2u}\,e^{-\Delta\, u}\sum_{n\ge -r} D^{d+2}_{n,\vec s} \,e^{-n\, u}
\end{align}
where in the second line we've shifted $n$ by $r$. Now let's focus on the spin-$s$ representation, i.e. $\vec s=(s, 0, \cdots, 0)$. In this case $D^{d+2}_{n, s}$ vanishes for $n\in\{-r, -(r-1), \cdots, -2\}$ (and also $n=s-1$ but this is irrelevant to our discussion) and hence the sum in eq. (\ref{comp}) effectively starts from $n=-1$
\begin{align}
\log Z_{s,\nu}=\int^\infty_0\frac{du}{2u}\,e^{-\Delta\, u}\sum_{n\ge -1}D^{d+2}_{n,s}  \,e^{-n\, u}
\end{align}
Compared to (\ref{dS}), it's clear that the only difference is the absence of $e^{-\bar \Delta u}$ because in AdS only one boundary mode is dynamical and the other one is identified as a source.

\section{Evaluation of various residues}\label{residues}
This appendix is a collection of  technical proofs and results about residues of certain functions appearing in the character integrals. The ultimate goal here is to compute the residue of
\begin{align}
F_{d, \nu}(u)= \frac{1}{2u}\frac{1+e^{-u}}{1-e^{-u}}\frac{e^{-(\frac{d}{2}+\nu) u}}{(1-e^{-u})^d}
\end{align}
at $u=0$ for even dimension $d$, which is closely related to the  one-loop partition functions in odd dimensional AdS. For most of the discussions in this section, we consider a general dimension $d$ and only restrict the result to even $d$ in the end. An intermediate step to $\text{Res}_{u\to 0}F_{d, \nu}(u)$ is the residue of  the following function
\begin{align}
G_{d, \nu}(u)\equiv \frac{e^{-(\frac{d}{2}+\nu) u}}{(1-e^{-u})^{d+1}}
\end{align}
which itself is also very interesting because we need it to verify the contour prescription proposed in section \ref{Contour}.

First we  show by induction that 
\begin{align}\label{indu}
\text{Res}_{u\to 0} G_{d, \nu}(u)=\frac{(-)^d}{d!}\frac{\Gamma(\nu+\frac{d}{2})}{\Gamma(\nu-\frac{d}{2})}
\end{align}
It's straightforward to check that eq. (\ref{indu}) holds for $d=1$. Assuming the induction condition (\ref{indu}), we show that it also works for $d+1$. Let $C_0$ be a small circle around $u=0$, i.e. it doesn't enclose any other poles of $G_{d,\nu}(u)$ except $u=0$. Then the residue of $G_{d+1,\nu}(u)$ at $u=0$ can be expressed as a contour integral along $C_0$ counterclockwisely
\begin{align}
\text{Res}_{u\to 0} G_{d+1, \nu}(u)=\oint_{C_0}\, \frac{du}{2\pi i}\frac{e^{-(\frac{d+1}{2}+\nu) u}}{(1-e^{-u})^{d+2}}
\end{align}
To use the induction condition, we should lower the power in the denominator which can be realized by integration by part:
\begin{align}\label{rec}
\text{Res}_{u\to 0} G_{d+1, \nu}(u)&=-\frac{1}{d+1}\oint_{C_0}\, \frac{du}{2\pi i}e^{-(\frac{d}{2}+\nu-\frac{1}{2}) u} \frac{d}{du} \frac{1}{(1-e^{-u})^{d+1}}\nonumber\\
&=-\frac{\frac{d-1}{2}+\nu}{d+1}\oint_{C_0}\, \frac{du}{2\pi i}\frac{e^{-(\frac{d}{2}+\nu-\frac{1}{2}) u}}{(1-e^{-u})^{d+1}}=-\frac{\frac{d-1}{2}+\nu}{d+1}\text{Res}_{u\to 0} G_{d, \nu-\frac{1}{2}}(u)
\end{align}
Applying the induction condition (\ref{indu}) to eq. (\ref{rec}) yields
\begin{align}
\text{Res}_{u\to 0} G_{d+1, \nu}(u)=\frac{(-)^{d+1}}{(d+1)!}\frac{\Gamma(\nu+\frac{d+1}{2})}{\Gamma(\nu-\frac{d+1}{2})}
\end{align}
 This confirms that (\ref{indu}) holds for all $d$.
 
 To bridge the gap between $\text{Res}_{u\to 0} G_{d, \nu}(u)$ and $\text{Res}_{u\to 0} F_{d, \nu}(u)$, we need to define another function 
\begin{align}
H_{d, \nu}(u)\equiv \frac{1+e^{-u}}{1-e^{-u}}\frac{e^{-(\frac{d}{2}+\nu) u}}{(1-e^{-u})^d}=G_{d, \nu}(u)+G_{d, \nu+1}(u)
\end{align}
It's direct to write down the residue of $H_{d, \nu}(u)$ at $u=0$ by using its relation with the $G$-functions
\begin{align}\label{resH}
\text{Res}_{u\to 0} H_{d, \nu}(u)=\frac{2(-)^{d}}{d!}\nu \frac{\Gamma(\nu+\frac{d}{2})}{\Gamma(\nu+1-\frac{d}{2})}
\end{align}
which is a polynomial in $\nu$ for positive integer $d$. In addition, the R.H.S of eq. (\ref{resH}) is an even function in $\nu$ when $d$ is even and an odd function when $d$ is odd.  More explicitly, for $d=2r$, 
\begin{align}\label{resHeven}
\text{Res}_{u\to 0} H_{d, \nu}(u)=\frac{2}{d!}\,\prod_{j=0}^{r-1}(\nu^2-j^2)
\end{align}
and for $d=2r+1$, 
\begin{align}\label{resHodd}
\text{Res}_{u\to 0} H_{d, \nu}(u)=-\frac{2\nu}{d!}  \,\prod_{j=0}^{r-1}\left(\nu^2-\left(j+\frac{1}{2}\right)^2\right)
\end{align}
To achieve our original goal, the residue of $F_{d, \nu}(u)$ at $u=0$ for even $d$, we need the following differential relation between $F_{d, \nu}(u)$ and $H_{d, \nu}(u)$
\begin{align}
\partial_\nu F_{d, \nu}(u)=-\frac{1}{2} H_{d, \nu}(u)
\end{align}
which yields
\begin{align}
\text{Res}_{u\to 0} F_{d, \nu}(u)=\frac{(-)^{d+1}}{d!}\int_0^\nu\, dx\,x \frac{\Gamma(x+\frac{d}{2})}{\Gamma(x+1-\frac{d}{2})}+\text{const}
\end{align}
The unknown constant can be easily fixed for even $d$ without any extra effort. This claim follows from the observation that $F_{2r, 0}(u)$ is an even function in $u$. Thus $\text{Res}_{u\to 0} F_{2r, \nu}(u)$ vanishes when $\nu=0$ and the integration constant has to be zero:
\begin{align}\label{resTeven}
\text{Res}_{u\to 0} F_{2 r, \nu}(u)&=-\frac{1}{(2r)!}\int_0^\nu\, dx\,\prod_{j=0}^{r-1}(x^2-j^2)\nonumber\\
&=-\frac{1}{(2r)!}\sum_{n=1}^r\frac{a_n(r)}{2n+1} \nu^{2n+1}
\end{align}
where the numerical coefficients $\{a_n(r)\}$ are defined through the following generating function
\begin{align}\prod_{j=0}^{r-1}(x-j^2)=\sum_{n=1}^r\, a_n(r) \,x^n\end{align}

\section{Various coordinate systems in Euclidean/Lorentzian AdS}\label{coodAdS}
We begin with the embedding space representation of Lorentzian $\text{AdS}_{d+1}$ of unit radius 
\begin{align}
-(X^0)^2+(X^1)^2+\cdots+(X^d)^2 - (X^{d+1})^2=-1
\end{align}
By Wick rotation $X^{d+1}\to -i X^{d+1}$, we obtain Euclidean AdS in embedding space
\begin{align}
-(X^0)^2+(X^1)^2+\cdots + (X^{d+1})^2=-1
\end{align}
The global coordinate for Euclidean AdS is chosen to be
\begin{align}\label{globalcoord}
X^0=\cosh\eta, \,\, X^a=\sinh\eta\,\Omega^a_{d}, \,\,\,\,\ 1\le a\le d+1
\end{align}
where $\eta\ge 0$ and $\Omega_{d}$ denotes a point on  $S^d$. In this coordinate, the metric is given by 
\begin{align}\label{globalmetric}
ds^2_{\text{EAdS}_{d+1}}=d\eta^2+\sinh^2\eta\, d\Omega^2_{d}
\end{align}
In particular when $d=1$, choosing $\Omega_1=(\cos\varphi, \sin\varphi)$, the metic is $ds^2_{\text{EAdS}_{2}}=d\eta^2+\sinh^2\eta\, d\varphi^2$. Under the Wick rotation $\varphi\to i t$, we transform back to Lorentzian signature. In embedding space,  it means we choose the  following coordinate systems on two patches that cover different portions of Lorentzian AdS
\begin{align}\label{BHcoord}
\text{Southern }: \begin{cases}
X^0=\rho\\ X^1=\sqrt{\rho^2-1}\cosh t_S\\ X^2=\sqrt{\rho^2-1}\sinh t_S
\end{cases}, \,\,\,\, \text{Northern }: \begin{cases}
X^0=\rho\\ X^1=-\sqrt{\rho^2-1}\cosh t_N\\ X^2=\sqrt{\rho^2-1}\sinh t_N
\end{cases}
\end{align}
where I've replaced $\cosh\eta$ by $\rho\ge 1$. The metric of southern/northern patch can be expressed as
\begin{align}
ds^2_{\text{AdS}_{2}}=-(\rho^2-1)dt^2+\frac{d\rho^2}{\rho^2-1}, \,\,\,\,\, t=t_S, t_N
\end{align}
which describes a black hole solution with a point-like horizon at $\rho=1$, the  intersection of the southern and northern patches. The temperature of this black hole is $T=\frac{1}{2\pi}$.

When $d\ge 2$, there exists a similar Wick rotation that describes a spacetime of the same temperature. Notice that the global coordinate system (\ref{globalcoord}) realizes a $S^d$ foliation of $\text{EAdS}_{d+1}$. By using the Wick rotation between de Sitter static patch and sphere, we  obtain a dS foliation of $\text{AdS}_{d+1}$. More explicitly, we choose the following coordinate system for $\Omega_d$
\begin{align}
\Omega_d=(r \,\Omega_{d-2}, \sqrt{1-r^2}\,\cos\varphi, \sqrt{1-r^2}\,\sin\varphi), \,\,\,\, 0\le r\le 1
\end{align}
where $\Omega_{d-2}$ denotes the usual spherical coordinates of $S^{d-2}$. Upon a Wick rotation $\varphi\to i t$,  $\Omega_d$ becomes a point on $\text{dS}_d$ and (\ref{globalmetric}) becomes the Rindler-AdS metric \cite{Parikh:2012kg}. As before, the Wick-rotated coordinate system describes two patches of Lorentzian AdS
\begin{align}
\text{Southern}: \begin{cases}
X^0=\cosh\eta \\ X^{\bar i}=r \sinh\eta\, \Omega^{\bar i}_{d-2}\\ X^d=\sinh\eta\cosh t_S \sqrt{1-r^2}\\ X^{d+1}=\sinh\eta\sinh t_S \sqrt{1-r^2}
\end{cases}, \,\,\,\,\, \text{Northern}: \begin{cases}
X^0=\cosh\eta \\ X^{\bar i}=r \sinh\eta\, \Omega^{\bar i}_{d-2}\\ X^d=-\sinh\eta\cosh t_N \sqrt{1-r^2}\\ X^{d+1}=\sinh\eta\sinh t_N \sqrt{1-r^2}
\end{cases}
\end{align}
in either of which  the metric  is
\begin{align}
ds^2_{\text{AdS}_{d+1}}=d\eta^2+\sinh^2\eta\left(-(1-r^2)dt^2+\frac{dr^2}{1-r^2}+r^2 d\Omega^2_{d-2}\right), \,\,\,\, t=t_S, t_N 
\end{align}
The two patches intersect at the horizon $r=1$ which has the geometry of $\text{EAdS}_{d-1}$. 

Finally, let's also introduce the global coordinate of  AdS
\begin{align}
\text{Global}: \begin{cases} X^0=\sqrt{R^2+1} \cos t_G\\ X^i=R\, \Omega_{d-1}^i\\ X^{d+1}=\sqrt{R^2+1}\sin t_G\end{cases}
\end{align}
At time $t_G=t_S=t_N=0$, i.e. $X^{d+1}=0$, the southern and northern patches cover the $X^d\ge 0$ and $X^d\le 0$ parts of the global spatial slice respectively.

\section{$\SO(2,d)$ Harish-Chandra characters}\label{comchar}
$\text{SO}(2,d)$ is the isometry group of $\text{AdS}_{d+1}$. In conventions in which the generators of $\text{SO}(2, d)$ are hermitian operators $L_{MN}, 0\le M,N\le d+1$, they are subject to commutation relations
\begin{align}
[L_{MN}, L_{PQ}]=i(\eta_{MP}L_{NQ}+\eta_{NQ}L_{MP}-\eta_{MQ}L_{NP}-\eta_{NP}L_{MQ})
\end{align}
where $\eta_{MN}=\text{diag}(-,+,\cdots, +, -)$. The physical interpretation of this algebra will be clear using the following Cartan-Weyl type basis
\begin{align}\label{Cartan}
H=L_{0,d+1}, \,\,\,\,  L^\pm_i=L_{i0}\mp iL_{i, d+1}, \,\,\,\, M_{ij}=L_{ij}
\end{align}
with commutation relations 
\begin{align}\label{Cartancom}
[H, L_i^{\pm}]=\pm L_i^{\pm}, \,\,\,\,\, [L_i^-, L_j ^+]=2\delta_{ij} H-2i M_{ij}, \,\,\,\,\, [L_{ij}, L_k^\pm]=i(\delta_{ik}L_j^\pm-\delta_{jk}L_i^{\pm})
\end{align}
where the trivial commutation relations are omitted. While acting on the $\text{AdS}_{d+1}$ quantum Hilbert space, $H$ can be identified with the Hamiltonian which generates time translation in global coordinates,  and $M_{ij}$ can be identified with angular momentum operators. The $L_i^\pm$ can then be viewed as raising/lowering operators for energy eigenstates. We're mainly interested in single-particle  Hilbert space $\CH_\Delta$ built from a primary state $|\Delta\rangle$ (also known as the lowest energy state), i.e. $H|\Delta\rangle=\Delta |\Delta\rangle,  L_i^-|\Delta\rangle=0$. By construction the Hilbert space $\CH_\Delta$ furnishes a representation of $\so(2,d)$ \footnote{Rigorously speaking, $\CH_\Delta$ cannot be lifted to an $\SO(2,d)$ representation for $\Delta\notin\mathbb Z$ because the identity element $e^{2\pi i H}$ in $\SO(2,d)$ is mapped to $e^{2\pi i \Delta}\not=1$ in this representation. Actually, the exponential map of Lie algebra yields a representation of the universal covering group of $\SO(2, d)$ which in bulk corresponds to unwrapping the periodic AdS time (thus $e^{2\pi i H}$ is no longer an identity). For simplicity of notation, we still call $\CH_\Delta$ an $\SO(2,d)$ representation and the corresponding character an $\SO(2,d)$ character.}.

In most of the physics literature  \cite{Basile:2016aen,Witten:1987ty, Dolan:2005wy, Dobrev:1991zb}, the $\SO(2,d)$ character of representation $\CH_{\Delta}$ is computed with respect to a compact Cartan algebra, in particular Hamiltonian and rotations. Here, for our purpose of thermal interpretations in section \ref{th}, we illustrate in (unitary) scalar primary representations how to compute $\SO(2,d)$ character associated to a noncompact generator, i.e. generator of boost in $\text{AdS}_{d+1}$. 

\subsection{$\SO(2,1)$ character}
As shown in eq. (\ref{Cartan}) and (\ref{Cartancom}), the Lie algebra $\so(2,1)$ is generated by $\{H, L^\pm\}$ with commutation relations $[H,L^\pm]=2 L^\pm, [L^-, L^+]=2H$. Starting from the primary state $|\Delta\rangle$, we build a tower of descendants $|n\rangle\equiv (L^+)^n |\Delta\rangle, n\ge 0$ which is a basis of the Hilbert space $\CH_\Delta$. Then the character associated to a noncompact generator, say $L_{12}=\frac{i}{2}(L^+-L^-)$, is defined as 
\begin{align}
\Tr_{\CH_\Delta} e^{it L_{12}}\equiv \sum_{n\ge 0}\frac{\langle n| e^{it L_{12}}|n\rangle}{\langle n| n\rangle}, \,\,\,\,\, t\in\mathbb{R}
\end{align}
Though defined through a simple and transparent way physically, it's technically very hard to figure out the character by computing this sum. Therefore, we'll use a different realization of the same representation that makes the same computation doable.

\vspace{10pt}

\noindent{}\textbf{Disc realization}: In \cite{Witten:1987ty}, by using the standard coadjoint orbit method, Witten showed  that the Hilbert space $\CH_\Delta$ can be mapped to the space of normalizable holomorphic function $f(z)$ on disc $D=\{z\in \mathbb{C}: |z|<1\}$ with inner product 
\begin{align}\label{inner1}
||f||^2=\int_D \,|f(z)|^2\,(1-z\bar z)^{2(\Delta-1)}d^2 z
\end{align}
On these holomorphic function, the generators of $\so(2,1)$ act as 
\begin{align}
H=z\,\partial_z+\Delta, \,\,\,\, L^-=-i \partial_z, \,\,\,\,\,L^+=-i(z^2\partial_z+2\Delta z)
\end{align}
The normalizable function $f(z)=z^k$ is an eigenfunction of $H$ with eigenvalue $\Delta+k$ and hence the character associated to $H$ is 
\begin{align}
\Tr_{\CH_\Delta} q^H=\sum_{k=0}^\infty \, q^{\Delta+k}=\frac{q^\Delta}{1-q}, \,\,\,\,\, 0<|q|<1
\end{align}

\vspace{10pt}

\noindent{}\textbf{Upper half-plane realization}: Using a fractional linear transformation $z\to w=\frac{1-i z}{z-i}$, we can map the disc $D$ to the upper half-plane  $\mathbb{H}=\{x+i y\in\mathbb{C}: y>0\}$ and the new Hilbert space consists of normalizable holomorphic functions $f(w)$ on  $\mathbb{H}$ with inner product \cite{Knapp, David}
\begin{align}\label{inner2}
||f||^2=\int_{\mathbb{H}} \frac{dx dy}{y^2} \, y^{2\Delta}\, |f(w)|^2, \,\,\,\,\, w=x+iy,\,\, y>0
\end{align}
When acting on a holomorphic function $f(w)$ on $\mathbb{H}$, the algebra $\so(2,1)$ is realized as 
\begin{align}\label{Uaction}
H=-i\left(\frac{1+w^2}{2}\partial_w+\Delta\, w\right),\,\,\,\,\, L_{10}=i\left(\frac{w^2-1}{2}\partial_w+\Delta\, w\right), \,\,\,\, L_{12}=i(w\partial_w+\Delta)
\end{align}
%where we've switched back to the hermitian basis (hermitian with respect to the inner product (\ref{inner2})). %
The eigenfunctions  of $H$ are $\phi_k(w)\equiv (w-i)^k(w+i)^{-2\Delta-k}, k\in\mathbb{N}$, with $H\phi_k=(\Delta+k)\phi_k$.
% These eigenfunctions are derived by noticing that any function in the representation space transforms as a form of degree $\Delta$ under conformal transformations, i.e. $f(z) dz^\Delta$ is invariant. In particular, taking $f(z)=z^k$ and using the holomorphic transformation $z\to w=\frac{1-i z}{z-i}$, we can obtain $\phi_k(w)$. 

\vspace{10pt}

\noindent{}$\mathbf{\mathbb{R}_+}$ \textbf{realization \cite{David} and evaluation of character}:  Given a holomorphic function $f(w)$ on the upper half-plane, we can define a new function $F$ on $\mathbb{R}_+$:
\begin{align}\label{fF}
F(\xi)=\int_{\Im(w)=\text{const}}\, \frac{dw}{2\pi}\, f(w)^{-iw\xi},\,\,\,\,\, \xi\in\mathbb{R}_+
\end{align}
and the inverse transformation is given by
\begin{align}\label{rtou}
f(w)\equiv \int_0^\infty\, d\xi\, F(\xi) e^{i w \xi }, \,\,\,\,\, \Im (w)>0
\end{align}
Under the integral transformation (\ref{fF}), the inner product (\ref{inner2}) is mapped to 
\begin{align}\label{inner3}
||F||^2=\int_0^\infty\, d\xi \, \xi^{1-2\Delta} |F(\xi)|^2
\end{align}
and the $\so(2,1)$ action (\ref{Uaction}) is mapped to 
\begin{align}\label{action}
H=\frac{1}{2}\xi(1-\partial_\xi^2)&+(\Delta-1)\partial_\xi, \,\,\,\, L_{10}=\frac{1}{2}\xi(1+\partial_\xi^2)+(1-\Delta)\partial_\xi\nonumber\\
&L_{12}=-i(\xi\,\partial_\xi+1-\Delta)
\end{align}
To find all eigenfunctions of $H$ in (\ref{action}), let's start from the primary state $\phi_0(w)=(w+i)^{-2\Delta}$ in the upper half plane realization. $\phi_0$ can be expressed as a Schwinger parameterization:
\begin{align}\label{Schw}
(w+i)^{-2\Delta}=\frac{1}{i^{2\Delta}\Gamma(2\Delta)}\int_{0}^\infty \frac{d\xi}{\xi}\, \xi^{2\Delta} e^{-\xi(1-iw)}
\end{align}
Comparing (\ref{Schw}) with (\ref{rtou}), we immediately get that the dual function of $\phi_0(w)$ in $\mathbb{R}_+$ is $G_0(\xi)=\xi^{2\Delta-1}e^{-\xi}$ (dropping unimportant normalization constants). For the dual function of $\phi_k(w)$, we use the ansatz $G_k(\xi)=G_0(\xi)P_k(\xi)$, where $P_k(\xi)$ is a polynomial in $\xi$. Then the eigenequation $H G_k=(\Delta+k)G_k$ yields a second order differential equation of $P_k(\xi)$
\begin{align}
\frac{\xi}{2}\,P_{k}(\xi)''+(\Delta-\xi)\,P_k(\xi)'+k\,P_k(\xi)=0
\end{align}
whose polynomial solution is the generalized Laguerre polynomial $P_k(\xi)=L^{(2\Delta-1)}_k(2\xi)$. Thus the spectrum of $H$ is given by
\begin{align}
H G_k=(\Delta+k)G_k,\,\,\,\, G_k(\xi)=\xi^{2\Delta-1}\,L^{(2\Delta-1)}_k(2\xi)\,e^{-\xi}
\end{align}
By using the recurrence relation of Laguerre polynomial, we can also show that $L^\pm$ indeed behaves like lower/raise operator
\begin{align}\label{recurL}
L^- G_k(\xi)=(1-2\Delta-k) G_{k-1}(\xi), \,\,\,\,\, L^+ G_k(\xi)=-(k+1) G_{k+1}(\xi)
\end{align}
As another self-consistency check of this representation,  we show  the Fourier/Laplace transformation (\ref{rtou}) of $G_k(\xi)$ is $\phi_k(w)$ up to normalization factors. To do this, we need the series expansion of a generalized Laguerre polynomial $L^{(\alpha)}_n(x)=\sum_{\ell=0}^n\frac{(-)^\ell}{\ell!}\binom{n+\alpha}{n-\ell}x^\ell$, which yields
\begin{align}
\int_{0}^\infty\, d\xi\, G_k(\xi) e^{i\,\xi\,w}&=\sum_{\ell=0}^k\frac{(-2)^\ell}{\ell!}\binom{2\Delta+k-1}{k-\ell}\int_{0}^\infty\,\frac{d\xi}{\xi} \,\xi^{2\Delta+\ell}e^{-(1-iw)\xi}\nonumber\\
&=\frac{\Gamma(2\Delta+k) i^{2\Delta}}{k!(w+i)^{2\Delta}}\sum_{\ell=0}^k \binom{k}{\ell} \left(\frac{-2}{1-i w}\right)^\ell=\frac{\Gamma(2\Delta+k) i^{2\Delta}}{k!}\phi_k(w)
\end{align}

Finally, we are at a stage of actually evaluating the character associated with $L_{12}$ by using the new basis $G_k(\xi)$ of $\CH_\Delta$: 
\begin{align}
\Tr e^{it L_{12}}&=\sum_{k\ge 0}\frac{(G_k, e^{it L_{12}} G_k)}{(G_k, G_k)}
\end{align}
where $(e^{it L_{12}} G_k)(\xi)=e^{(1-\Delta)t}G_k(e^t\xi)$ is obtained by exponentiating the action of $L_{12}$ in (\ref{action}) and $(G_k, G_k)=\frac{\Gamma(2\Delta+k)}{2^{2\Delta} \, k!}$ is a result of the orthogonality of generalized Laguerre polynomial:
\begin{align}
\int_0^\infty  x^\alpha  L^{(\alpha)}_n\,L^{(\alpha)}_m e^{-x}\,dx=\frac{\Gamma(n+\alpha+1)}{n!}\delta_{nm}
\end{align} 
Thus the character $\Tr e^{it L_{12}}$ can be expressed as
\begin{align}
\Tr e^{it L_{12}}=e^{\Delta t}\sum_{k\ge 0}\frac{k!}{\Gamma(2\Delta+k)}\int_0^\infty\, d\xi \xi^{2\Delta-1}e^{-\frac{1+e^t}{2}\xi} L^{(2\Delta-1)}_k(\xi) L^{(2\Delta-1)}_k(e^t \xi)
\end{align}
If we switch the order of summation and integration mindlessly, the sum is not convergent. To makes sense of this procedure, we introduce a factor $(1-\delta)^k, \delta>0$ 
\begin{align}
\sum_{k\ge 0}&\frac{k!}{\Gamma(2\Delta+k)}L^{(2\Delta-1)}_k(\xi) L^{(2\Delta-1)}_k(e^t \xi)(1-\delta)^k\nonumber\\
&=\frac{(1-\delta)^{\frac{1}{2}-\Delta}}{\delta}\frac{e^{(\frac{1}{2}-\Delta) t}}{\xi^{2\Delta-1}}e^{-(\frac{1}{\delta}-1)(1+e^t)\xi} I_{2\Delta-1}\left(\frac{2e^{\frac{t}{2}}\sqrt{1-\delta}}{\delta}\xi\right)
\end{align}
This equation is called ``Hardy-Hille formula'' \cite{Szego}. With the summation regularized and evaluated, the remaining integral can be computed by using the result on page 91 of \cite{mathphys}
\begin{align}
\Tr e^{it L_{12}}&=\frac{(1-\delta)^{\frac{1}{2}-\Delta}}{\delta}e^{\frac{t}{2}}\int_0^\infty d\xi e^{-A \,\xi} I_{2\Delta-1}(B \xi)\nonumber\\
&=(1-\delta)^{\frac{1}{2}-\Delta}e^{\frac{t}{2}}\left(\frac{B}{A+\sqrt{A^2-B^2}}\right)^{2\Delta-1}\frac{1}{\delta\sqrt{A^2-B^2}}
\end{align}
where 
\begin{align}
A=\left(\frac{2}{\delta}-1\right)\cosh(t/2)e^{\frac{t}{2}}, \,\,\,\,\, B=\frac{2\sqrt{1-\delta}}{\delta}e^{\frac{t}{2}}
\end{align}
Expanding around $\delta=0$ and keeping the leading term yield
\begin{align}
\Tr e^{it L_{12}}=\frac{(\cosh(t/2)+\sinh(|t|/2))^{1-2\Delta}}{2\sinh(|t|/2)}
\end{align}
When $t>0$, it's reduced to $\Tr e^{it L_{12}}=\frac{e^{-\Delta t}}{1-e^{-t}}$ and when $t<0$, it's $\Tr e^{it L_{12}}=\frac{e^{\Delta t}}{1-e^{t}}$. Altogether, the character associated to the noncompact generator $L_{12}$ can be summarized as
\begin{align}\label{SO21}
\Tr e^{i tL_{12}}=\frac{e^{-\Delta |t|}}{1-e^{-|t|}}=\Tr e^{-|t|H}
\end{align}
This equation also holds if $L_{12}$ is replaced by $L_{10}$ since they are related by a conjugation of $H$.

\subsection{$\SO(2, d)$ character}
In higher dimensional, we fix an $\so(2,1)$ subalgebra, i.e. generated by $\{H, L_1^\pm\}$ and decompose the $\so(2, d)$-invariant Hilbert space $\CH_\Delta^{\text{AdS}_{d+1}}$ into $\so(2,1)$-invariant subspaces where the formula  (\ref{SO21}) can be used. For concreteness, we use $\so(2,2)$ scalar representations to illustrate how this decomposition procedure works. Let $|\Delta\rangle$ be the scalar primary state of an $\so(2,2)$ representation
\begin{align}
H|\Delta\rangle=\Delta|\Delta\rangle, \,\,\,\, L^-_{1}|\Delta\rangle=0, \,\,\,\, L^-_{2}|\Delta\rangle=0, \,\,\,\,M_{12}|\Delta\rangle=0
\end{align}
Decomposition of  $\CH_\Delta^{\text{AdS}_{3}}$ into $\so(2,1)$-invariant subspaces  is equivalent to finding $\so(2,2)$ descendants that are $\so(2,1)$ primary. Suppose $|\psi_n\rangle=\sum_{k=0}^n c_k (L_1^+)^{n-k}(L_2^+)^k|\Delta\rangle$ with $c_n=1$ is such a state. Then the $\so(2,1)$  primary condition $L_1^-|\psi\rangle=0$ imposes a nontrivial recurrence relation relation on $c_k$
\begin{align}
(n-k)(2\Delta+n+k-1)c_k=(k+1)(k+2) c_{k+2}
\end{align}
which fixes the coefficients $\{c_k\}$ completely, for example 
\begin{align}
& |\psi_1\rangle=L_2^+|\Delta\rangle, \,\,\,\, |\psi_2\rangle=(L_2^+)^2|\Delta\rangle+\frac{1}{2\Delta+1}(L_1^+)^2|\Delta\rangle\nonumber\\
& |\psi_3\rangle=(L_2^+)^3|\Delta\rangle+\frac{3}{2\Delta+3}(L_1^+)^2\,L_2^+ |\Delta\rangle\nonumber\\
& |\psi_4\rangle=(L_2^+)^4|\Delta\rangle+\frac{6}{2\Delta+5}(L_1^+)^2\,(L_2^+)^2 |\Delta\rangle+\frac{3}{(2\Delta+3)(2\Delta+5)}(L^+_1)^4|\Delta\rangle
\end{align}
Each $|\psi_n\rangle$ induces an $\so(2,1)$ representation $\CH^{\text{AdS}_2}_{\Delta+n}$ of scaling dimension $\Delta+n$.  Thus we have 
\begin{align}\label{inc}
\CH^{\text{AdS}_3}_{\Delta}\supseteq\bigoplus_{n\ge0}\CH^{\text{AdS}_2}_{\Delta+n}
\end{align}
Counting the dimension of $H$-eigenspace with eigenvalue $\Delta+K$ for any $K\in\mathbb{N}$ on the two sides of (\ref{inc}), we find it should be an isomorphism of vector spaces
\begin{align}\label{decom1}
\CH^{\text{AdS}_3}_{\Delta}\cong\bigoplus_{n\ge0}\CH^{\text{AdS}_2}_{\Delta+n}
\end{align}
Applying the $\SO(2,1)$ character formula (\ref{SO21}) to the decomposition (\ref{decom1}) yields the $\SO(2,2)$ character associated with $L_{10}$
\begin{align}
\Tr_{\CH^{\text{AdS}_3}_{\Delta}} e^{i t L_{10}}=\sum_{n=0}^\infty \Tr_{\CH^{\text{AdS}_2}_{\Delta+n}} e^{i t L_{10}}=\frac{e^{-\Delta|t|}}{(1-e^{-|t|})^2}
\end{align}
In higher dimensions, the decomposition formula (\ref{decom1}) is generalized to 
\begin{align}\label{decom2}
\CH^{\text{AdS}_{d+1}}_{\Delta}\cong\bigoplus_{n\ge0}\left(\CH^{\text{AdS}_2}_{\Delta+n}\right)^{\oplus D_n}, \,\,\,\, D_n=\binom{d+n-2}{n}
\end{align}
and thus the corresponding character becomes 
\begin{align}
\Tr_{\CH^{\text{AdS}_{d+1}}_{\Delta}} e^{i t L_{10}}=\sum_{n=0}^\infty D_n \, \Tr_{\CH^{\text{AdS}_2}_{\Delta+n}} e^{i t L_{10}}=\frac{e^{-\Delta|t|}}{(1-e^{-|t|})^d}
\end{align}
Though we've only computed character of a noncompact generator for scalar representations in this appendix, we believe
\begin{align}
\Tr_{\CH_{[\Delta, \vec s]}} e^{i t L_{10}}=\Tr_{\CH_{[\Delta, \vec s]}} q^H, \,\,\,\,\, q=e^{-|t|}
\end{align}
holds for any unitary representation $[\Delta, \vec s]$, massive or massless.

\section{Physics of $\SO(2, d)$ character}\label{GG}
We will try to build up some physical intuitions about the character $\chi(t)\equiv \Tr e^{i t L_{d, d+1}}$ that is computed by brutal force  in the last appendix.
In section \ref{QNM}, we construct quansinormal modes in Rinder-AdS and show that they are counted by the character $\chi(t)$. In section \ref{coarse}, we compute the density of $L_{21}$ eigenstates numerically in $\text{AdS}_2$ by imposing an upper bound on the eigenvalues of the global Hamiltonian $H$ and compare it with the density of states defined as the Fourier transformation of $\chi(t)$.

\subsection{Quasinormal modes in Rindler-AdS}\label{QNM}
It is clear that the  character $\Tr e^{ i t H}, \Im t>0$ counts normal modes in $\text{AdS}_{d+1}$ because $H$ is the Hamiltonian in global coordinate. However, the same interpretation does not hold for $\Tr e^{i t L_{d, d+1}}$ because $L_{d, d+1}$ is not  a positive definite opeator. Instead,  as we will show in the following, it counts resonances/quasinormal modes in Rindler-AdS.

To construct quasinormal modes in an efficient algebraic way \cite{Sun:2020sgn}, it's convenient to define the following dS-type conformal generators:
\begin{align}
D= -i L_{d, d+1},\,\,\,\, P_\mu=i( L_{\mu, d}+L_{\mu, d+1}),\,\,\,\, K_\mu= i(L_{\mu, d}-L_{\mu, d+1})
\end{align}
subject to commutation relations (we only show the nontrivial ones that will be used in the derivation):
\begin{align}
[D, P_\mu]=P_\mu, \,\,\,\,\, [D, K_\mu]=-K_\mu
\end{align}
where $0\le \mu \le d-1$. 
In terms of  embedding space coordinates, the differential operator realization of $D, P_\mu, K_\mu$ is 
\begin{align}
D=-(X^d\partial_{X^{d+1}}&+X^{d+1}\partial_{X^d}), \,\,\,\, P_\mu= X_\mu(\partial_{X^d}+\partial_{X^{d+1}})+(X^{d+1}-X^d)\partial_{X^\mu},\,\,\,\,\nonumber\\
&K_\mu= X_\mu(\partial_{X^d}-\partial_{X^{d+1}})-(X^{d+1}+X^d)\partial_{X^\mu}
\end{align}
Consider a scalar field of scaling dimension $\Delta$. Then its ``primary mode'' i.e. eigenfunction  of $D$ with eigenvalue $\Delta$ which is annihilated by $K_\mu$, is given by 
\begin{align}
\psi_\Delta (X) \equiv \left(X^d+X^{d+1}\right)^{-\Delta}
\end{align}
    In the  southern Rindler coordinate of AdS \footnote{In the $\text{AdS}_2$ case, we should actually use the black hole coordinate (\ref{BHcoord}) and then the ``primary mode'' becomes $\psi_\Delta(t_S, \rho)=(\rho^2-1)^{-\Delta/2}\, e^{-\Delta \, t_S}$.}, this ``primary mode'' $\psi_\Delta (X)$ descends to
\begin{align}
\psi_\Delta(\eta, t_S, r)=(\sinh\eta)^{-\Delta} (1-r^2)^{-\frac{\Delta}{2}} e^{-\Delta t_S}
\end{align}
    $\psi_\Delta$ has $e^{-\Delta\eta}$-type fall-off at the future boundary and satisfies in-going boundary condition at horizon with quasinormal frequency identified as $i\omega=\Delta$. The other quasinormal modes are  descendants of $\psi_\Delta$\footnote{There are two different definitions  of descendants depending on the choice of   Hamiltonian: either $L_{0, d+1}$ or $L_{d+1, d}$. Since  the Hamiltonian in Rindler-AdS is $L_{d+1,d}$, the descendants are obtained by acting $P_\mu$ on $\psi_\Delta$.} because the equation of motion and in-going boundary condition are invariant under the action of $\SO(2,d)$. At level $n$, there are $\binom{n+d-1}{d-1}$ linearly independent quasinormal modes whose quasinormal frequency $\omega_n$ is related to their scaling dimension under $D$ by $i\omega_n = \Delta+n$. Notice that 
\begin{align}
\Tr e^{-t D}=\frac{e^{-\Delta t}}{(1-e^{-t})^d}=\sum_{n\ge 0}\binom{n+d-1}{d-1}e^{-i\omega_n t}, \,\,\,\, t>0
\end{align}
and thus the character $\Tr e^{-t D}$ counts quasinormal modes. The same construction can be easily generalized to higher spin fields, either massive of massless.

\subsection{Numerical computation of density of state}\label{coarse}
For the southern Rindler-AdS Hamiltonian $L_{d+1,d}$, the associated density of (single-particle) states can be formally defined as 
\begin{align}
\rho(\omega)=\tr \delta(L_{d+1, d}-\omega), \,\,\,\,\, \omega>0
\end{align}
which is also a Fourier transformation of the character $\chi(t)=\Tr e^{i t L_{d,d+1}}$
\begin{align}
\rho(\omega)=\int\frac{dt}{2\pi}\, \chi(t) \, e^{i\omega t}=\int_{0}^\infty \frac{dt}{2\pi}\, \chi(t)\left(e^{i\omega t}+e^{-i\omega t}\right)
\end{align}
The Fourier transformation above is UV-divergent but it can be easily regularized by using a hard cutoff $\frac{1}{\Lambda}$ for the lower bound of the $t$ integral. For example, for a scalar field of scaling dimension $\Delta$ in $\text{AdS}_2$, this regularization yields 
\begin{align}\label{conrho}
\rho_\Lambda(\omega)=\frac{1}{\pi}\log\left(e^{-\gamma_E} \Lambda\right)-\frac{1}{2\pi}\sum_{\pm}\psi(\Delta\pm i\omega)
\end{align}
where $\gamma_E$ is the Euler constant and $\psi(x)=\Gamma'(x)/\Gamma(x)$ is the digamma function.

\begin{figure}[t]
\centering
\begin{subfigure}{.4\textwidth}
  \centering
  \includegraphics[width=1.0\linewidth]{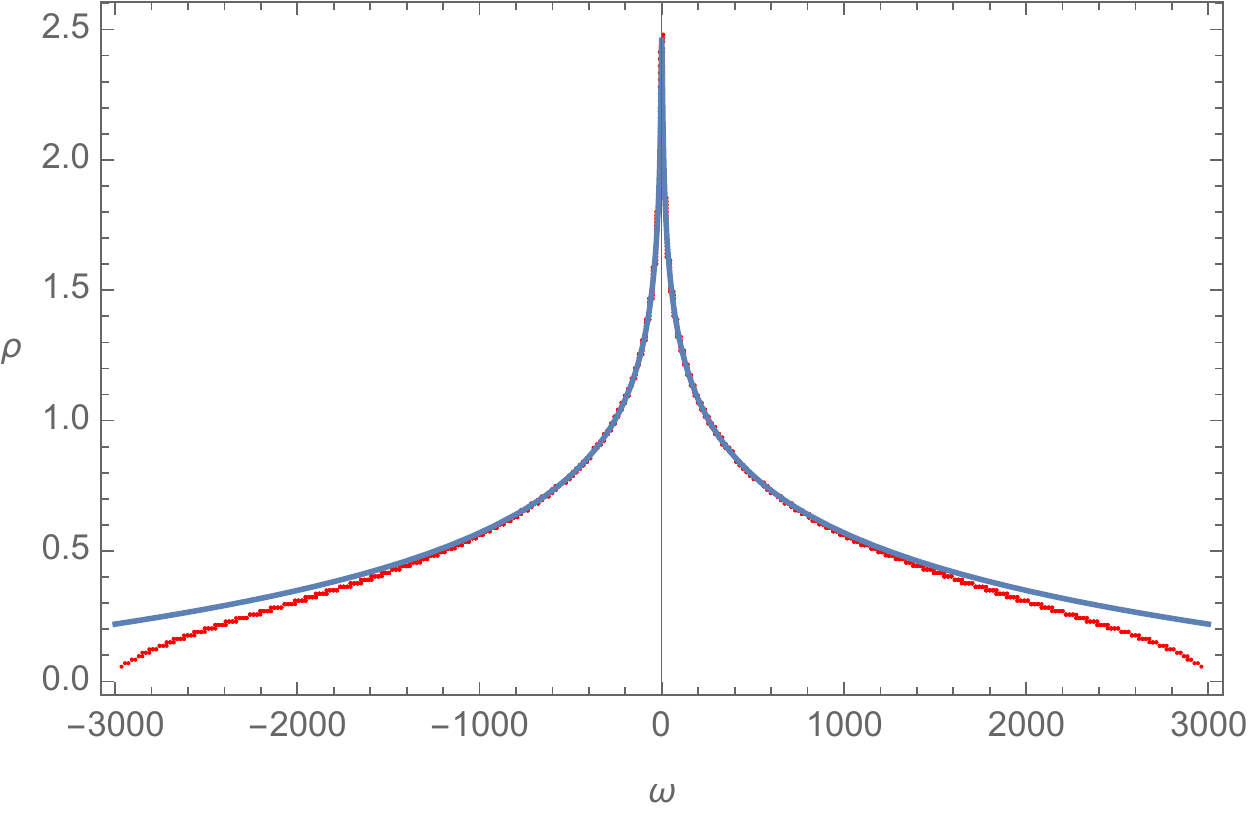}
  %\caption{odd $d$}
  \label{fig:com1}
\end{subfigure}\,\,\,\,\,
\begin{subfigure}{.4\textwidth}
  \centering
  \includegraphics[width=1.0\linewidth]{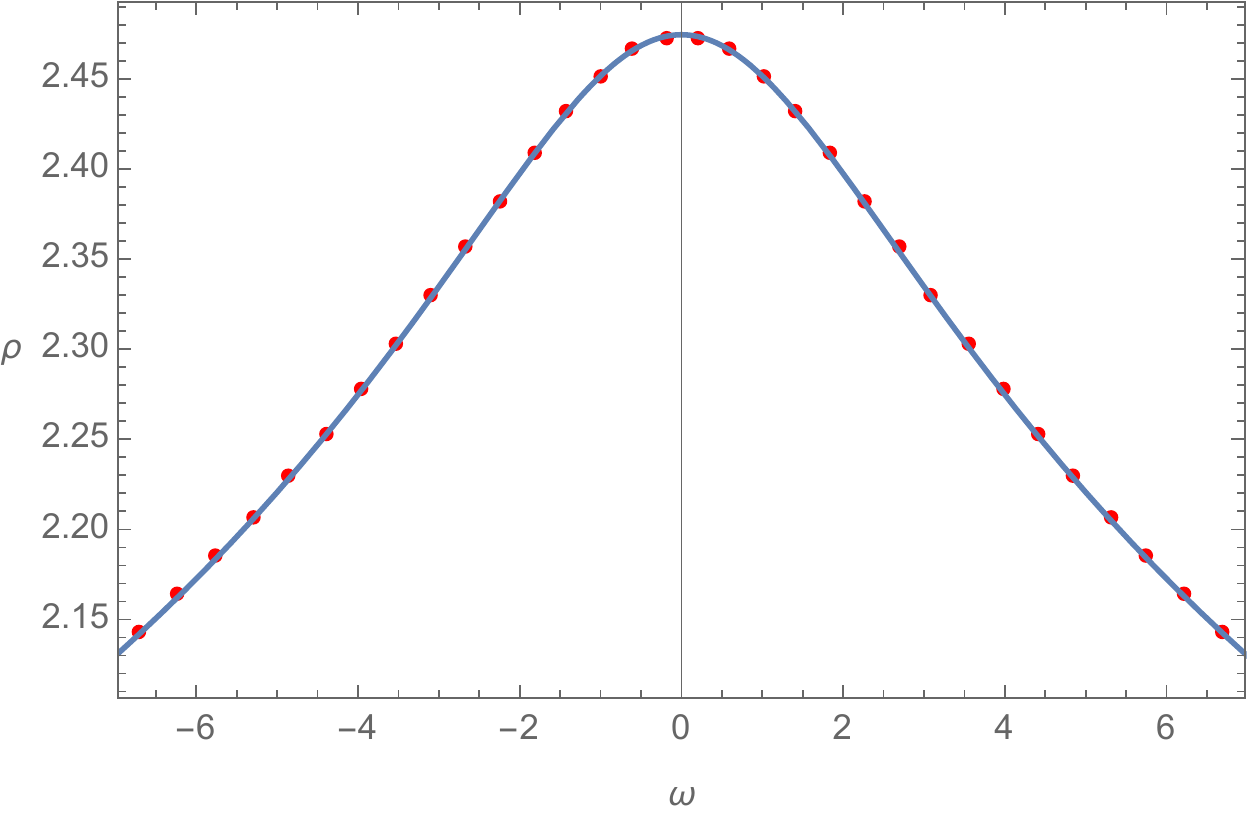}
 % \caption{even $d$}
  \label{fig:com2}
\end{subfigure}
\caption{Density of states for a $\Delta=3$ scalar in $\text{AdS}_2$. The red dots show the coarse-grained density of states $\bar\rho_K(\omega)$, cf. (\ref{disrho}), of the truncated model with global energy cut-off $K=2999$. The blue lines show the character induced density of states $\rho_\Lambda(\omega)$, cf. (\ref{conrho}),  with the UV cut-off being $e^{-\gamma_E}\Lambda\approx 5981$.  The plot on the left shows the two densities for the full spectrum of the truncated model while the plot on the right zooms in on the deep IR region.}
\label{fig:comparison}
\end{figure}

On the other hand, approximating $\rho_\Lambda(\omega)$ by a model of finite dimensional Hilbert space would provide a more physical interpretation for it. Such an approximation can be easily implemented by imposing a UV cutoff on the spectrum of $H$.  For example in the $\text{AdS}_2$ case, consider a truncated Hilbert space $\CH_K$ generated by $G_k(\xi)$ with $0\le k\le K$ and thus the highest energy of $H$ is $\Delta+K$. Normalizing $G_k(\xi)$ and using the recurrence relations (\ref{recurL}),  $L^\pm$ are realized as finite dimensional matrices in $\CH_K$
\begin{align}
L^+_{k+1, k}=-\sqrt{(k+1)(k+2\Delta)}, \,\,\,\,\, L^-_{k, k+1}=-\sqrt{(k+1)(k+2\Delta)}
\end{align}
Altogether, in this truncated model, the noncompact ``Hamiltonian'' $L_{21}=\frac{i}{2}(L^--L^+)$ is a sparse $(K+1)\times (K+1)$ matrix, which admits an efficient numerical diagonalization. With the eigenspectrum $\{\omega_k\}_{0\le k\le K}$ (which is ordered such that $\omega_{k+1}\ge \omega_k$) obtained from diagonalization, a coarse-grained density of eigenstates can be defined as 
\begin{align}\label{disrho}
\bar\rho_K(\omega_k)\equiv \frac{2}{\omega_{k+1}-\omega_{k-1}}
\end{align}
To compare the character induced density $\rho_\Lambda$ and the discretized density $\bar\rho_K$ for a fixed $K$, we adjust the UV cut-off $\Lambda$ such that they coincide around $\omega\approx 0$, i.e. more precisely at the lowest non-negative eigenvalue of $L_{21}$.  Such a comparison for $\Delta=3$ and $K=2999$ is shown in fig. \ref{fig:comparison}. They agree fairly well in the IR region and hence the UV truncated model is a pretty good approximation for computing density of states.

\def\refskip{

%%%%% CLEAR DOUBLE PAGE!
\newpage{\pagestyle{empty}\cleardoublepage}

\end{document}